\documentclass{jfm}

\usepackage{graphicx}
\usepackage{newtxtext}
\usepackage{newtxmath}
\usepackage{natbib}
\usepackage{hyperref}
\usepackage{amsmath}

\usepackage[final]{changes}
\definechangesauthor[name={Dominique Heitz}, color=orange]{dh}

\hypersetup{
    colorlinks = true,
    urlcolor   = blue,
    citecolor  = black,
}

\newcommand{\RomanNumeralCaps}[1]
\linenumbers

\makeatletter
\def\absfooterflag{}
\def\pagelimitfooter{}
\makeatother

\title{Physics-informed coherent motions to predict Lagrangian trajectories}

\author{Ali R Khojasteh\aff{1,2}
  \corresp{\email{a.r.khojasteh@tudelft.nl}},
   \and Dominique Heitz\aff{1}}

\affiliation{

\aff{1}INRAE, OPAALE, 17 avenue de Cucill\'e, 35044, Rennes, France

\aff{2}Laboratory for Aero and Hydrodynamics, Delft University of Technology, 2628 CD Delft, \\
The Netherlands
}

\begin{document}

\maketitle

\begin{abstract}

Accurate prediction of Lagrangian trajectories in turbulent flow remains challenging due to limited temporal information in transport functions. This paper shows that surrounding coherent motions sharing the same dynamics carry enough information to provide highly probable trajectories even from sparse temporal observations. The proposed \emph{coherent predictor} builds on Lagrangian coherent structures (LCSs), the advective transport barriers that govern the cohesive motion of neighbouring particles. Coherent trajectories are quantified using a local segmentation with the finite-time Lyapunov exponents (FTLE). The coherent predictor incorporates information from the particle's position history and neighbouring coherent velocity and acceleration into a novel cost function to predict its trajectory. The proposed cost function follows a physics-informed approach where the position history acts as a data fidelity term and the coherent velocity and acceleration act as physics-based regularisation constraints. We assess our proposed approach using both three-dimensional (3D) synthetic and experimental data of the wake behind a smooth cylinder and two-dimensional (2D) homogeneous isotropic turbulent (HIT) flow. The coherent predictor is deemed generic due to its consistent behaviour regardless of flow dimensions, Reynolds number, and flow topology. Our results show that the optimal cost function parameters can be modelled from the measurement uncertainties, giving lower prediction error and uncertainty than current methods. We see direct signatures of flow topology on the prediction error map, including the cylinder leading edge boundary layer, the sideward shear layers, and the vortex formation structures. These topologies are marked by high Lagrangian gradients and 3D directional motions.

\end{abstract}

\begin{keywords}
Authors should not enter keywords on the manuscript, as these must be chosen by the author during the online submission process and will then be added during the typesetting process (see \href{https://www.cambridge.org/core/journals/journal-of-fluid-mechanics/information/list-of-keywords}{Keyword PDF} for the full list).  Other classifications will be added at the same time.
\end{keywords}

{\bf MSC Codes }  {\it(Optional)} Please enter your MSC Codes here

\section{Introduction}
\label{sec:intro}
The transport of tracer particles, which forms Lagrangian trajectories, is a fundamental problem in fluid dynamics, and \textit{Lagrangian particle tracking} (LPT) is now a standard tool for quantifying the motion of these fluid particles. Applications of Lagrangian trajectories span various fields in physics and engineering, including the mixing process and diffusion properties of tracer particles \citep{Viggiano2021LagrangianJet}, vortex shedding topology analysis \citep{Gold2023LagrangianUnderwater}, turbulent bubbly jet \citep{Kim2022ExperimentalWater}. All these applications require trajectory reconstruction that precisely represents Lagrangian motion before undertaking any further analysis. For decades, LPT studies have focused on reconstructing tracer particle trajectories by treating them as isolated signals, separate from their surroundings \citep{Schroder20233DMechanics}. While the use of predictive models has expanded the spatial and temporal capabilities of reconstructing dense and complex Lagrangian trajectories \citep{Schroder20233DMechanics,Schanz2016Shake-The-Box:Densities}, the open problem is to precisely predict the motion of tracer particles as representatives of Lagrangian dynamics in turbulent and chaotic flows. Trajectory prediction refers to a function that receives tracer positions and estimates the positions of the next time step to reconstruct the Lagrangian trajectory. These techniques rely solely on mathematical models without incorporating physics-based constraints and keep treating particles as isolated signals. Physics-informed approaches, where physical observations regularise a prediction model, have shown improvements across computational physics \citep{Karniadakis2021PIML,Raissi2019PINN} and have recently entered the particle tracking field for dense velocity reconstruction \citep{Cai2024PINNPTV}, but such a framework has not yet been applied to Lagrangian trajectory prediction itself. The main reason why Lagrangian trajectories are subject to error and uncertain reconstructions is the temporal resolution in experiments. \citet{Khojasteh2021LagrangianInitialization} showed the temporal resolution could be over an order of magnitude larger than the Kolmogorov time scale due to experimental limitations. Consequently, we miss numerous small-scale motions before the next observation (i.e., camera image) is recorded, resulting in a loss of Lagrangian information between two observations. Thus, depending on position history alone is insufficient to fully recover the Lagrangian trajectory with certainty. Lagrangian prediction extends beyond LPT experiments where explicit time-stepping methods, such as the Euler and Runge Kutta schemes, are used in ocean analyses to determine the transport of tracer particles based on Eulerian simulations \citep{vanSebille2018LagrangianPractices,ValdiviesoDaCosta2004LagrangianAssessment}. Due to the high computational cost and complexities of performing \textit{direct numerical simulation} (DNS), researchers often use coarse models with sparse temporal resolutions such as the Oceanic General Circulation Model for the Earth Simulator \citep[OFES][]{Masumoto2004ASimulator}. As \citet{Qin2014QuantificationModel} showed, this can become a significant source of error in the temporal interpolation of simulated velocity fields. This is due to a lack of temporal information in the transport function, highlighting a common issue in Lagrangian studies across various fields. Yet, even with sparse temporal observations, there might be enough information from surroundings to provide highly probable trajectories. If successful particle position predictions are possible in turbulent flows, they can be obtained thanks to suitable coherent neighbours. This is the main motivation of the paper. Accordingly, all predictions are single-step forecasts.

Knowing that trajectories are characterised by \textit{Lagrangian coherent structures} (LCS) acting as barriers between regions in which tracer particles have different kinematics, we propose a novel approach considering groups of coherent particles, rather than an isolated approach, to estimate Lagrangian particle positions. Numerous techniques have been developed to diagnose LCS barriers and understand fluid transport where particles are generally advected by the flow, from trajectory-based coherent segmentation to vector displacement-based techniques \citep{Haller2023TransportData, Mowlavi2022DetectingData,Hadjighasem2017ADetection,Green2007DetectionTurbulence}. The concept of coherency in Lagrangian trajectories refers to regions with similar dynamics while the flow evolves. \citet{Haller2000LagrangianTurbulence} introduced direct Lyapunov exponents and their ridges as indicators of hyperbolic LCSs. This finding led to more investigations into applying nonlinear dynamical system theory to fluid flows. Later on, \citet{Shadden2005DefinitionFlows} studied how passive \textit{finite-time Lyapunov exponents} (FTLE) ridges are the same as LCS material lines and boundaries of coherent motions. FTLE is a quantitative metric that determines how to extract coherent structures in the Lagrangian framework. These coherent structures are objective and have the most robust representations of trajectories that act as skeletons of flow \citep{MacMillan2021TheStructures}. Objective means the material responses of their surface (or line) transport barriers are independent of the observer \citep{Haller2023TransportData,Haller2015LagrangianStructures}. The further theory was introduced by \citet{Haller2020ObjectiveFields} to objectively quantify the transport of active vector fields such as momentum and vorticity. We propose a prediction of Lagrangian particle positions that uses LCS boundaries to control the transport of tracer particles. Using LCS, we identify coherent and non-coherent neighbour trajectories and local geometric separatrices, and we then estimate the dynamics of trajectories in their coherent surroundings. Specifically, we use FTLE to quantify the boundaries between separated regions, since the method has prior experimental support \citep[see, e.g.,][]{Eisma2021DoLayer} and has been shown to identify coherent structures reliably \citep{Balasuriya2016HyperbolicStretching,Hadjighasem2017ADetection}. 

We adopt a local segmentation approach instead of computing the global FTLE field to classify neighbours as coherent or non-coherent. The local FTLE approach offers a more targeted analysis of the particle's immediate surroundings, reduces computational and three-dimensional (3D) transport barrier quantification complexities, and provides a precise understanding of the coherent motions in the vicinity of particles. We introduce primary and secondary coherent neighbours in turbulent flows. Primary coherent neighbours travel along a similar path as the target particle within the same time interval. In contrast, secondary coherent neighbours exhibit a phase delay and are situated ahead in their history relative to the target particle. This secondary information, representing future knowledge for the target particle, provides valuable prior information for predicting Lagrangian trajectories. We, therefore, define a physics-informed cost function called \emph{coherent predictor} that considers the history of trajectories and local coherent motions for prediction with the least biased and low uncertainty level compared to the state-of-the-art. We show that the coherent predictor can be modelled as a function of measurement uncertainty regardless of the flow case. That means the proposed approach remains generic to the best of available test cases in the present study. By "generic," we mean that the behaviour of the coherent predictor remains independent of flow dimensions (2D or 3D), Reynolds number, and flow topology. The coherent predictor offers valuable insights into Lagrangian physics by examining groups of coherent motions, leading to a deeper understanding of how effectively they share the same dynamics even in low temporal resolutions. Precise prediction also allows for the reconstruction of longer trajectories instead of split tracklets throughout the entire observation event, which provides valuable temporal information for Lagrangian statistics analysis.

This paper is structured as follows. In \S~\ref{sec:p2}, we introduce physics-based coherent terms as an additional constraint to the mathematical prediction model in the form of a cost function, followed by local segmentation using the FTLE metric. Then \S~\ref{sec:p3} presents the numerical and experimental flow configurations used in this study, including DNS for two-dimensional (2D) \textit{homogeneous isotropic turbulent} (HIT) flow at a Reynolds number of $3000$, the DNS of the cylinder wake flow at a Reynolds number of $300$, and both DNS and experimental investigations for the 3D wake flow behind a smooth cylinder at a Reynolds number of $3900$. We discuss the cost function and its minimisation procedure in \S~\ref{sec:p41} and present the sensitivity and confidence analyses of Lagrangian position estimation based on coherent motions in \S~\ref{sec:p42} and \S~\ref{sec:p43}, respectively. In \S~\ref{sec:p42b}, we quantify the contribution of the secondary coherent neighbours on top of the primary ones. We then test our proposed approach on a real particle tracking experiment in \S~\ref{sec:p44}, where we find the smallest deviation from the optimised positions. Finally, we provide a summary of our conclusions in \S~\ref{sec:p5}, and three appendices collect the cost function derivation (Appendix~\ref{Appendix_A}), the 3D-wake robustness analyses (Appendix~\ref{appB}), and the physics-informed neural network with temporal collocation (Appendix~\ref{appC}).


\section{Lagrangian position estimation}
\label{sec:p2}
 In the Lagrangian framework, we study fluid motion from the perspective of individual fluid particles as they move through space and time. Within this framework, we can build prediction functions that capture the complex dynamics of particles within a fluid. The simplest prediction approach is using a polynomial function, suggested by \citet{Schanz2013ShakePositions}, resulting in reasonable 3D trajectory reconstructions in simple flows \citep{Schroder2015AdvancesHills,Schanz2014ShakeFlows,Schanz2013ShakePositions,Schanz2016Shake-The-Box:Densities}. Significant misprediction occurs in the case of flow associated with complexities such as high turbulence level and high Reynolds number \citep{Tan2020IntroducingTracking}. In such conditions, even by increasing the polynomial order from three to ten, mispredictions remain \citep{Tan2020IntroducingTracking}. As an alternative, \citet{Schroder2015AdvancesHills} introduced optimal temporal filtering of particle positions, such as the Wiener filter, in Lagrangian tracking experiments. Since then, this concept has become widely adopted in the \textit{time-resolved particle tracking velocimetry} (4D-PTV) studies \citep{Tan2020IntroducingTracking,Schroder2015Near-wallShake-The-Box}. The Wiener filter showed robust behaviour in prediction with complex flows, such as inside the turbulent boundary layer \citep{Knopp2021ExperimentalGradient}. However, it still suffers from high-motion gradients because the temporal filter cancels them unless an additional constraint is imposed. This implies that the prediction function lacks sufficient information to estimate particle dynamics. Unlike tracer particles, Lagrangian fibres provide broad details of their dynamics, including orientation and rotation rates \citep{Alipour2021LongFlow}. This additional information on Lagrangian dynamics helps the prediction. For individual tracer particles, our knowledge is limited to their histories. Despite implementing filtering and smoothing schemes such as \textit{Shake-the-Box} (STB) using Wiener filter \citep{Schroder2015Near-wallShake-The-Box}, our prediction is limited by the history of the target particle, ignoring every Lagrangian trajectory is spatially and temporally coherent with a specific group of other tracer particles following the same behaviour. It is not quantitatively studied after what turbulence intensity level or velocity gradient the predictor function fails. This isolated prediction approach yields spurious trajectory reconstructions and fails to represent true Lagrangian dynamics. Therefore, any further analysis would yield uncertain and inaccurate results, particularly in complex flow regions such as wakes, impinging jets, and regions near solid boundaries \citep{Khojasteh2021LagrangianInitialization,Kernel,Novara2023TwoPulseSTB}. 

\begin{figure}
\centering
  \includegraphics[width=0.85\textwidth]{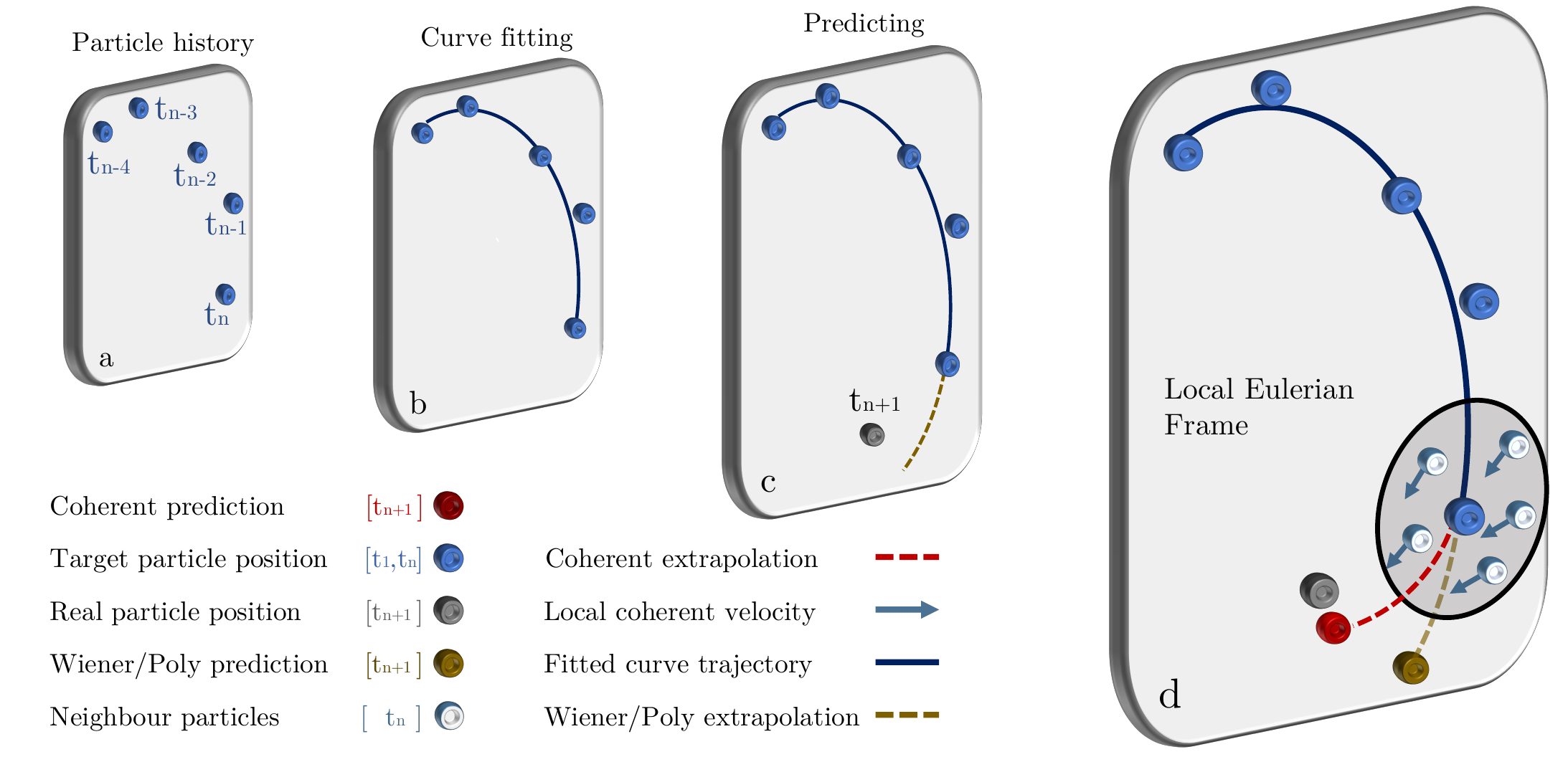}
\caption{Particle position prediction scenario from $t_{n}$ to $t_{n+1}$. (a) Known particle positions from history starting from $t_{n-4}$ up to $t_{n}$. (b) Trajectory (dark blue line) obtained from filtered curve fitting of known particle positions. (c) Prediction based on extrapolating the fitted trajectory (golden dashed line) from $t_{n}$ to $t_{n+1}$. (d) Modified prediction (red dashed line) using velocity and acceleration information of coherent neighbours of the target particle at $t_{n}$.}
\label{fig:p1}       
\end{figure}

The prediction function comprises a set of coefficients and independent variables. The coefficients are used to predict the outcome of a dependent variable, which can be written as a cost function. Therefore, the objective is to minimise the cost function to reach the optimal trajectory prediction at the next time step. Essentially, position histories of trajectories are the main input ingredients of the prediction process, as shown in figure~\ref{fig:p1}.a.b. The predictor subsequently fits a smooth curve over the noisy history $\left\{t_1,\cdots,t_n\right\}$ of the target particle and estimates its possible position at time step $t_{n+1}$ (see figure~\ref{fig:p1}.c). At this stage, we categorise recently available and the proposed prediction functions based on the input information they require into position-based and Lagrangian coherent cost functions, discussing the former in \S~\ref{sec:p21} and the latter in \S~\ref{sec:p22}. The Lagrangian coherent cost function seeks a solution in which the estimated position satisfies neighbouring coherent motions (see figure~\ref{fig:p1}.d). By bringing Lagrangian physics into the prediction function, we show that even a weak signal from the motion of just one coherent neighbour can improve the particle prediction accuracy in complex flow regions, which improves our understanding of the interactions and transport in such flows. 

\subsection{Position-based cost function}
\label{sec:p21}

Two main position-based methods have been reported in particle tracking algorithms: polynomial and Wiener filter predictors \citep{Kernel,Tan2020IntroducingTracking,Schanz2016Shake-The-Box:Densities}. Both techniques rely solely on the history of the target particle. The polynomial predictor aims to minimise the least mean square error of the history to find the optimal polynomial coefficients and then extrapolate the function with the same coefficients to the next time step. In the Wiener filter approach, we design and adjust filter parameters based on history and then shift the designed filter solution to the next time step. Both polynomial and Wiener filter predictor functions require two parameters to set: the polynomial or Wiener filter order and the length of history to perform the Linear regression analysis. The integral time scale can be used to set the duration of the regression window (history). Then, we should select the order of the polynomial or Wiener filter. 

Polynomial coefficients must be optimally determined by minimising the mean square error, so that the corresponding polynomial curve with an order of $\ell$ best fits the given positions. The polynomial model can be written as
\begin{equation}
\begin{aligned}
\mathbf{X}_n=\sum_{j=0}^{\ell}\mathbf{a}_{j}.t_n^{j},
\end{aligned}
\label{Eq:p21}       
\end{equation}
where $\left\{\mathbf{a}_{0}, \cdots, \mathbf{a}_{\ell}\right\}$ are unknown coefficients of the predictor function and $\mathbf{X}_n=\mathbf{X}(t_n)$ is the estimation of the polynomial position at the time step $n$. Therefore, the least square cost function is
\begin{equation}
\begin{aligned}
\mathcal{J}=\underbrace{\frac{1}{k}\sum_{i=n-k+1}^{n}\lVert\mathbf{X}_i-\mathbf{y}_i\rVert^2}_\text{particle history},
\end{aligned}
\label{Eq:p22}       
\end{equation}
where $\mathbf{y}_{i}$ are the last $k$ observed positions of the particles in question, i.e., their history up to the current position at time $n$ (with $k\geqslant \ell+1$). 
The objective is to estimate unknown polynomial coefficients $\mathbf{\hat{a}}=\left\{\mathbf{\hat{a}_{0}}, \cdots, \mathbf{\hat{a}_{\ell}}\right\}$ by minimising the cost function in~\eqref{Eq:p22} as
\begin{equation}
\begin{aligned}
\left\{
    \begin{array}{l}
        \mathbf{\hat{a}}=\underset{\mathbf{a}}{\arg\,\min}~\mathcal{J}(
        \mathbf{X},\mathbf{y}), \\
        \text{such that~} \eqref{Eq:p21}.
    \end{array}
\right.
\end{aligned}
\label{Eq:p23}       
\end{equation}
Then, based on this history-driven polynomial model, estimated from the last $k$ observed positions up to the time $n$, we predict $\mathbf{X}_{n+1}=\mathbf{X}(t_{n+1})$, the particle positions at time $n+1$.

In the finite impulse response Wiener filter approach, as a short-term linear prediction model, we first design a linear estimator (filter) for the history of the target particle. Consider the signal $\mathbf{X}_{n-j}$ (i.e., history) is given to a Wiener filter of order $\ell$ as
\begin{equation}
\begin{aligned}
\mathbf{X}_{n}=\sum_{j=1}^{\ell}\mathbf{w_j}.\mathbf{X}_{n-j},
\end{aligned}
\label{Eq:p24}       
\end{equation}
where $\mathbf{w_j}$ are the filter parameters and the filter output is indicated by $\mathbf{X}_{n}$. Similar to the polynomial solution in~\eqref{Eq:p23}, the objective is to find $\left\{\mathbf{w_{1}}, \cdots, \mathbf{w_{\ell}}\right\}$ filter parameters that minimise a quadratic cost function with the mean square error. The resulting Wiener filter in~\eqref{Eq:p24} is a linear minimum mean square error estimator. Then we predict the future signal value (i.e., particle position) with the designed filter parameters at $t_{n+1}$. Predicting a signal from its past samples depends on the auto-correlation function in~\eqref{Eq:p24}, or equivalently, the signal's bandwidth and power spectrum. \citet{Schanz2016Shake-The-Box:Densities} showed that the Wiener filter can predict Lagrangian trajectories by using the auto-correlation functions mentioned above. On the other hand, a Wiener filter can forecast the amplitude of a signal over a short time interval using a linearly weighted combination of past samples. 

\subsection{Lagrangian coherent cost function}
\label{sec:p22}

Particle trajectories in fluid flows are highly influenced by the coherent structures present in the flow. To improve the prediction of particle positions, it is essential to consider the effects of these coherent structures. The Lagrangian coherent cost function incorporates information from temporal and local spatial coherent motions in the form of extra constraints for the position-based cost function \eqref{Eq:p22}. Each Lagrangian trajectory carries information about position, velocity, and acceleration. We impose the dynamics of neighbouring coherent particles at time $t_n$ into the prediction function through the coherent velocity and acceleration values noted $\dot{\mathbf{y}}_{{\rm c},n}$ and $\ddot{\mathbf{y}}_{{\rm c},n}$, respectively. The imposed coherent velocity and acceleration terms create additional constraints to~\eqref{Eq:p22}. Therefore, the proposed weighted cost function called coherent predictor can be written as
\begin{equation}
\begin{aligned}
\mathcal{J}=\frac{1}{k}\sum_{i=n-k+1}^{n}\lVert\mathbf{X}_i-\mathbf{y}_i\rVert^2+{\alpha_1}\lVert\dot{\mathbf{X}}_n-\dot{\mathbf{y}}_{{\rm c},n}\rVert^2+{\alpha_2}\lVert\ddot{\mathbf{X}}_n-\ddot{\mathbf{y}}_{{\rm c},n}\rVert^2,
\end{aligned}
\label{Eq:p25}       
\end{equation}
where $\alpha_1$ and $\alpha_2$ are dimensional velocity and acceleration weights, respectively. The weighted cost function~\eqref{Eq:p25} contains three terms. The first term is the least mean square minimisation problem of the polynomial predictor based on position history observations $\mathbf{y}_i$. The second and third terms involve neighbouring coherent velocity $\dot{\mathbf{y}}_{{\rm c},n}$ and acceleration $\ddot{\mathbf{y}}_{{\rm c},n}$ observations, whose estimations are detailed in \S~\ref{sec:p221}. The coherent velocity and acceleration terms enter the cost function as Tikhonov-style penalties rather than hard constraints, so the polynomial fit remains feasible when the coherent information is weak or absent.

The cost function~\eqref{Eq:p25} adds up different squared quantities (i.e., differences in position, velocity, and acceleration) with distinct dimensions. $\alpha_1$ and $\alpha_2$ values are directly related to the dimensions of the particular problem considered. They will, therefore, change according to the configurations studied. At this stage, it is difficult to determine the behaviour of their optimal values, i.e., leading to the most accurate position prediction according to the different situations. Our first objective is to reformulate the function by explaining it in relation to the turbulent characteristic scales of the problem so that the weighting parameters become dimensionless. In a second step, in \S~\ref{sec:p41}, we propose to model their behaviour. To obtain a general cost function, we sized the variables with the following scales
\begin{equation}
\begin{aligned}
\mathbf{X}'=\frac{\mathbf{X}}{D} ~,~ {\mathbf{y}}'=\frac{\mathbf{y}}{D} \\
\dot{\mathbf{X}}'=\frac{\dot{\mathbf{X}}}{U} ~,~ \dot{\mathbf{y}}'=\frac{\dot{\mathbf{y}}}{U} \\
\ddot{\mathbf{X}}'=\ddot{\mathbf{X}}~\frac{D}{U^2} ~,~ \ddot{\mathbf{y}}'=\ddot{\mathbf{y}}~\frac{D}{U^2},
\end{aligned}
\label{Eq:p26}       
\end{equation}
where $D$ is an integral length scale, and $U$ is the velocity reference. The integral length scale represents the size of the largest coherent structures in the flow, while $U$ represents the characteristic velocity of the flow. These reference scales non-dimensionalise the cost function, so it can be applied to a wide range of flow configurations. Therefore, the modified cost function becomes

\begin{equation}
\begin{aligned}
\mathcal{J}={D^2}~\left(\frac{1}{k}\sum_{i=n-k+1}^{n}\left(\mathbf{X}_i'-\mathbf{y_i}'\right)^2\right)+{\alpha_1}{U^2}~\left(\dot{\mathbf{X}}_n'-\dot{\mathbf{y}}_{{\rm c},n}'\right)^2+{\alpha_2}{\frac{U^4}{D^2}}~\left(\ddot{\mathbf{X}}_n'-\ddot{\mathbf{y}}_{{\rm c},n}'\right)^2,
\end{aligned}
\label{Eq:p27}       
\end{equation}
which can be simplified as
\begin{equation}
\begin{aligned}
\mathcal{J}'=\underbrace{\frac{1}{k}\sum_{i=n-k+1}^{n}\left(\mathbf{X}_i'-\mathbf{y}_i'\right)^2}_\text{particle history}+\underbrace{{\alpha_1'}~\left(\dot{\mathbf{X}}_n'-\dot{\mathbf{y}}_{{\rm c},n}'\right)^2}_\text{coherent velocity}+\underbrace{{\alpha_2'}~\left(\ddot{\mathbf{X}}_{n}'-\ddot{\mathbf{y}}_{{\rm c},n}'\right)^2}_\text{coherent acceleration},
\end{aligned}
\label{Eq:p28}       
\end{equation}
where
\begin{equation}
\begin{aligned}
{\alpha_1'}={\alpha_1}{(\frac{U}{D})^2} ~,~ {\alpha_2'}={\alpha_2}{(\frac{U}{D})^4}~,~{\rm and}~ \mathcal{J}'=\frac{\mathcal{J}}{D^2}.
\end{aligned}
\label{Eq:p29}       
\end{equation}

Both non-dimensional $\alpha_1'$ and $\alpha_2'$ weights determine how much velocity and acceleration signals can constrain the cost function minimisation process, knowing that the position history has a weight equal to one in~\eqref{Eq:p28}. If both parameters are set for any range below one, the history will have the most significant impact on the prediction function. The cost function~\eqref{Eq:p28} can be written compactly as
\begin{equation}
\mathcal{J}' = \mathcal{L}_{\rm data} + \mathcal{L}_{\rm physics},
\label{Eq:p28b}
\end{equation}
where $\mathcal{L}_{\rm data} = \frac{1}{k}\sum_{i=n-k+1}^{n}(\mathbf{X}_i'-\mathbf{y}_i')^2$ is the data fidelity loss over the measured position history and $\mathcal{L}_{\rm physics} = \alpha_1'\,\mathcal{L}_{\rm vel} + \alpha_2'\,\mathcal{L}_{\rm acc}$ groups the coherent velocity and acceleration terms as physics-based regularisation penalties. This decomposition mirrors the loss structure used in physics-informed neural networks \citep{Raissi2019PINN}, where observed data and physical constraints are combined in a single objective. The distinction is that our physics term originates from coherent neighbour observations rather than governing equation residuals and the minimisation is carried out analytically rather than through gradient descent. Appendix~\ref{appC} tests this interpretation further by replacing the polynomial basis with a neural field while retaining the same data-plus-physics structure. The weights $\alpha_1'$ and $\alpha_2'$ control the regularisation strength, where higher values force the solution to comply more with the coherent neighbour observations at the expense of the fit to the position history. The minimum of the cost function can be found by solving $\frac{\partial \mathcal{J}'}{\partial a}=0$, where its mathematics is addressed in Appendix \ref{Appendix_A}. We can use a minimum of four-time step histories of particles to minimise the cost function and predict the next step. The solution is not only smooth on the history of the target particle but also satisfies the local coherent dynamics of the flow. In the worst-case scenario where there is no coherent neighbour information, the prediction function is just a simple polynomial predictor without additional constraints. To obtain two coherent $\dot{\mathbf{y}}_{{\rm c},n}$ and $\ddot{\mathbf{y}}_{{\rm c},n}$ observations, neighbour trajectories need to be identified as coherent or non-coherent with the target particle. In \S~\ref{sec:p221}, we discuss how to use the LCS metric locally for particle segmentation and coherent neighbour identification over sparse trajectories. 

\begin{figure}
\centering
  \includegraphics[width=0.75\textwidth]{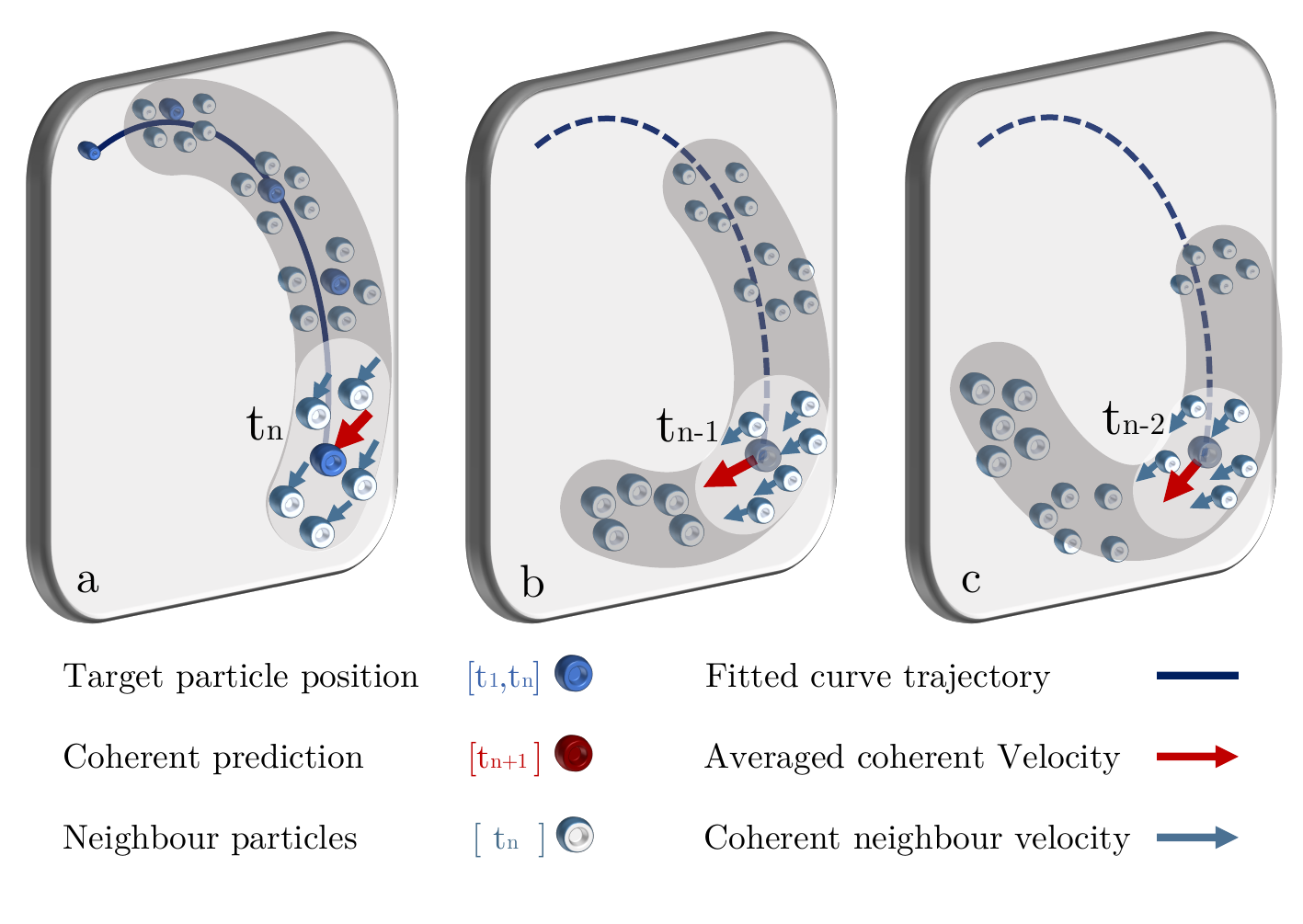}
\caption{Prediction using primary and secondary coherent particles. (a) Primary coherent neighbours sharing the trajectory of the target particle at the current time $t_n$. (b) Secondary coherent neighbours with a one-time-step phase delay, whose history at $t_{n-1}$ matches the target's neighbourhood at $t_n$. (c) Secondary coherent neighbours with a two-time-step phase delay, whose history at $t_{n-2}$ matches the target's neighbourhood at $t_n$. The dashed segments mark the phase-delayed history accessed by the secondary identification.}
\label{fig:p2}       
\end{figure}

\subsection{Local coherent dynamics}
\label{sec:p221}
\begin{figure}
\centering
  \includegraphics[width=0.9\textwidth]{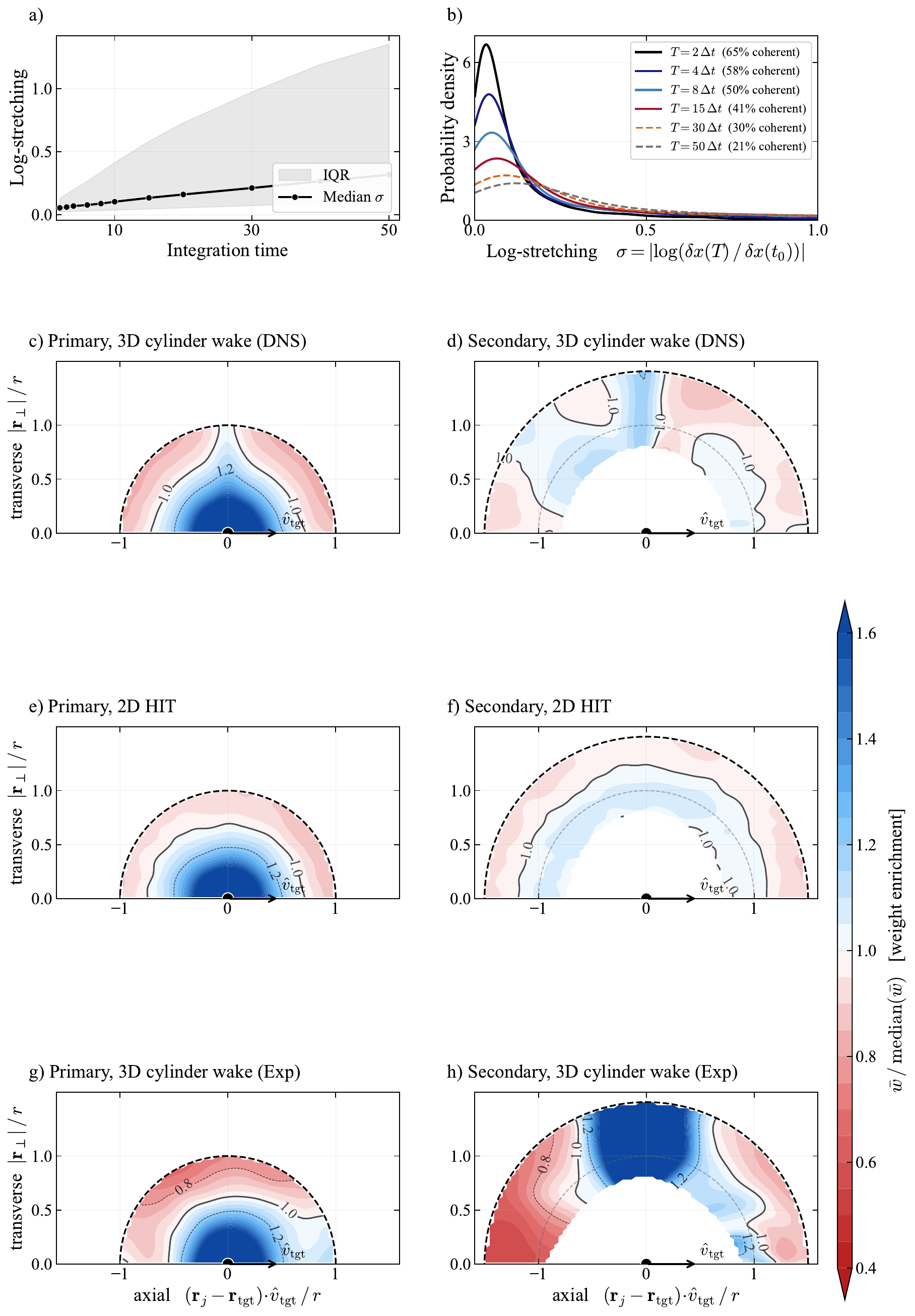}
\caption{Effect of FTLE integration time and spatial distribution of coherent neighbours, DNS $Re=3900$ wake flow case. (a) Median $\Lambda$ and interquartile range vs integration time $T$. (b) PDF of $\Lambda$ at $T = 2, 4, 8, 15, 30, 50 \, \Delta t$. Bottom rows: weighted neighbour density in target-centred cylindrical coordinates (axial coordinate aligned with target velocity). Left column: primary coherent neighbours at $t_n$. Right column: secondary coherent neighbours at $t_{n-\tau}$. Row~1: DNS $Re=3900$ ($278\,837$ targets). Row~2: 2D HIT ($65\,000$ particles). Row~3: experimental $Re=3900$.}
\label{fig:p_ftle_time}
\end{figure}

%
%
This section explains how to segment the neighbourhood of a target particle into coherent and non-coherent groups using a local rate-of-separation diagnostic. For a short enough time window, a fraction of the neighbouring particles shares its kinematics with the target \citep[see, e.g.,][]{Khojasteh2021LagrangianInitialization}. This coherent fraction depends on the local flow regime and the seeding concentration, and it decays as the integration window increases (see figure~\ref{fig:p_ftle_time}, top panels), especially in turbulent flows where the fluid content of coherent structures is rapidly exchanged \citep{Bross2023InteractionAPG}.

\subsubsection{Rate-of-separation quantification}

Here, we locally quantify coherency using the FTLE operator, with a methodology similar to that of \citet{kasten2009localized} \citep[see also][]{Fang2019PRF}, to determine the rate of separation locally and index neighbour trajectories as coherent or non-coherent.
The use of finite-time Lyapunov exponents for detecting hyperbolic LCSs traces back to \citet{Haller2000LagrangianTurbulence}, with further developments by \citet{Shadden2005DefinitionFlows}. We implement FTLE as an operator to highlight the strongest short-time repelling/attracting regions (i.e., rate of separation) rather than strict material barriers \citep[see, e.g.,][]{kasten2009localized}, not as impenetrable transport barriers as defined in LCS.
A global study from \citet{Hadjighasem2017ADetection} compared twelve techniques to detect coherent Lagrangian structures in 2D flows. It was found that FTLE is a simple  algorithm with suitable performance in capturing hyperbolic LCS. However, the technique becomes unreliable in elliptic LCSs. In the present study, we are looking for hyperbolic actions (i.e, local separation) between the target particle and its neighbours. 

We adopt a unique approach to compute the 
rate-of-separation
metric for local segmentation. Instead of calculating the global FTLE field, we define a local frame around each target particle (see figure~\ref{fig:p1}). Our goal is to classify all neighbours within this area as coherent or non-coherent concerning the target particle. The frame remains fixed during a series of time steps, offering a local Lagrangian view of neighbourhood behaviour (see figure~\ref{fig:p2}). The advantage of computing the local
rate-of-separation
over a global computation is that it aligns with our objective of finding coherent neighbours both temporally and spatially. There is no need to classify other clusters of coherent motions when predicting the trajectory of the target particle. Such an objective needs a local segmentation approach. It targets the particle's immediate surroundings, reduces both the computational cost and the difficulty of 3D transport-barrier quantification, and provides a precise picture of the dynamics around the target particle. We use a local frame since trajectories far from the target particle are less likely to be coherent and would only add computational cost. The maximum displacement of the target particle determines the size of the local frame in each direction. If the 2D (or 3D) displacements are equal in all directions, the shape forms a circle (or sphere) around the target particle. We then use the FTLE metric to quantify the stretching rate for each neighbouring trajectory and the target particle. 

FTLE examines the deformation of the flow maps of the nearest neighbours over time. Our approach simplifies to computing the spatial displacements between the target particle and its neighbours over a finite time. Essentially, two neighbour trajectories can stretch, contract, or rotate. The flow map $\Phi_{t_0}^{T}$ of a single particle can be formulated as
\begin{equation}\label{Eq:p210}
\Phi_{t_0}^{T}\left(\mathbf{x}\right): \mathbf{x}(t_0) \rightarrow \mathbf{x}(T),
\end{equation}
where $\mathbf{x}(t_0)$ is the starting position of the interval time $T$, and $\mathbf{x}(T)$ is the final position. Let $\mathbf{x}_0$ and $\mathbf{x}_0^{\text{N}}$ denote the initial positions of a target and its neighbouring particle. The Lyapunov exponent is a term that highlights regions with the most stretching if the interval time is positive (known as forward). Differences between the flow maps of the target particle $\Phi_{t_0}^t\left(\mathbf{x}_0\right)$ and its neighbour $\Phi_{t_0}^{t}\left(\mathbf{x}_0^{\text{N}}\right)$ would result in a vector displacement where we define
%


%

\begin{equation}\label{Eq:p214}
\Delta=\bigl(\nabla\Phi_{t_0}^t(\mathbf{x}_0)\bigr)^{\!\rm T}\nabla\Phi_{t_0}^t(\mathbf{x}_0).
\end{equation}

$\Delta$ is a symmetric positive definite matrix, also known as the right Cauchy-Green deformation tensor 
%
%
with three real and positive eigenvalues in a 3D domain over finite time. The maximum eigenvalue (the largest singular value) of the Cauchy-Green tensor $\lambda_{\rm max}(\Delta)$ shows the maximum amount that can be possibly stretched (expansion or separation) in finite time, representing how the flow field is sensitive to a perturbation. The eigenvector corresponding to $\lambda_{\rm max}(\Delta)$ represents the direction of the separation. Ultimately, the  rate of separation value $\Lambda_{t_0}^T$ is defined by scaling the magnitude of the maximum displacement as 
\begin{equation}\label{Eq:p217}
\Lambda_{t_0}^T=\frac{1}{\left|T-t_0\right|}log(\sqrt{\lambda_{\rm max}\left(\Delta\right)})=\frac{1}{\left|{T-t_0}\right|}log\left (\frac{\delta x(T)}{\delta x(t_0)}\right ).
\end{equation}

\subsubsection{Trajectory segmentation}

To compute the rate of separation, we assume that the local flow map corresponds to the particle trajectory over the finite time from $t_0$ to $T=k \times \Delta t$ (in this study, $(t_0-T) \geqslant 4 \times \Delta t$). We plotted the effect of integration time from $k=2$, which corresponds to the minimum track length, up to $k=50$ on the DNS $Re=3900$ wake test case. The top-left panel of figure~\ref{fig:p_ftle_time} shows that the median rate of separation $\Lambda$ grows monotonically with the integration time, and its interquartile range widens accordingly. This is expected in turbulent flows, as particles that are locally close eventually separate, and therefore the cumulative log-stretching increases with $T$. The top-right panel of figure~\ref{fig:p_ftle_time} shows the probability density function of $\Lambda$ at six integration times. For short windows, the distribution is sharply peaked near zero, meaning that most neighbours remain coherent with the target particle. As $T$ increases, the distribution dilutes, broadens and shifts towards higher values of $\Lambda$. The fraction of coherent neighbours decays from $65\%$ at $T=2\,\Delta t$ to $21\%$ at $T=50\,\Delta t$. This indicates that the local rate-of-separation is a short-time diagnostic, and a short integration window (in our analysis $T = 5\,\Delta t$) is sufficient to identify coherent neighbours, and longer windows do not improve the discrimination but only reduce the coherent population available for the predictor. In terms of typical PTV experiments, this window length remains well below the Lagrangian integral timescale, so it is accessible even at moderate temporal resolutions.

We use a spherical neighbourhood of radius \(r = \,U_{\text{loc}}\Delta t\) around the target, where \(U_{\text{loc}}\) corresponds to the max velocity of the target particle and therefore max expected displacement. This makes the frame adapt to local advection.

Let \(\mathcal{N}_c(n)\) denote the set of \(N_c\) coherent neighbours of the target at time step \(n\).
Each neighbour \(j\in\mathcal{N}_c\) is assigned a weight
$w_j = \bigl(|\Lambda_j|/\langle|\Lambda|\rangle_{\mathcal{N}_c} + \varepsilon_\Lambda\bigr)^{-1} + \alpha_w\bigl(d_j/r + \varepsilon_d\bigr)^{-1}$,
where $|\Lambda_j|$ is the magnitude of the neighbour's rate of separation relative to the target, $\langle|\Lambda|\rangle_{\mathcal{N}_c}$ is its mean over the coherent-neighbour set, $d_j$ is the Euclidean distance to the target particle, and $\varepsilon_\Lambda = \varepsilon_d = 10^{-15}$ prevent division by zero. Any $w_j$ that remains non-finite after this regularisation is replaced by unity before the weights are normalised by $\sum_{k\in\mathcal{N}_c} w_k$. When a neighbour has $|\Lambda_j|$ much smaller than the mean over $\mathcal{N}_c$, the corresponding weight grows large and dominates the normalised average. This is intentional, since a near-coherent neighbour is exactly the one we want the predictor to follow. The subscript~$w$ is used to distinguish this weighting coefficient from the cost-function weights $\alpha_1$ and $\alpha_2$ introduced in~\eqref{Eq:p25}.
$\alpha_w$ is a tuning parameter.
The idea behind this weighting is to give more weight to neighbours that are closer to the target and have a smaller rate of separation. Indeed the $w_j$ function could be optimised with additional coefficients, which goes beyond the interest of the current manuscript.


The bottom rows of figure~\ref{fig:p_ftle_time} show the spatial distribution of primary and secondary coherent neighbours around the target particle for our three flow configurations. For each dataset, we accumulated the weighted neighbour density in a target-centred cylindrical coordinate system, where the axial direction is aligned with the target's velocity vector and the transverse axis measures the perpendicular distance. The primary coherent neighbours, shown in the left column, are concentrated close to the target particle with an enrichment of $2.2$--$3.7$ times the background density at $R < 0.2\,r$. This enrichment decays to near-background levels at the frame boundary. The concentration is nearly axisymmetric in all three flow types, which is expected since the distance term in the weighting function dominates at short separations. The secondary coherent neighbours, shown in the right column at their past positions $t_{n-\tau}$, form a band at $r \approx r_{\rm sec}$. The angular distribution of this band is, however, different depending on the flow type. In the 3D cylinder wake (rows~1 and~3), the secondary band is elongated in front of the target particle along its velocity direction. This is consistent with mean advection that displaces coherent neighbours ahead of the target over the phase delay $\tau$. In the 2D homogeneous isotropic turbulence (row~2), the secondary band forms a nearly uniform ring with no preferred direction, since the isotropic flow field has no mean advection to break the symmetry. This contrast between the anisotropic wake and the isotropic turbulence shows that the secondary neighbours capture information about the underlying flow physics and not only expand the search radius. The integration time scan and the weighting parameter calibration discussed here are reproduced in the \texttt{03\_ftle\_evaluation.ipynb} notebook of the companion repository (see Code availability statement).


By using the Lagrangian framework instead of the Eulerian velocity field, one can directly measure the flow map, as the spatial derivative $\delta x(T)/\delta(t_0)$ can be directly obtained from the trajectories. Forward-FTLE corresponds to a positive interval time, where we assess forward flow maps. Conversely, a negative interval time leads to backward-FTLE values, indicating that particles stretching more in backward time accumulate material in forward time. Both backward and forward values provide essential information on the Lyapunov exponent field. Regions that stretch the most in forward and backward times determine stable and unstable manifolds over finite time. High FTLE field values indicate the presence of ridges that divide the local area into different clusters of coherent particles. Particles that start on one side of a ridge tend to stay on the same side, meaning that FTLE ridges act like invariant manifolds that particles cannot pass through. Lower FTLE values imply that neighbouring particles behave similarly, showing no signs of separation from the target particle over the finite time. This process segments the phase space into different coherent regions, allowing for the indexing of a group of neighbour particles as coherent or non-coherent with the target particle.


The coherent predictor indexes neighbour trajectories as coherent or non-coherent within the interval time from $t_{1}$ to $t_n$. In the present study, these trajectories are called primary spatial/temporal coherent neighbours. As shown in figure~\ref{fig:p2}.a, primary coherent neighbours follow a similar path with the target particle during the same period (same phase). However, a secondary group of coherent particles with a phase delay is still possible. 
Secondary refers to neighbours whose history was coherent with the target particle with a phase delay. The history of the secondary coherent neighbours is \textit{a priori} knowledge for the target particle. Schematics of secondary coherent neighbours with one and two-time step phase delays are shown in figure~\ref{fig:p2}.b.c. 
One and two-time step phase delays refer to groups of particles that were spatially located in the neighbourhood of the target particle at $t_{n-1}$ and $t_{n-2}$, respectively. Then the FTLE function determines secondary coherent neighbours between the target particle at $t_n$ and the secondary particles at $t_{n-\tau}$ where $\tau$ is the phase delay. 
In steady flow scenarios, secondary coherent particles show the exact motion of the target particle. However, as flow unsteadiness increases, the uncertainty of relying on information from secondary coherent neighbours also increases. 
%
%
Further study is needed to track coherent particle clusters in real experiments.
For future work, the rate of separation function could be replaced with techniques like \textit{Coherent Structure Colouring} \citep[CSC,][]{SchlueterKuck2017CSC, Martins2021DetectionColouring, Harms2024LGR}, which is beyond the scope of the present study. As a result of local segmentation, we can first identify coherent neighbour trajectories and then obtain the weighted averaged values of velocity $\dot{\mathbf{y}}_{{\rm c},n}$ and acceleration $\ddot{\mathbf{y}}_{{\rm c},n}$ over coherent neighbours based on their rate of separation and inverse of their Euclidean distances to the target particle.




\begin{table}
\centering

\begin{tabular}{l  l  l  l  l}

Case & 2D HIT & 3D wake flow & 3D wake flow & 3D wake flow \\

&
\multicolumn{1}{l}{\includegraphics[width=2cm]{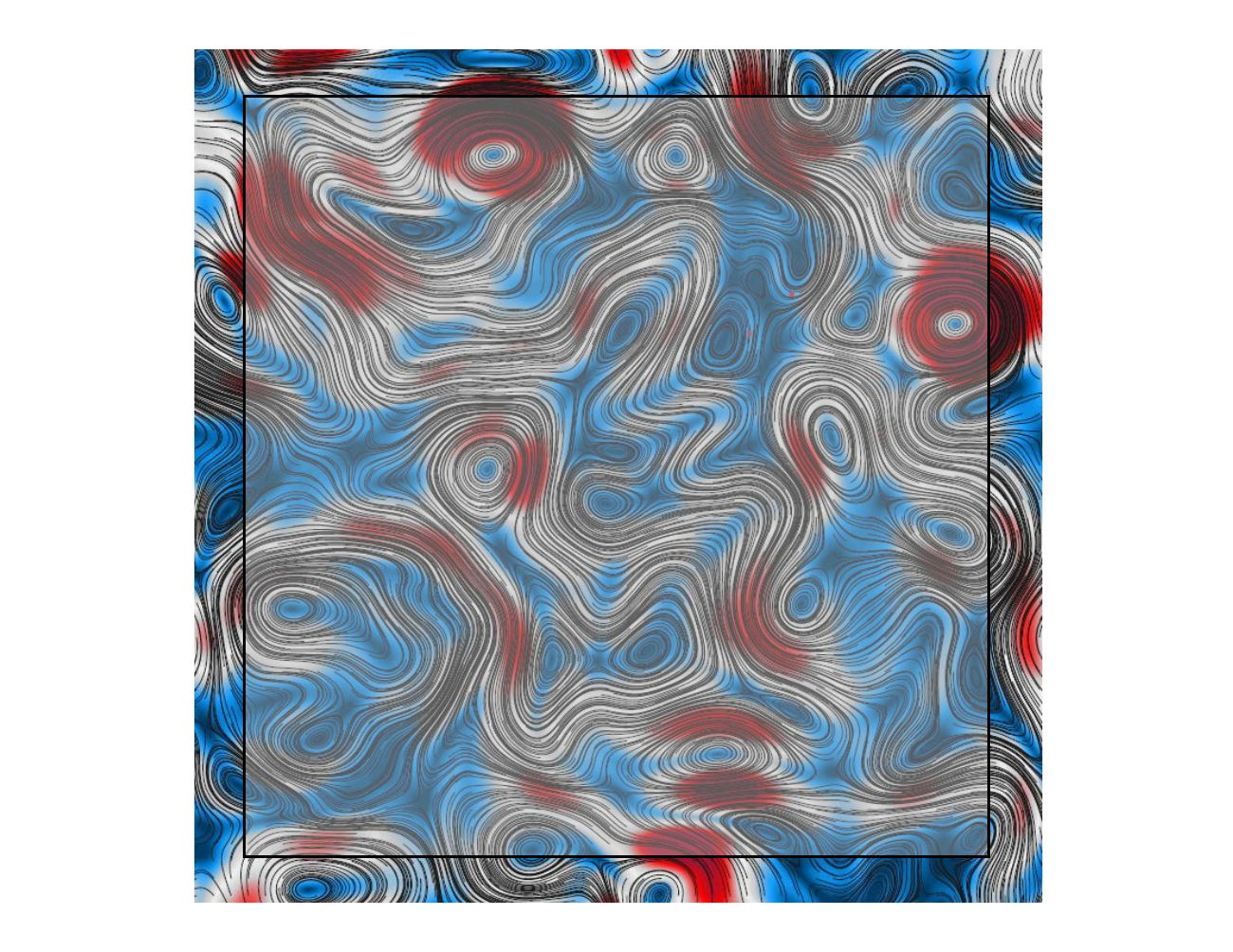}} &
\multicolumn{1}{l}{\includegraphics[width=2cm]{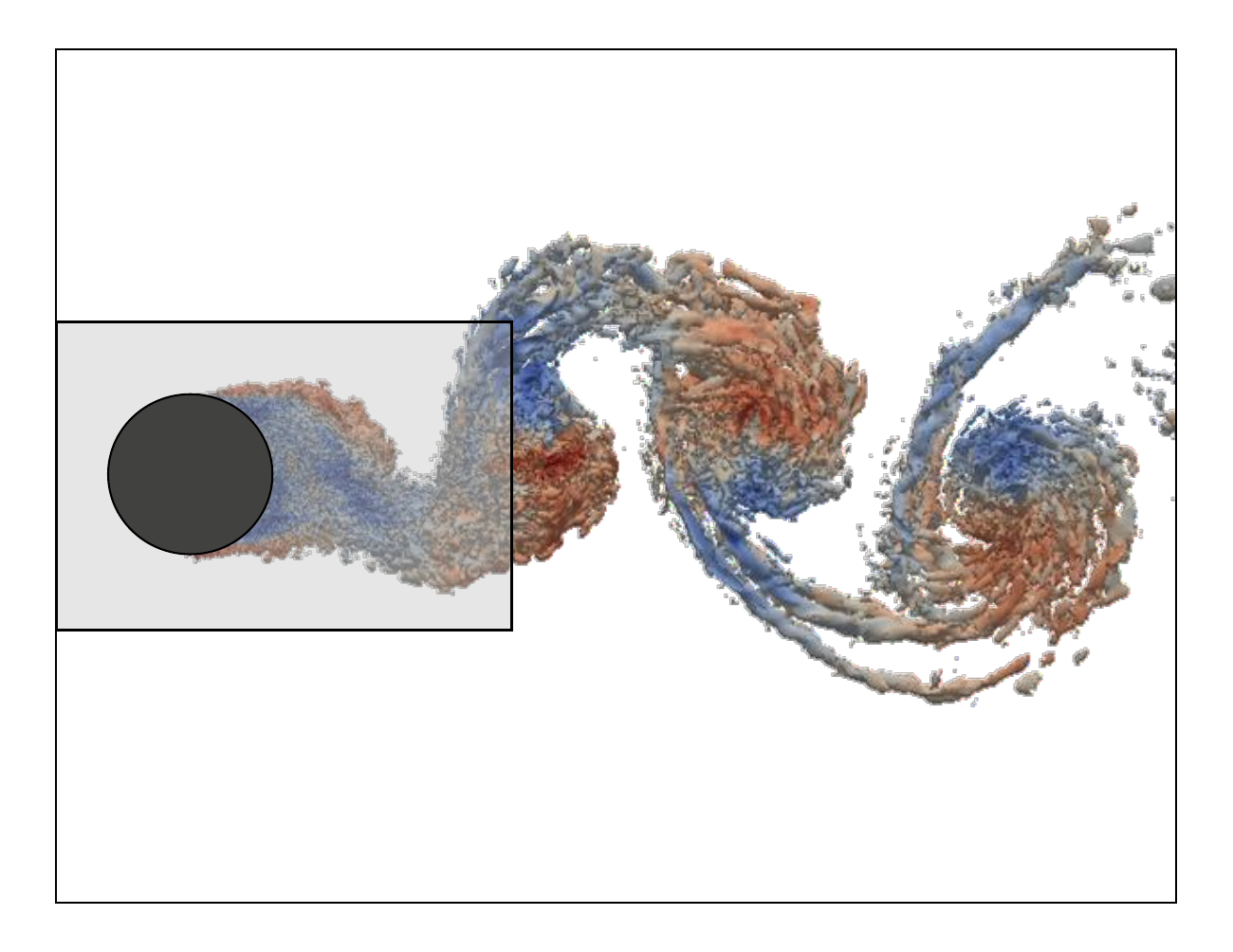}} &
\multicolumn{1}{l}{\includegraphics[width=2cm]{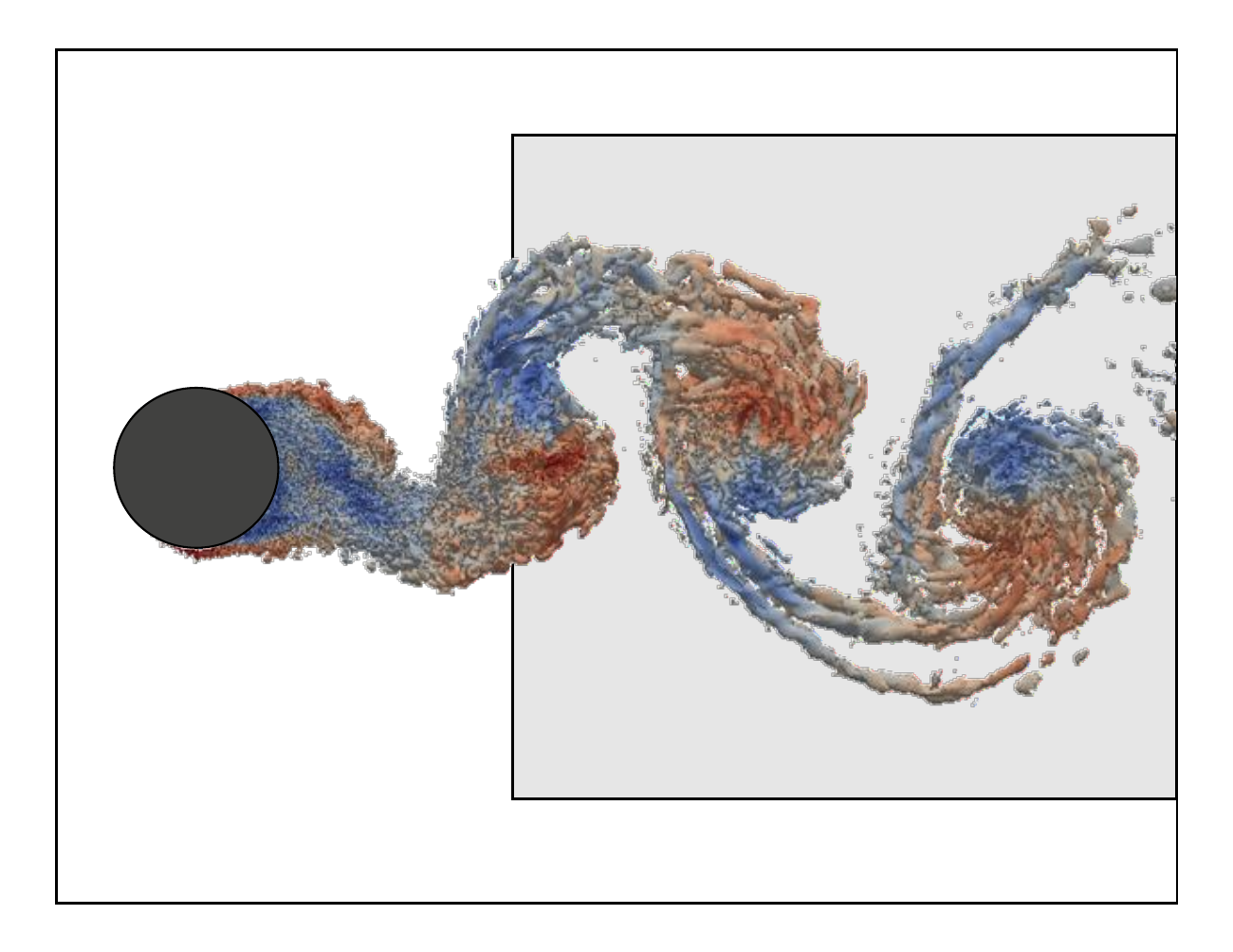}} &
\multicolumn{1}{l}{\includegraphics[width=2.5cm]{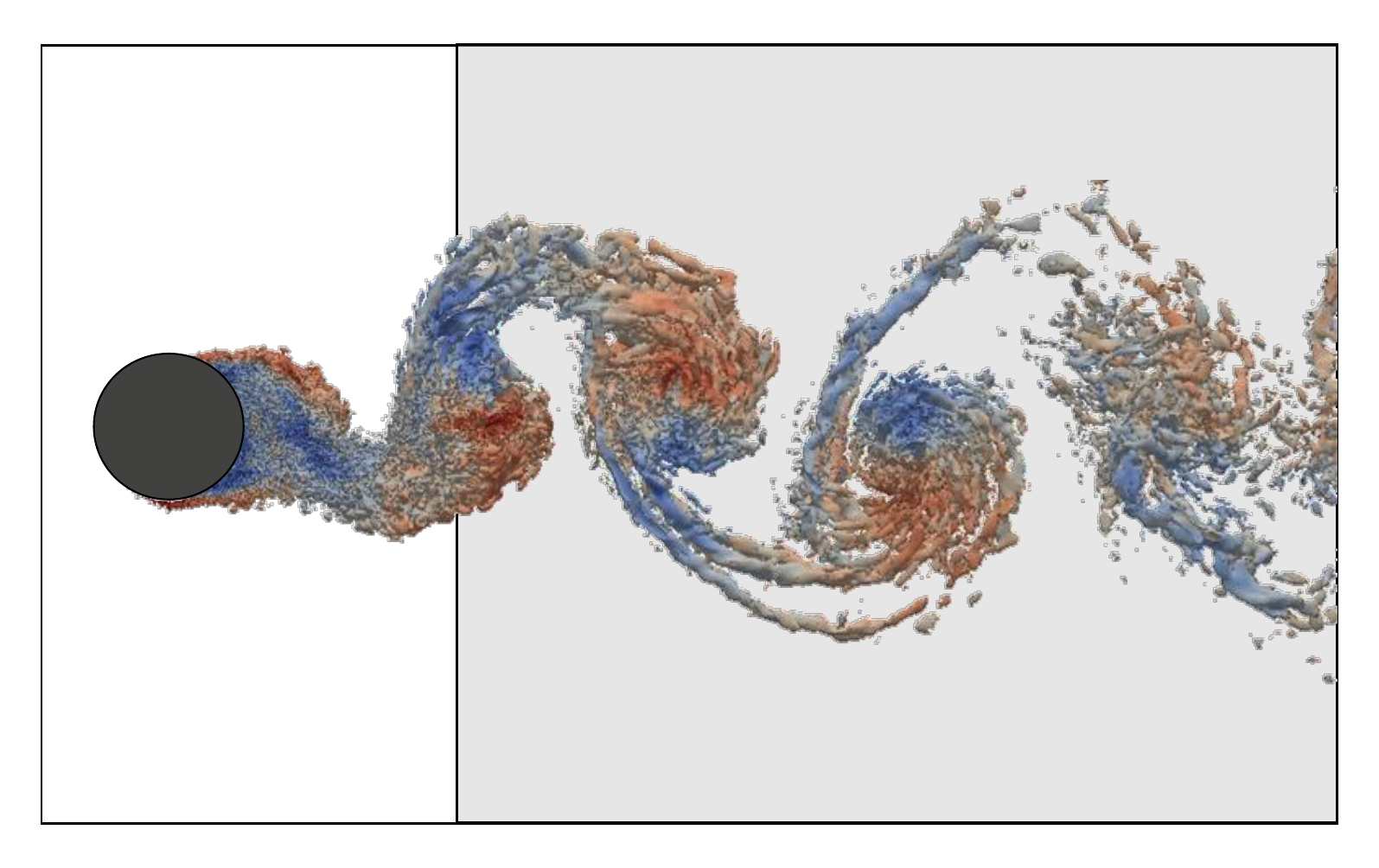}} \\

Approach & DNS & DNS & DNS & 4D-PTV\\
Reynolds & $3000$ & $3900$ & $300$ & $3900$\\
Dimension & $2 \pi \times 2 \pi$ & 4D $\times$ 2D $\times$ 2D &  8D $\times$ 8D $\times$ 6D & 24D $\times$ 14D $\times$ 4D\\
Grid size & $256 \times 256$ & $308 \times 328 \times 87$ & $145 \times 145 \times 108$ & - \\
Time step & $0.01 \ D/U$ & $0.00075 \ D/U$ & $0.005 \ D/U$ & $0.115 \ D/U$  \\
Camera resolution & - & - & - & $1280 \times 800$ \\
Number of particles & $65536$ & $150000$ & $150000$ & $\approx12000$ \\
Number of snapshots & $1000$ & $1000$ & $1000$ & $4000$ \\
$\eta$ & $0.0103$ & $0.0058$ & $0.0398$ & $0.0058$ \\
$\langle d_{\rm nn}\rangle / \eta$ & $1.8$ & $7.8$ & $1.9$ & $10.4$ \\
\end{tabular}
\caption{Details of numerical (DNS) and experimental (EXP) test cases. The grey squares indicate the regions that were recorded. The figures are schematic and not to scale. The flow structures are adopted from \citet{Lehmkuhl2014UnsteadyNumbers}. The Kolmogorov length scale $\eta = (\nu^3/\varepsilon)^{1/4}$ and the mean nearest-neighbour spacing $\langle d_{\rm nn}\rangle$ are reported in the bottom two rows. For the 3D cylinder wake at $Re = 3900$, $\varepsilon$ is computed from the volume-averaged strain-rate tensor of the DNS Eulerian field. For the 2D HIT, $\varepsilon$ is computed from $\varepsilon = \nu\langle|\nabla \mathbf{u}|^2\rangle$. The $Re = 300$ value is scaled from the $Re = 3900$ result using $\eta_{300} = \eta_{3900}(3900/300)^{3/4}$. The experimental case shares the DNS $Re = 3900$ value of $\eta$. For the 2D HIT case the integral scale is $D = 2\pi$, matching the non-dimensional domain size.}
\label{tab:p1}
\end{table}

\section{Numerical and experimental flow configurations}
\label{sec:p3}
To validate and assess our proposed cost function in turbulent flows, we used four numerical synthetic and experimental velocimetry measurements. Direct numerical simulations (DNS) were performed for 2D homogeneous isotropic turbulent (HIT) flow at a Reynolds number of $3000$, as well as the 3D wake behind a circular cylinder at Reynolds numbers of $3900$ and $300$. The specifics of these numerical cases can be found in table~\ref{tab:p1}. A variety of flow dynamics, including vortices, saddle points, and shear flows, are present in the 2D-HIT case, complicating the accurate prediction of flow motion (see figure~\ref{fig:p3}). For the 3D wake behind the cylinder at a Reynolds number of $3900$, we captured DNS snapshots of the formation region where unstable sideward shears collapse into the wake. This region, denoted by grey squares in table~\ref{tab:p1}, is particularly challenging for identifying true trajectories due to the emergence of complex 3D directional trajectories with high gradients \citep{Khojasteh2021LagrangianInitialization,Novara2023TwoPulseSTB}. We also examined the wake behind the cylinder at a lower Reynolds number of $300$ to assess the effectiveness of the proposed approach across various Reynolds numbers. In addition to the synthetic analyses, particle tracking experiments of the wake behind a circular cylinder at a Reynolds number equal to $3900$ were conducted in a wind tunnel. In the experiment, we captured the vortex-shedding behaviour after the formation region, where large-scale streamwise and spanwise coherent motions are significant. Having all these cases lets us evaluate the proposed approach across a wide range of complexities in the wake flow. 

\subsection{Direct numerical simulations (DNS)}
\label{sec:p31}

\subsubsection{Homogeneous isotropic turbulence}
\label{sec:p311}

\begin{figure}
\begin{center} 
  \includegraphics[width=1\textwidth]{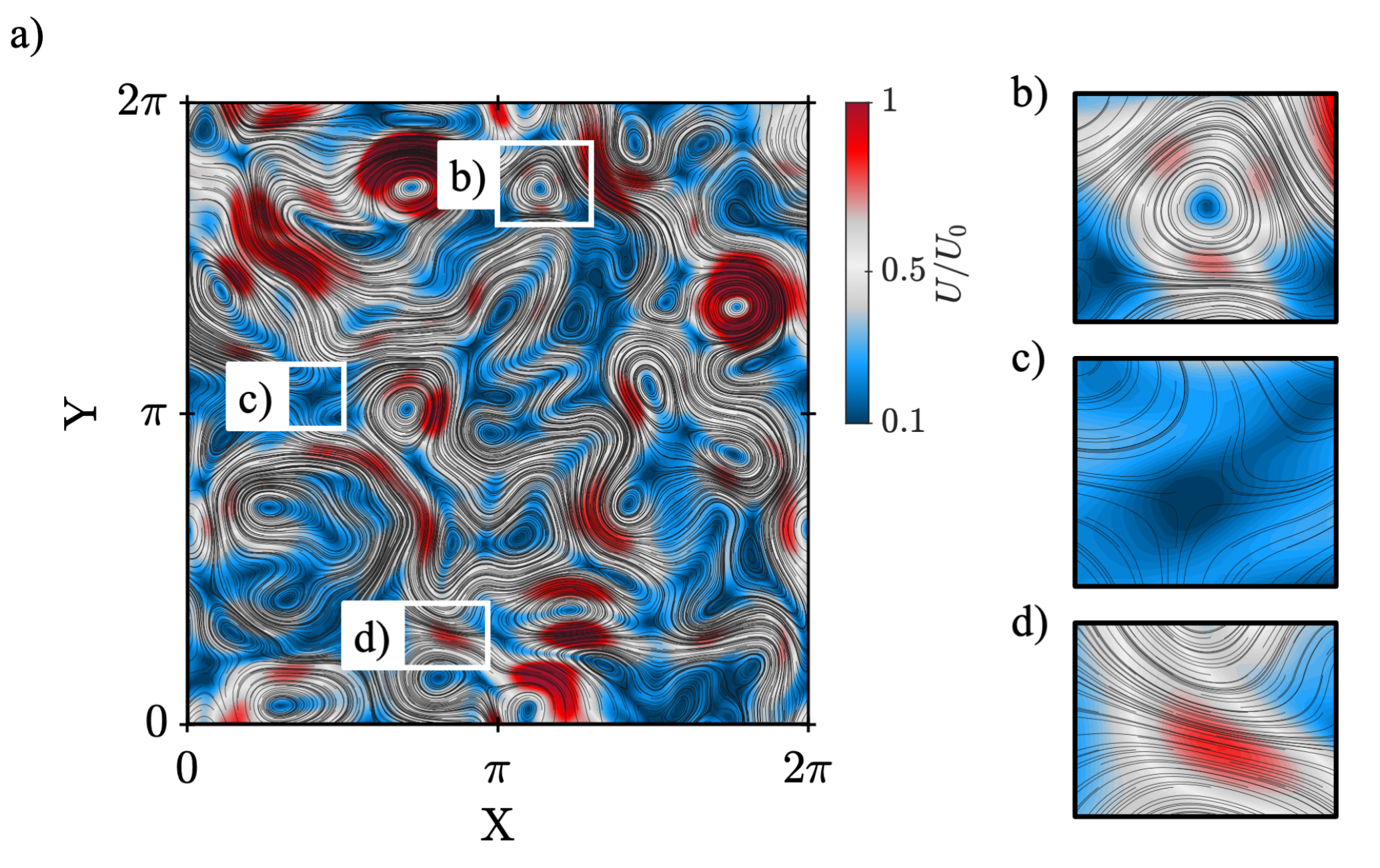}
\caption{\label{fig:p3} One snapshot of 2D homogeneous isotropic turbulent (HIT) flow.  }
\end{center} 
\end{figure}

We transported synthetic particles on 2D-HIT flow obtained from DNS at a Reynolds number equal to $3000$. The DNS snapshots were obtained from the FLUID European project \citep{Heitz2007DeliverableIST}. The Navier-Stokes equation was solved using incompressible condition ($\boldsymbol{\nabla} \cdot \boldsymbol{u}=0$). The vorticity and scalar equations are solved in Fourier space using de-aliased Fourier expansions in two directions with periodic boundary conditions. The time integration is the third-order Runge Kutta scheme. The square non-dimensional domain size of $2 \pi \times 2 \pi$ was discretised into $256 \times 256 $ node elements with periodic boundary conditions in four boundaries. This means that a particle enters in one side of the domain if a particle leaves on the opposite side at the same mirrored position. The dimensional DNS time step was set at $0.01 \ D/U$, where $D$ is the integral scale and $U$ is the reference velocity. We collected particle transport of $10000$ DNS time steps. Figure~\ref{fig:p3} shows instantaneous velocity and streamlines of the computational domain. We used the 2D-HIT synthetic data as the ground truth particle positions to examine how adding Lagrangian coherency can improve the prediction accuracy and robustness. 

\subsubsection{Wake behind a circular cylinder}
\label{sec:p312}
To evaluate our position prediction approach in a 3D complex flow, we used DNS simulations of the wake behind a smooth cylinder at Reynolds numbers equal to $3900$ and $300$ (based on the diameter $D$ of the cylinder and the free-stream velocity $U$) computed by an open-access code called \textit{Incompact3d} \citep{Laizet2011Incompact3d:Cores}. In both cases, the dimensional DNS time step was set to $0.00075 D/U$ and $0.005 D/U$, respectively, and $150000$ particles were transported every $10$ DNS time steps using fourth-order Runge Kutta temporal and trilinear spatial schemes. Particle trajectories in the synthetic case are smooth and predictable when the synthetic temporal resolution is in the same order as the DNS time step due to the short travelling distance between two time steps (less than the Kolmogorov timescale).

For the Reynolds number 3900 case, the flow field was sampled at $1000$ time steps for a domain of $4D\times2D\times2D$, with the cylinder located within the domain at ($1D,1D$). Details of the simulation and synthetic transport of particles can be found in \citet{Khojasteh2022Lagrangian3900}. For the Reynolds number $300$ case, a cylinder wake flow was generated from an initial null velocity field within the reference domain of size $20D\times12D\times6D$, discretised into $360\times217\times108$ nodes. Turbulent structures were induced by imposing an inflow velocity $u_{in}/U = 1$ at each DNS time step. The cylinder was modelled using an Immersed Boundary Method \citep{Parnaudeau2008Experimental3900,Parnaudeau2004CombinationGeometry}. We imposed free-slip conditions on the y-axis borders and periodic conditions along the cylinder axis. The outflow was modelled by a first-order advection model. We sampled $1000$ time steps of data within a domain of $8D\times8D\times6D$, situated far downstream from the cylinder, which was located at ($4D,4D$). These two cases let us evaluate the proposed approach under different flow conditions and Reynolds numbers.

\subsection{Time-resolved particle tracking velocimetry (4D-PTV)}
\label{sec:p32}
We performed an experimental study of the cylinder wake flow at a Reynolds number equal to $3900$ (the same as the synthetic data). Experiments were carried out in the INRAE low-speed wind tunnel equipped with a centrifugal fan, a diffuser, a plenum chamber with honeycomb and grids, a contraction section decreasing by four, and an area with transparent walls for testing \citep{Chandramouli2019FastObservations}. With the aid of hot wire anemometry, the velocity profile at the wind tunnel entrance was checked to ensure uniformity. The free-stream turbulence intensity level was found to be less than $0.2~\%$. The cross-section of the testing zone is square, with a width of $28~\text{cm}$ and a length of $100~\text{cm}$. It has a slightly tilted upper wall to reduce longitudinal pressure gradients. The setup allowed for continuous flow velocity selection between $1$ and $8~\text{m/s}$.

\begin{figure}
\centering
  \includegraphics[width=1\textwidth]{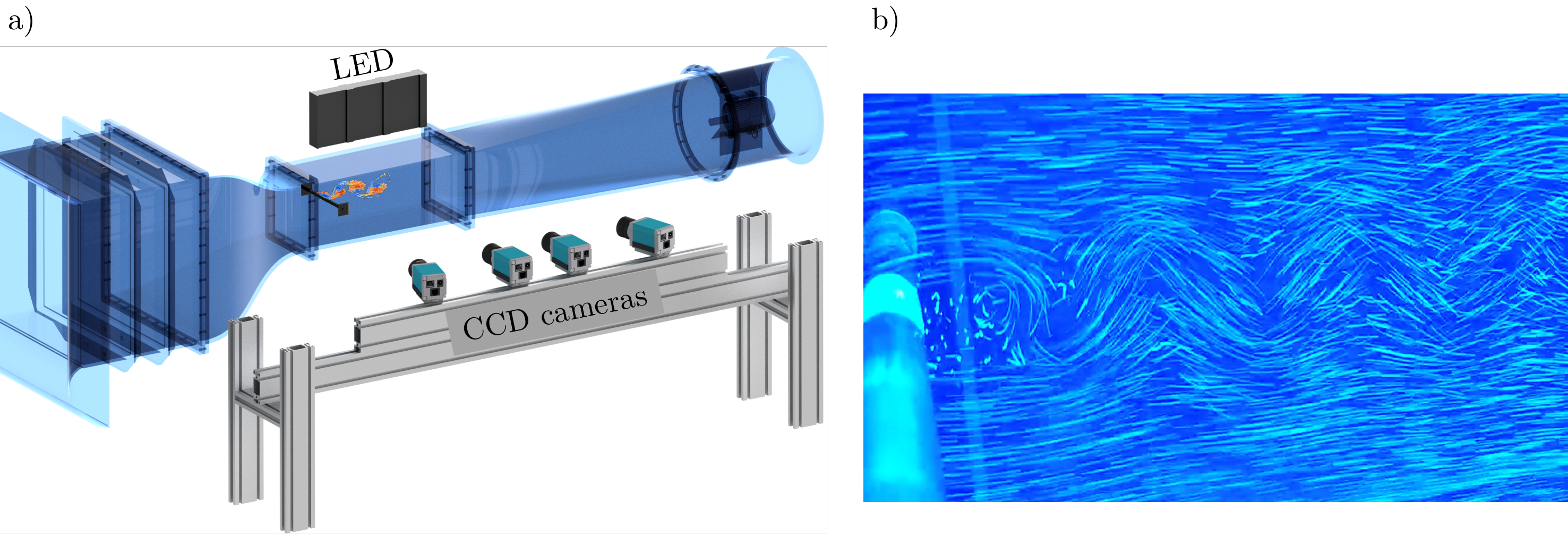}
 \caption{Experiment setup design for the cylinder wake flow at Reynolds $3900$. (a) Schematic of the camera setup design. (b) One snapshot of HFSB tracer particles passing downstream of the cylinder. }
\label{fig:p4}       
\end{figure}

We designed an experimental setup for time-resolved volumetric measurement, as shown in figure~\ref{fig:p4}.a. Four CMOS SpeedSense DANTEC cameras with a resolution of $1280 \times 800$ pixels and a maximum frequency of $3.25 \ \text{kHz}$ are used. We equipped the cameras with Nikon $105 \ \text{mm}$ lenses. For off-axis acquisitions, we mounted the lenses on Scheimpflug adapters (LaVision GmbH). The first two cameras are positioned in backward light scattering, while the second two cameras receive maximum intensity signal in forward scattering. The calibration error was lower than $0.06$ pixel and reduced to $0.04$ after the volume self-calibration in which each $\text{mm}$ is equal to $6.28$ pixels. The inter-frame particle shift was kept below $12$ pixels, following the recommendation of \citet{Schanz2016Shake-The-Box:Densities}. The volume of interest was $280 \ \text{mm} \times\ 160 \ \text{mm} \times\ 46 \ \text{mm}$ starting from roughly $4D$ downstream of the cylinder. The aperture was set at $8$ to achieve $46 \ \text{mm}$ depth of focus. We used an LED system on the top and a mirror at the bottom of the test section to illuminate the volume of interest. The acquisition was long enough to observe dynamic evolutions of the von Kármán vortex streets downstream of the cylinder.

Helium Filled Soap Bubbles (HFSB) are the main tracer particles used for large-scale volumetric experiments in air. Soap bubbles are the only tracers with a size significantly larger than $10~ \mu m$, leading to sufficient light scattering in the Large-scale volumetric measurements. Smaller particles, such as oil droplets, can create a very dense particle concentration and follow the flow accurately, but they scatter very little light. For this reason, volumetric measurements using these small seeding particles are restricted to small measurement volumes. Therefore, in the present experiment, the seeding particles were HFSB, resulting in desired intensity signal with appropriate particle size. In a similar experiment, \citet{Scarano2015OnExperiments} studied the application of using HFSB in the wake flow past a cylinder in a volume of $20\times20\times12~\text{cm}^3$ ($4800~\text{cm}^3$). Based on a study carried out by \citet{Caridi2016HFSB-seedingTunnels}, HFSB was determined to have several orders of magnitude higher intensity than fog droplets. Particle response time is a value that determines how particles follow the flow with fidelity. Time response is directly linked to the particle diameter and mass density discrepancy to the air in wind tunnel experiments. Due to this reason, large fog droplets ($>10-20~\mu \text{m}$) do not follow the flow with enough fidelity because of their poor time-response value. However, the mass density discrepancy is close to zero since HFSB particles are filled with Helium (lighter than air). As a result, the particle time response for HFSB becomes small enough to follow the flow with fidelity. Accordingly, \citet{Scarano2015OnExperiments} reported that the HFSB time-response is maintained well below $100~\mu \text{s}$, which means that particles should adequately follow the flow in low-speed experiments. So, large particles with favourable time-response values provide the ability to perform large-scale measurement volumes \citep{Schneiders2016Large-scaleTracers}. However, during the experiment, we noticed that bubbles in wind tunnel experiments are limited by three primary factors: generation rate, lifetime, and image glare points. Due to these limitations, HFSB for large-scale volumetric measurements inside the wind tunnel leads to low particle concentration (approximately $1~\text{particle}/\text{cm}^3$). One of the earliest studies of using HFSB reported $\text{ppp}<0.01$ for the seeding density, which could only resolve the shedding large-scale structures and was unable to capture turbulent small scales \citep{Scarano2015OnExperiments}. In 2018, \citet{Gibeau2018AMeasurements} reached to $1.6~\text{particle}/\text{cm}^3$ with the idea of having a multi-wing seeding system. The impact of the multi-wing generator on the flow stream is found to be at most $1.9~\%$ of the turbulence intensity with a negligible deficit on the mean flow. \citet{Gibeau2020EvaluationGenerator} reached $0.02~\text{ppp}$ over a volume of $2000~\text{cm}^3$ using a full-scale HFSB generator with $48$ nozzles. Recent advances in HFSB scalability have further extended volumetric measurement sizes \citep{Morias2024HFSBscalability}.

\begin{figure}
\centering
  \includegraphics[width=1\textwidth]{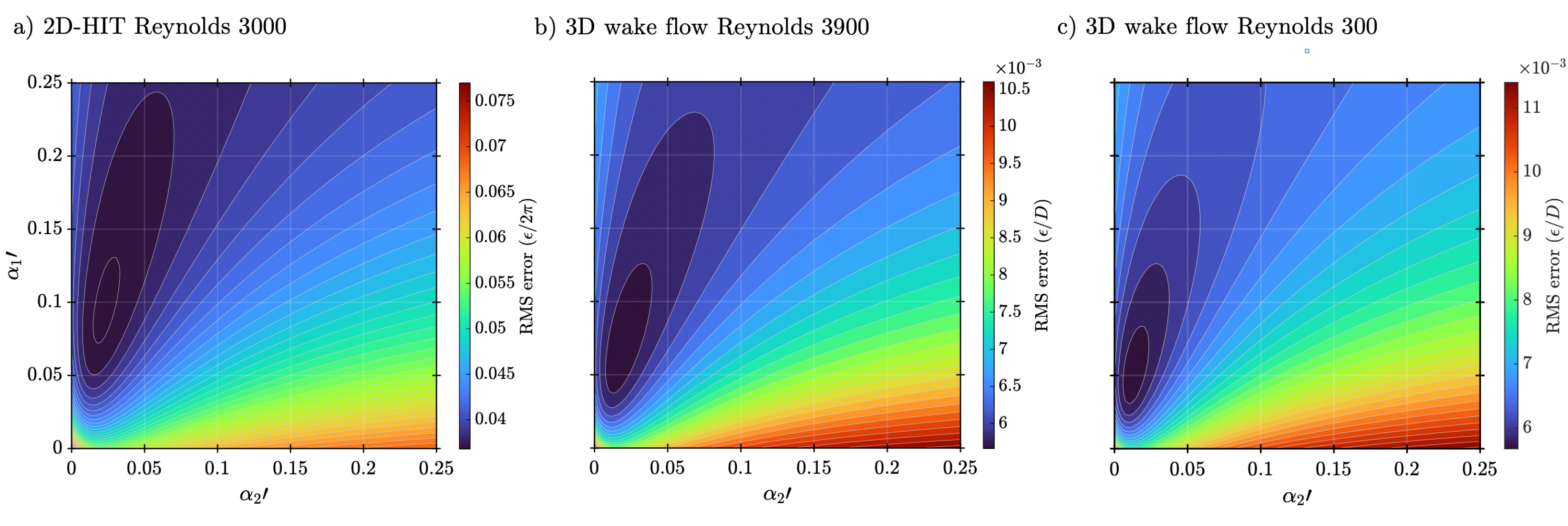}
\caption{Contour plot of the position estimation RMS error as a function of $\alpha_1'$ and $\alpha_2'$ weighting terms:  (a), 2D homogeneous isotropic turbulent flow at a Reynolds number equal to $3000$; (b),(c), 3D wake behind a cylinder at Reynolds numbers equal to $3900$ and $300$, respectively. The synthetic trajectories were generated based on the 1st LPT challenge estimation errors.} 
\label{fig:p5}       
\end{figure}

In the present study, we placed $50$ bubble generator nozzles with airfoil-shaped structures inside the wind tunnel chamber. Figure~\ref{fig:p4}.a shows homogeneous distribution of tracer particles during the experiment. The nozzles were far upstream of the measurement section in the settling chamber to ensure a sufficient number of bubbles were created, and the flow field was not disturbed by the existence of nozzles. The bubble lifetime is very short (around $2-3$ minutes) inside the wind tunnel, mainly because they explode by passing through honeycomb layers. To overcome this issue, we injected bubbles inside the chamber for a few minutes when the wind tunnel was off before starting the acquisition. We found that particles larger than three pixels create two glare points on two sides of the bubble when the illumination is LED. This requires more image treatments before running the 4D-PTV algorithm to avoid false particle reconstruction. However, the intensity of two glare points can diffuse and merge if the particle size is around two pixels. Therefore, we adjusted the camera magnification to reach particles with two pixel image sizes on average to surpass the glare point issue.

\section{Coherent predictor results}
\label{sec:p4}
We now apply the newly developed prediction approach, based on coherent velocity and acceleration priors as presented in \S~\ref{sec:p2}, to numerical and experimental flow configurations described in \S~\ref{sec:p3}. The predictor function involves two weighting parameters that need to be optimally tuned. In \S~\ref{sec:p41}, the behaviour of these parameters is analysed and modelled as a function of the uncertainties on the predicted position, coherent velocity and acceleration. In \S~\ref{sec:p42}, using the 2D HIT test case, we address how sensitive the proposed coherent predictor and the state-of-the-art approaches are to the change of particle concentration, noise level, and temporal resolution, including the sensitivity of the polynomial baseline to its order and history length. In \S~\ref{sec:p42b}, we quantify the contribution of the secondary coherent neighbours on top of the primary ones. We then quantify the proposed model's instantaneous and time-averaged bias error in the 3D wake flow test case context, along with uncertainty quantification using Monte Carlo simulation in \S~\ref{sec:p43}. Following this, we present the experimental demonstration of the coherent predictor in \S~\ref{sec:p44}.

\subsection{Optimal weighting parameters}
\label{sec:p41}
In this subsection, we aim to determine the optimal values for the weighting parameters ${\alpha'}$ by examining their influence on the prediction RMS error. The optimal solution minimises the cost function, resulting in the lowest estimation error. With the non-dimensional cost function~\eqref{Eq:p28} in hand (see appendix \ref{Appendix_A}), we seek to determine the most suitable ${\alpha'}$ values for optimal trajectory estimation. To achieve this, we conducted predictions across a range of ${\alpha'}$ weights. Real experiments carry uncertainties and inaccuracies, so we incorporate them into the three observation terms within the cost function, specifically the positions $\mathbf{y}_i'$, the coherent velocity $\dot{\mathbf{y}}_{{\rm c},n}$, and the coherent acceleration $\ddot{\mathbf{y}}_{{\rm c},n}$.

Using data from the 1st LPT challenge \citep{Sciacchitano2021MainChallenge,Leclaire2021FirstBenchmark,Schroder20233DMechanics} as a reference, we estimated the inaccuracies introduced into the predictor function. This challenge assessed the position estimation accuracy of six state-of-the-art time-resolved tracking algorithms, including our coherency-based track initialisation \citep{Khojasteh2021LagrangianInitialization}, for particle densities from $0.005$ to $0.2 ~\text{ppp}$. We, therefore, converted averaged RMS dimensional errors into non-dimensionalised position ${\epsilon_X}/D$, velocity ${\epsilon_{\dot{X}}}/U$, and acceleration ${\epsilon_{\ddot{X}}}D/{{U^2}}$ errors to create the synthetic data at each particle density. For example, the average of the reported RMS position error was $0.005 ~\text{mm}$ where the integral scale was $D = 10 ~\text{mm}$ at the particle density of $0.12 ~\text{ppp}$, which gives ${\epsilon_X}/D=5.10^{-4}$. The averaged velocity and acceleration errors at the same particle density were found to be ${\epsilon_{\dot{X}}}/U=0.01$ and ${\epsilon_{\ddot{X}}}{D/{U^2}}=0.3$, respectively. The acceleration estimation has at least an order of magnitude higher error than other terms. For such a scenario, we can expect the optimal solution should have less ${\alpha'_2}$ weight than ${\alpha'_1}$ because of having more acceleration estimation error than other terms. We introduced these errors as input uncertainties into three ground truth Lagrangian trajectories of 2D-HIT and 3D cylinder wake flow cases and then computed the correlation of the final estimation error and ${\alpha'}$ weights. Data creation of these cases is addressed in \S~\ref{sec:p3}. The LPT-challenge accuracies cited above are obtained on centred, fitted trajectories. At the endpoint of a live track, only a backward-looking finite-difference stencil is available, and the associated noise amplification on the velocity and acceleration estimates is substantially larger than at mid-track. This endpoint-specific degradation is quantified in \S~\ref{sec:p43}.

\begin{figure}
\centering
  \includegraphics[width=1\textwidth]{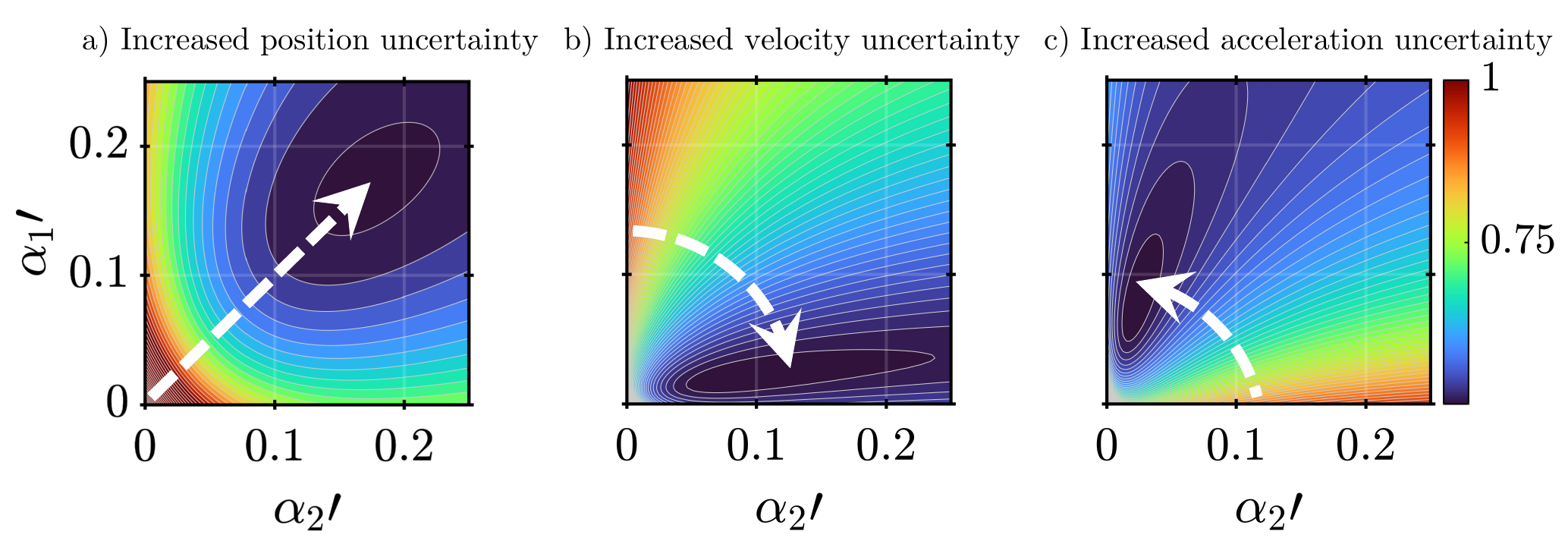}
\caption{Contour plot of the position estimation RMS error as a function of observations uncertainty levels for the 3D wake flow case at $Re = 3900$. (a) Increasing position uncertainty at fixed velocity and acceleration uncertainties. (b) Increasing velocity uncertainty at fixed position and acceleration uncertainties. (c) Increasing acceleration uncertainty at fixed position and velocity uncertainties. White arrows show the changing direction of optimal solutions (i.e., minimum estimation error) by increasing each observation uncertainty.}
\label{fig:p6}       
\end{figure}

Next, we plotted the prediction RMS error across a range of ${\alpha'}$ weights to identify the optimal solution in figure~\ref{fig:p5}. As acceleration errors exceeded other terms in both synthetic data sets, the minimum error corresponded to relatively lower ${\alpha'_2}$ values (i.e., ${\alpha'_1}/{\alpha'_2}>>1$). Both optimal weighting parameters were $\ll 1$, so the position history with unit weight remains the most informative signal in this regime. Comparing the estimated position error behaviour in the 2D-HIT and 3D wake flow cases revealed that all three cases required similar weighting parameters with ${\alpha'_1}/{\alpha'_2}\gg 1$ and small magnitudes ${\alpha'}\ll 1$. This implies that the proposed cost function remains generic in all three cases. Its independent nature might be related to the supplementary details provided by coherent neighbours, simplifying prediction complexities by indicating the appropriate direction and acceleration of Lagrangian trajectories. The only clear similarity between all three cases was the identical input uncertainty level. Consequently, we can hypothesise that the optimal values for the weighting parameters are directly connected to the uncertainty levels of the input observations. To validate this hypothesis, we designed three further parametric test cases. In figure~\ref{fig:p6}, we increased the uncertainty level of each parameter solely while other terms were fixed to assess how the optimal solution behaves concerning three position, velocity, and acceleration uncertainties. The optimal solution tends to linearly move away toward higher weighting magnitudes with increased position uncertainty (see figure~\ref{fig:p6}.a). If we fix the position uncertainty, the optimal solution rotates with the same distance around the coordinate centre, depending on which parameter is increased (see figures~\ref{fig:p6}.b and \ref{fig:p6}.c). This means that we can model the correlation between two weighting terms. Magnitudes of both gains increase by increased position uncertainty, and the slope for $\alpha_2/\alpha_1$ is a function of relative velocity and acceleration uncertainties. 

\begin{figure}
\centering
  \begin{minipage}[c]{0.62\textwidth}
    \centering
    \includegraphics[width=\textwidth]{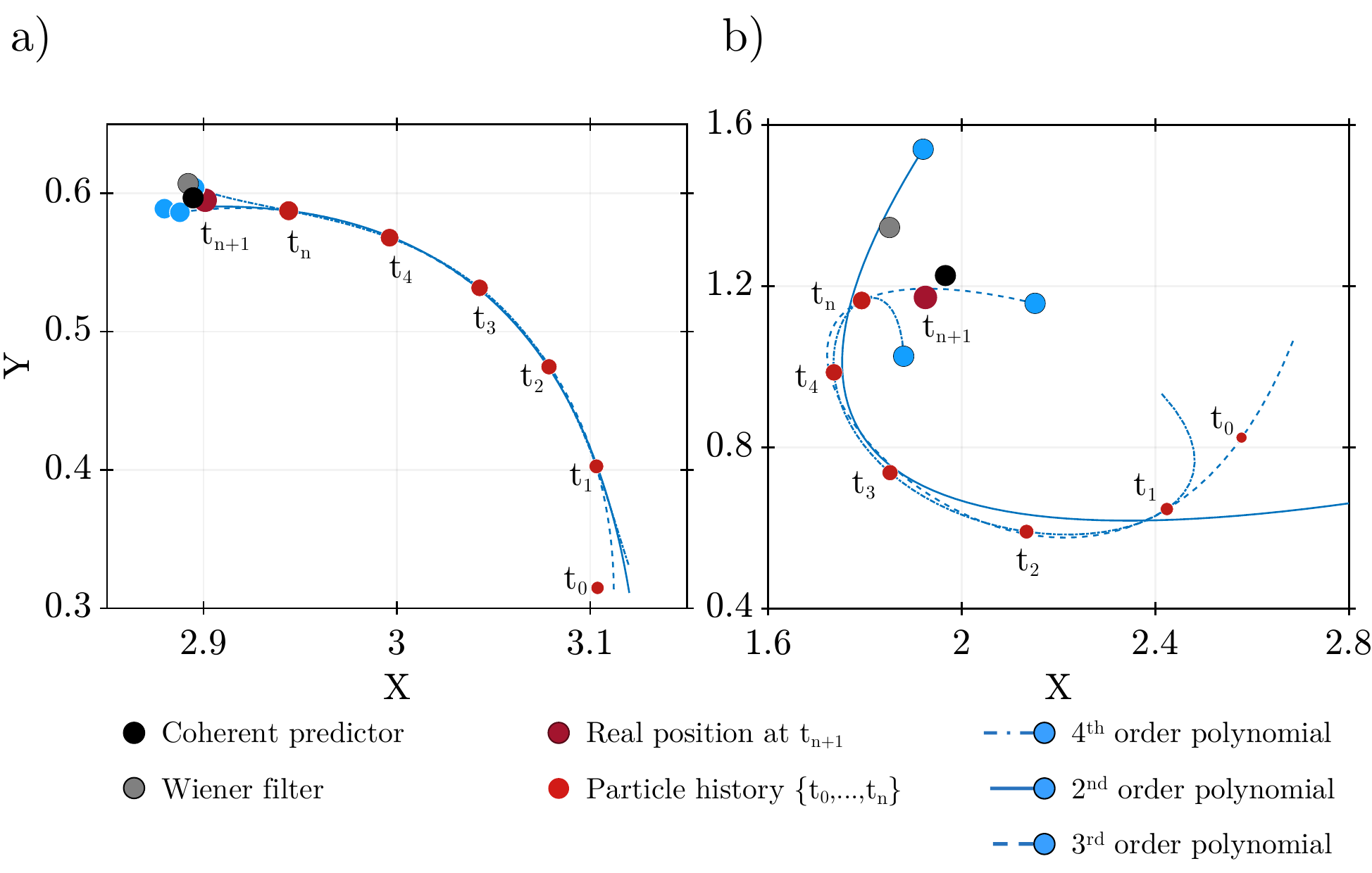}
  \end{minipage}\hfill
  \begin{minipage}[c]{0.34\textwidth}
    \centering
    \includegraphics[width=\textwidth,height=0.22\textheight,keepaspectratio]{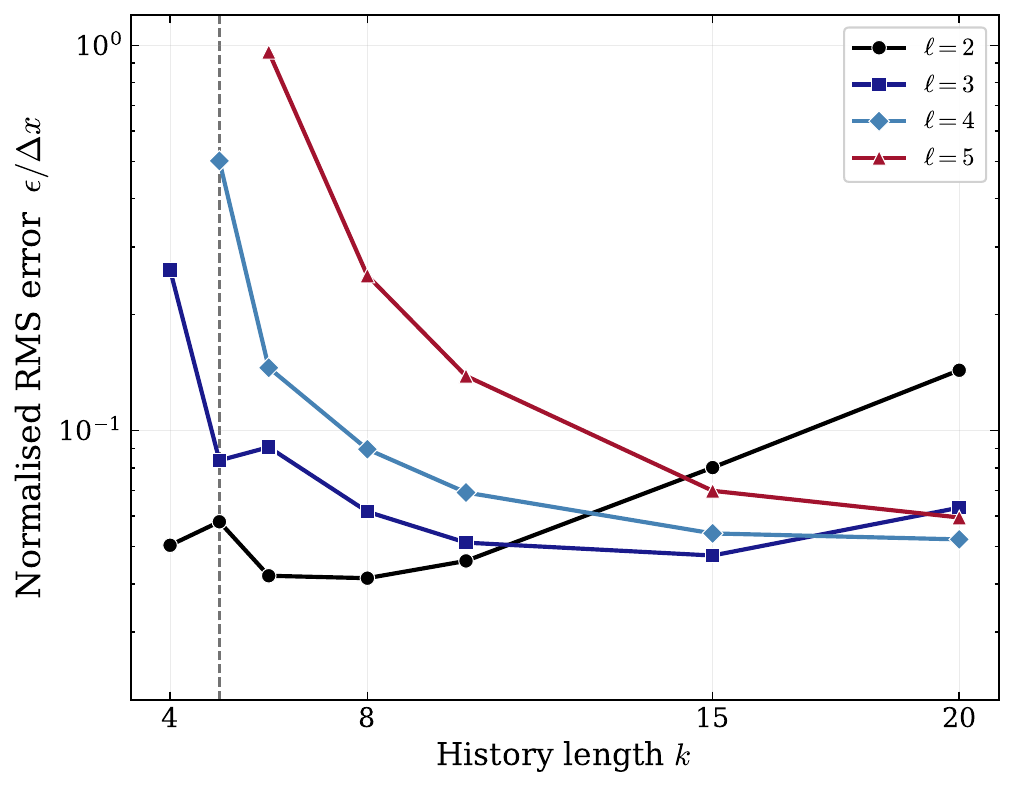}
  \end{minipage}
\caption{Prediction comparison for single trajectories in 2D-HIT: (a), Smooth trajectory; (b), Highly curved trajectory. (c) Position-prediction RMS error (noise-free) of the polynomial baseline as a function of history length $k$ for polynomial orders $\ell = 2$--$5$, pooled over four 3D cylinder-wake DNS datasets ($\approx 760\,000$ events). The vertical dashed line marks the working point $k = 5$.}
\label{fig:p10}       
\end{figure}

Assume $\sigma_p'$ is the dimensionless uncertainty of position, and $\sigma_u'$ and $\sigma_a'$ represent the dimensionless uncertainties of velocity and acceleration included in our observation terms. We use the same non-dimensionalisation as in~\eqref{Eq:p26}, where $\sigma_p'=\sigma_p/D$, $\sigma_u'=\sigma_u/U$, and $\sigma_a'=\sigma_a~D/U^2$. Based on figure~\ref{fig:p6}, it can be said that $(\alpha_1'^2+\alpha_2'^2)=c_1  \sigma_p'$ where $c_1$ is a constant, and it can be said that $\alpha_2'/\alpha_1'=c_2  \sigma_u'/\sigma_a'$. Consequently, a model for estimating $\alpha_1'$ and $\alpha_2'$ can be formulated as

\begin{equation}
\alpha_1' = \sqrt{\frac{c_1 \sigma_p'}{1 + \left(\frac{c_2 \sigma_u'}{\sigma_a'}\right)^2}}
\end{equation}

\begin{equation}
\alpha_2' = \alpha_1' \left(c_2 \frac{\sigma_u'}{\sigma_a'}\right).
\end{equation}

As discussed earlier, both $\alpha'$ weights must be less than one. This model offers an estimation for $\alpha_1'$ and $\alpha_2'$, dependent on $\sigma_p'$, $\sigma_u'$, and $\sigma_a'$. It is important to note, however, that the constants $c_1$ and $c_2$ should be determined through calibration using available synthetic data. In supervised learning, regularisation hyperparameters are typically found through cross-validation or grid search, which requires repeated evaluation on held-out data. Here, the optimal regularisation can be predicted directly from the noise characteristics of the input observations. This means that for a new experiment where the measurement uncertainties are known, the cost function can be configured without additional optimisation. This validates our hypothesis that the proposed cost function is correlated with introduced uncertainty levels. Therefore, an appropriate estimation of three uncertainty levels would provide enough information to set weighting parameters, regardless of the flow case (see figure~\ref{fig:p5}). We show that the proposed cost function is generic, as it remains independent of flow dimensions (2D or 3D), Reynolds number, and flow topology, with the same estimated position RMS error pattern for different cases. 

\subsection{Sensitivity analyses}
\label{sec:p42}
Having found the optimal parameters for our physics-based predictor in \S~\ref{sec:p41}, we first analyse its behaviour for different levels of trajectory complexity. To perform this analysis, we used 2D synthetic Lagrangian trajectories in the HIT flow case and compared the results of the coherent predictor with those obtained with classical predictor models. As illustrated in figure~\ref{fig:p3}, 2D-HIT carries a range of complex flow motions inside where numerous vortices interact, creating saddle points, strong shears, and vortical structures. Figure~\ref{fig:p10}.a shows one example of two trajectories extracted from 2D-HIT. The first one is for a smooth motion, where all predictor functions can estimate the next position with a negligible bias error. However, the bias error can significantly increase without knowing the surrounding flow motions as soon as a particle starts rotating with both velocity and acceleration changes. As shown in figure~\ref{fig:p10}.b, both lower and higher order polynomial functions, as well as the Wiener filter mispredict high variational dynamics of particles. The third-order polynomial predictor estimated the correct direction of the particle at time step $t_{n+1}$. However, the true direction estimation does not necessarily lead to a proper prediction. Misprediction in the third-order polynomial mainly resulted from the lack of correct acceleration estimation. For such a trajectory, the two other polynomial cases showed less accurate estimations. Unlike other methods, the coherent predictor has correct acceleration and direction estimations due to imposed additional constraints in its cost function. The velocity constraint governs the direction of the estimated position, while the acceleration constraint controls how far or near the prediction can go in the same direction when faced with acceleration variations. 

\begin{table}
  \begin{center}
\def~{\hphantom{0}}
\begin{tabular}{llllll}

Case & 1 & 2& 3 & 4 & 5  \\

\noalign{\smallskip}\hline\noalign{\smallskip}
Particle concentration ($\text{ppp}$) & 0.025 & 0.05 & 0.08 & 0.1 & 0.2 \\ 
Noise ratio ($\%$) & 0 & 15 & 30 & 45 & 60 \\ 
Temporal resolution  ($\times {\rm dt}_{DNS}$) & 20 & 40 & 60 & 80 & 100 \\

\end{tabular}

  \caption{Scenarios for 2D trajectory assessment.}
  \label{tab:p2}
  \end{center}
\end{table}

\begin{figure}
\centering
  \includegraphics[width=1\textwidth]{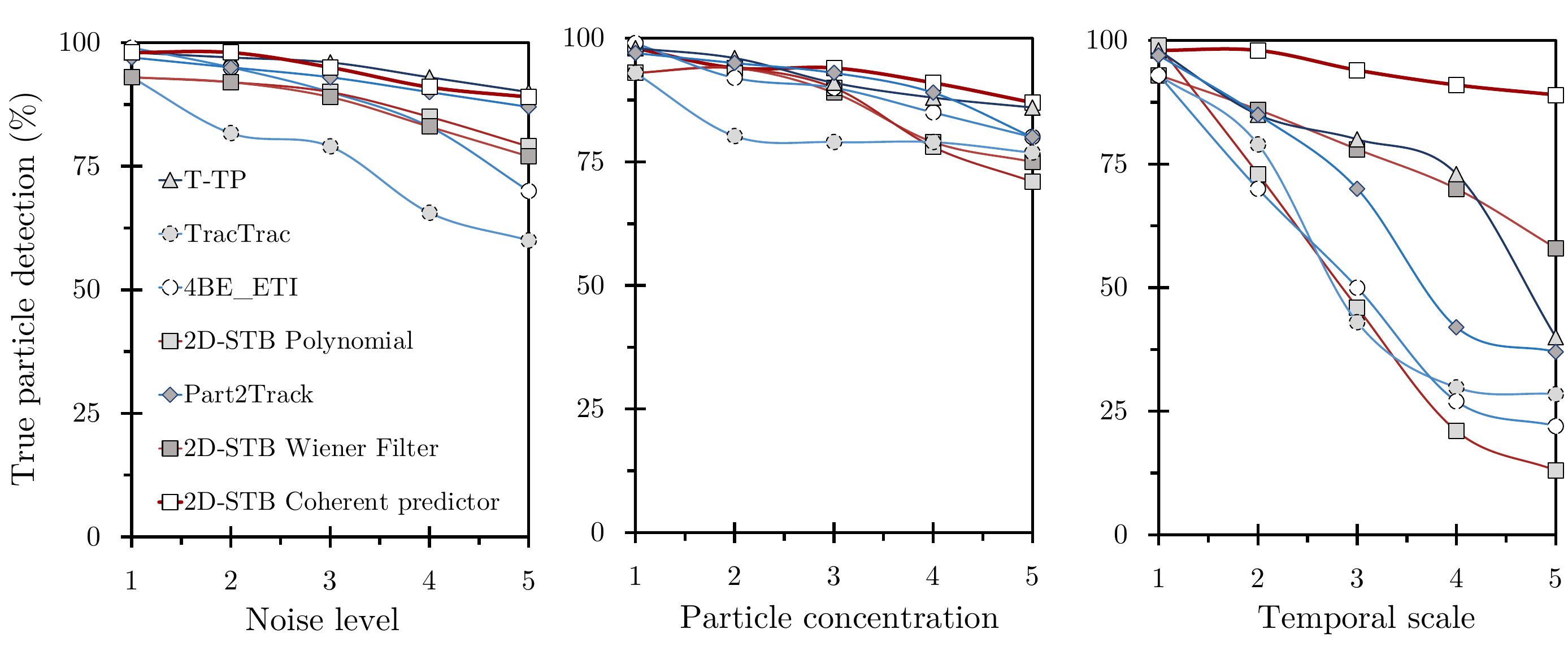}
\caption{Comparison performances of 2D Lagrangian reconstruction in terms of the fraction of true particle detection by changing each characteristic parameter. When one parameter is varied, the other two are held at noise $= 10\%$, ppp $= 0.05$, and $\Delta t = \Delta t_\mathrm{DNS}$. The noise ratio is defined as $\sigma_X / \langle|\Delta\mathbf{x}|\rangle$, the position noise standard deviation normalised by the mean single-step displacement.}
\label{fig:p11}       
\end{figure}

In the following, we evaluate the performance of the proposed prediction approach one step further with sequences of particle positions during their two-dimensional simulated trajectories. It allows not only to study the influence of parameters such as particle image concentration, noise level, and the temporal resolution of the image observations but also to compare the results with those obtained with state-of-the-art particle tracking methods. 
Noise is applied to the position, and therefore the derived velocity and acceleration values carry amplified noise levels under the same sampling.
We compared seven techniques in total. Four are open-access particle tracking codes that reconstruct 2D Lagrangian trajectories, and three are the single-track predictors (Polynomial, Wiener, Coherent) on the 2D-HIT benchmark. The first technique, TracTrac \citep{Heyman2019TracTrac:Estimation}, starts with the nearest neighbour approach and then searches for the best forward/backward match from $t_{n-1}$ to $t_{n+1}$. What sets this technique apart is the use of a predictor function based on the velocity field. TracTrac is capable of reconstructing over $1000$ tracks per second with $0.01$ pixel resolution accuracy \citep{Heyman2019TracTrac:Estimation} and was evaluated with PIV challenge cases \citep{Kahler2016MainChallenge}. Part2Track \citep{Janke2020Part2Track:Velocimetry}, the second technique, is a 2D polynomial predictive tracking method. This technique can be considered a simplified 2D version of STB, featuring a well-organised and robust code in terms of particle density. We also used 2D Enhanced Track Initialisation \citep[4BE-ETI][]{Clark2019ATracking}, representing four-frame-based techniques. 4BE-ETI examines all probabilities around the target particle and generates two consecutive predictions in the following four frames. Any particle close to the predicted position is considered for the prediction of the next frame. A solution is deemed reconstructed if a unique track is identified after four frames. Topology-based particle tracking \citep[T-PT][]{Patel2018RapidFields} is the last algorithm for the present assessment. The aforementioned technique generates particle descriptors called feature vectors of nearest neighbour particles for each frame. Subsequently, for each particle, groups of nearest neighbours are stored in the descriptor by binning these particles into a gridded bin, assuming that particles tend to remain within the same bin in the next time step. Eventually, it performs an iterative wrapping scheme to reconstruct Lagrangian trajectories. This method has been evaluated for biological flow motions \citep{Patel2018RapidFields}.

All the aforementioned techniques were assessed based on three characteristic parameters: particle concentration, temporal resolution and noise level, as detailed in table~\ref{tab:p2}. The aim is to determine how these characteristic parameters cause false Lagrangian reconstruction. When particle density is low, a simple optimisation approach would lead to building true trajectories. However, as the density increases and more particles interact and move close to each other, a more sophisticated algorithm is required to detect true tracks. In this context, we define particle density based on the number of particles per pixel ($\text{ppp}$). Adding noise level creates a realistic situation in synthetic image studies since noises are inevitable in experiments due to their nature. Therefore, the algorithm's robustness over different noise levels would provide valuable information to determine which technique is appropriate for a particular experiment. The DNS time step is smaller than the smallest timescale of turbulence, which is the order of the Kolmogorov timescale. This means the transport of particles between two DNS time steps is smooth enough to reach true prediction even with linear extrapolation. However, the experiment time step is multiple times higher than the DNS time step, depending on the acquisition setup. To reach a realistic condition, five time steps starting from every $20$ to $100$ DNS time steps are considered. All techniques receive the same starting position reconstruction. We increased each characteristic parameter individually within five scenarios (see  table~\ref{tab:p2}). Finally, we compared the deviation of the final detected positions with the ground truth data. The trajectory is reconstructed correctly if the estimation deviation is in the same order as the reconstruction accuracy.

\begin{figure}
\centering
  \includegraphics[width=0.75\textwidth]{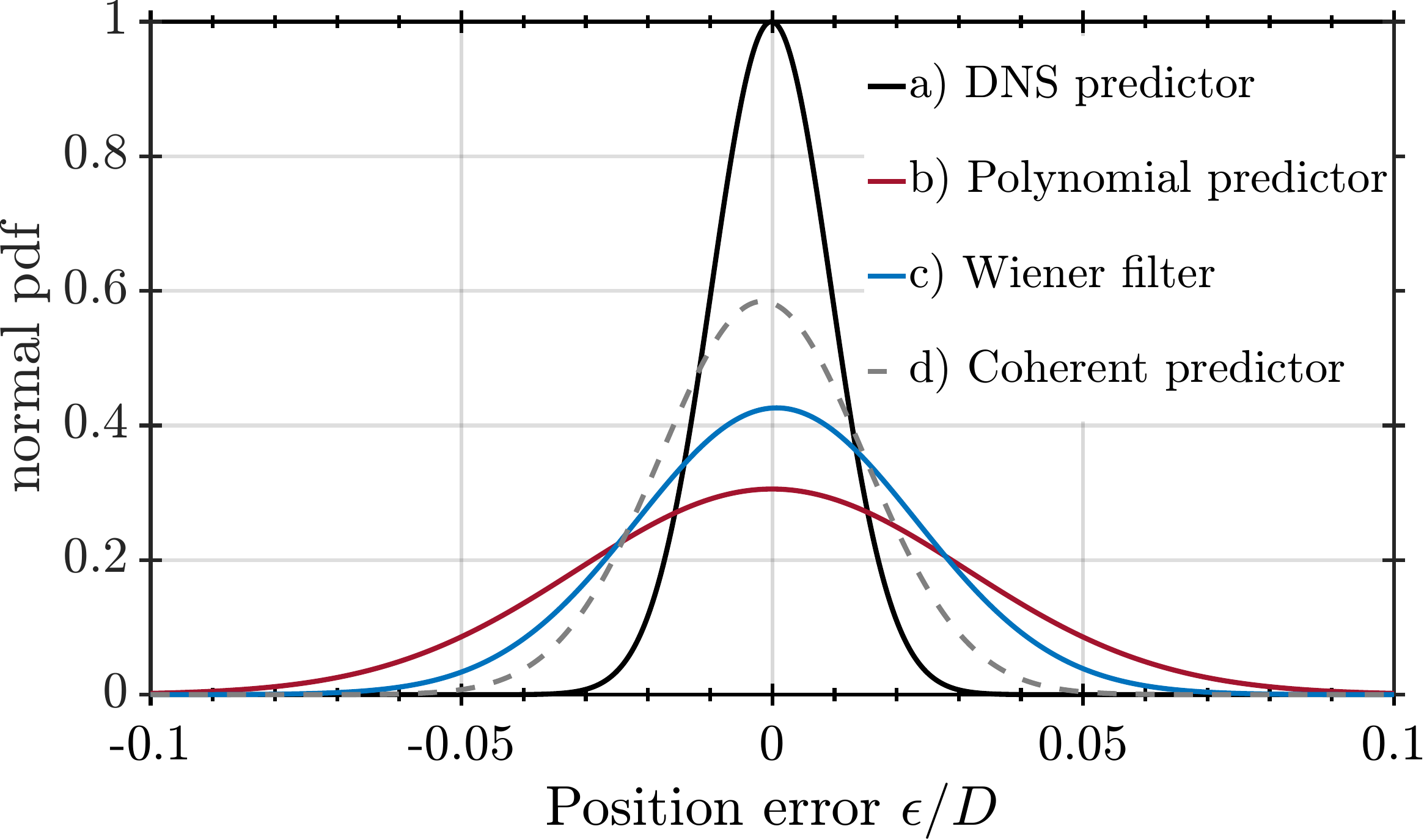}
\caption{Normal probability density function (pdf) of particle position estimation error in $x$~direction of four predictor functions.}
\label{fig:p12}       
\end{figure}

\begin{table}
\centering
\fontsize{7pt}{10pt}\selectfont 
\begin{tabular}{llll}
Method & Fit parameters & cost function & Prediction \\
& & & \\
a) DNS predictor & - & - & $y_{n+1}=\dot{\mathbf{X}}_{DNS} \cdot t_{n+1}$ \\ 
& & & \\

b) Polynomial predictor & ${X}_n=\sum_{j=0}^{\ell}a_{j}.t_n^{j}$ & $\mathcal{J}=\frac{1}{n}\sum_{i=1}^{n}\left(\mathbf{X}_i-\mathbf{y}_i\right)^2
$ & $y_{n+1}=\sum_{j=0}^{\ell}{a_{j}.t_{n+1}^{j}}$  \\ 
& & & \\

c) Wiener filter & ${X}_n=\sum_{j=1}^{\ell}w_j.u_{n}$ & $\mathcal{J}=\sum_{i=1}^{n}{{(\mathbf{X}_i-\mathbf{y}_i)}^2}
$ & $y_{n+1}=\sum_{j=2}^{\ell+1}w_j.u_{n}$\\
& & & \\

d) Coherent predictor &  $\mathcal{J}=\frac{1}{n} \sum_{i=1}^{n}\left[\left(a_{0}+a_{1} t_{i}+a_{2} t_{i}^{2}+a_{3} t_{i}^{3}\right)-\mathbf{y}_i'\right]^{2}$ & $\mathcal{J}'=\frac{1}{n}\sum_{i=1}^{n}\left(\mathbf{X}_i'-\mathbf{y}_i'\right)^2$  & $y_{n+1}=\sum_{j=0}^{\ell}{a_{j}.t_{n+1}^{j}}$\\
 & $~~~ +\alpha_{1}'\left[\left(a_{1}+2 a_{2} t_{n}+3 a_{3} t_{n}^{2}\right)-\dot{y}_{n}'\right]^{2}$  & $~~~ +{\alpha_1'}~\left(\dot{\mathbf{X}}_n'-\dot{\mathbf{y}}_{{\rm c},n}'\right)^2$ &  \\ 
 & $~~~ +\alpha_{2}'\left[\left(2 a_{2}+6 a_{3} t_{n}\right)-\ddot{y}_{n}'\right]^{2}$ & $~~~ +{\alpha_2'}~\left(\ddot{\mathbf{X}}_{n}'-\ddot{\mathbf{y}}_{{\rm c},n}'\right)^2$ &  \\
\end{tabular}
\caption{Prediction function formulations.}
\label{tab:p3}       
\end{table}

\begin{figure}
\centering
  \includegraphics[width=1\textwidth]{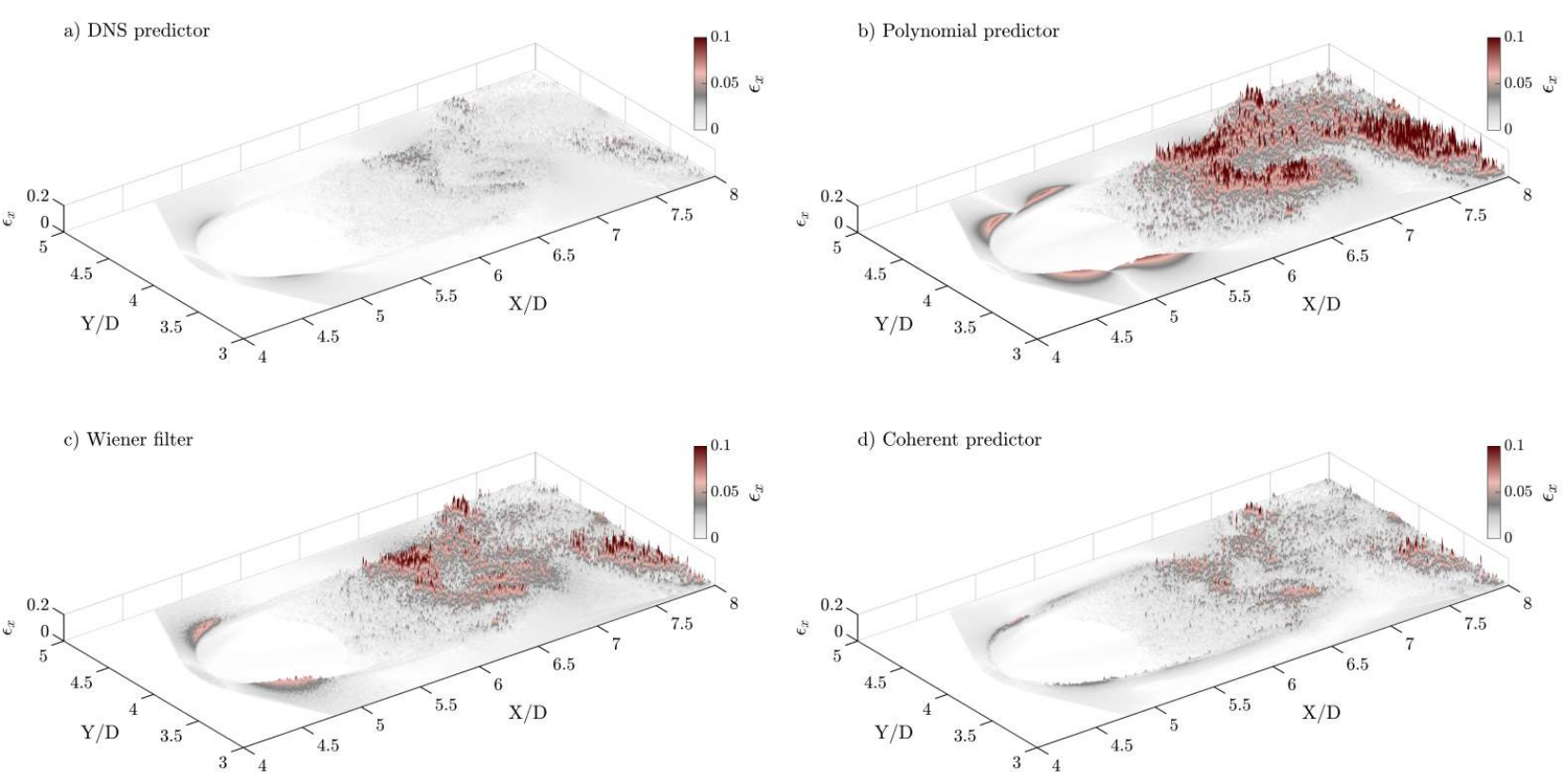}
\caption{Instantaneous position estimation error averaged in $z$~direction. (a) DNS predictor. (b) Polynomial Predictor. (c) Wiener filter. (d) Coherent predictor. Also shown in Supplementary Movie 1.}
\label{fig:p13}       
\end{figure}

Figure~\ref{fig:p11} shows the fraction of true reconstructed tracks over the total number of particles. By increasing the noise ratio up to $60~\%$ noisy reconstruction, Part2Track, T-PT, and coherent predictor tend to keep their robustness. Meanwhile, other techniques faced significant drops, losing nearly half of the true trajectories. Among all techniques, TracTrac has the most sensitivity to the noise ratio. Figure~\ref{fig:p11} shows that particle concentration has almost the same impact on all techniques. Their performance decreases in the same order as the concentration increases. By increasing the time step, we cannot observe small-scale motions of particles, so less temporal information is available. The number of true particles drops severely with increased temporal resolution. In all mentioned techniques, relying only on a single particle as a single signal to find its true Lagrangian motion, while missing the small-scale dynamics between two observations, causes more wrong trajectories. However, even a weak signal of coherent particle behaviour would lead to correct direction and prediction. Results showed that when adding spatial and temporal coherent information, the prediction function remained robust for up to $85~\%$ in all situations, while other techniques suffered from a lack of information. In summary, our sensitivity analysis shows that the proposed prediction function performs well across the tested scenarios. The 2D-HIT evaluation shows that our approach handles complex flow motions and vortical structures well, even at low temporal resolution.

Finally, we examine how the polynomial baseline itself depends on the order $\ell$ and the history length $k$ to make sure that the coherent predictor is compared against a properly tuned reference. Figure~\ref{fig:p10}.c shows the noise-free position-prediction RMS error (normalised by the characteristic displacement) as a function of $k$ for polynomial orders $\ell = 2$--$5$, pooled over all four 3D cylinder-wake DNS datasets ($\approx 760\,000$ prediction events)\footnote{The grid is restricted to $k \geqslant \ell+1$, so the point $(\ell,k) = (5,5)$ is excluded to keep the underlying Vandermonde system non-singular.}. We observe a crossover at $k \approx 10$. For short histories ($k \leqslant 8$), the quadratic polynomial ($\ell = 2$) achieves the lowest error because the limited data cannot constrain the cubic and quartic coefficients. For longer histories ($k \geqslant 15$), the cubic polynomial ($\ell = 3$) becomes clearly superior as it captures the trajectory curvature. At our working point $k = 5$, the performance of $\ell = 2$ and $\ell = 3$ is comparable (RMS $0.058$ vs $0.084$). We adopt $\ell = 3$ with $k = 5$ for three reasons. First, $k = 5$ is a realistic operating point for PTV, since real tracks are short due to particles entering and leaving the measurement volume. Second, the same $k$ is used for all three predictors (Polynomial, Wiener, Coherent), which ensures a fair comparison. Third, the coherent predictor requires cubic polynomial order to independently constrain velocity and acceleration in the cost function~\eqref{Eq:p28}. Therefore, $\ell = 3$ is the appropriate choice for this study. It is competitive at $k = 5$, it is required by the coherent cost function, and it becomes the clear optimum as more history becomes available.

\subsection{Contribution of secondary coherent neighbours}
\label{sec:p42b}

We introduced in \S~\ref{sec:p221} the concept of primary and secondary coherent neighbours and motivated the use of the secondary ones as additional prior knowledge about the target particle's future. In this subsection, we quantify to what extent the secondary coherent neighbours improve the prediction on top of the primary ones. For this purpose, we constructed three predictor variants. The first one is the polynomial predictor (Poly), taken as a reference. The second one is the coherent predictor using only the primary neighbours (P). The third one is the coherent predictor using a pooled set of primary and secondary neighbours, where the secondary set is constrained to exclude the primary ones (P\,+\,S). For all three cases, the same cost function presented in~\eqref{Eq:p28} is used, with the coherent velocity $\dot{\mathbf{y}}_{{\rm c},n}$ and acceleration $\ddot{\mathbf{y}}_{{\rm c},n}$ computed from the weighted average of the pooled set. The non-dimensional weights take the values $\alpha_1' = 0.1$ and $\alpha_2' = 20$, identified from the parametric study of \S~\ref{sec:p41}. The weighting function follows the same formulation introduced in \S~\ref{sec:p221}. Smoothed velocity and acceleration estimates are used for both primary and secondary neighbours to treat the two sets consistently. The test case is the DNS $Re=3900$ wake flow, where $600$ trajectories are evaluated over $15$ time steps, leading to $9000$ prediction samples. Unless stated otherwise, the percentage reductions reported in this subsection and in the remainder of the paper are single-step RMS position-error reductions pooled over all prediction samples, normalised by the polynomial-baseline RMS error on the same sample set.

\begin{figure}
\centering
  \includegraphics[width=1\textwidth]{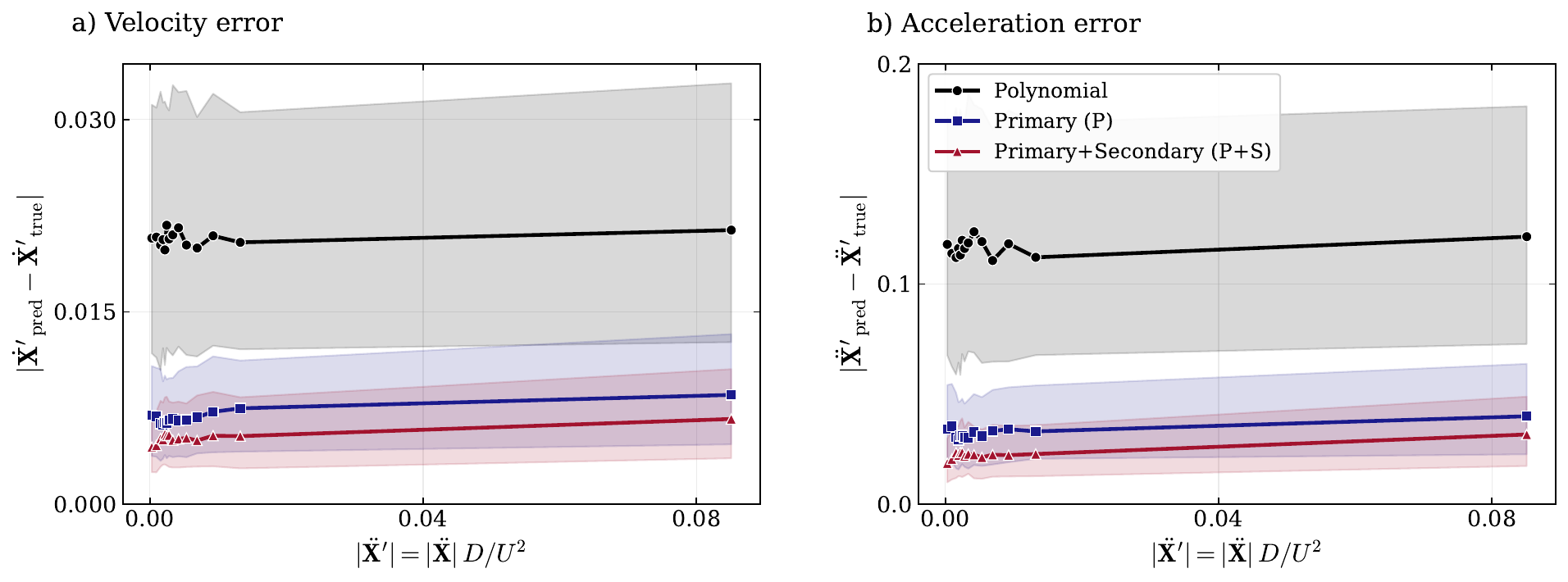}
\caption{Absolute prediction error as a function of the Lagrangian acceleration for the polynomial predictor (Poly), the primary coherent predictor (P), and the primary\,+\,secondary coherent predictor (P\,+\,S). DNS $Re=3900$ wake flow case, $9000$ prediction samples.}
\label{fig:p_pvsps}
\end{figure}

Figure~\ref{fig:p_pvsps} shows the absolute prediction error as a function of the Lagrangian acceleration for the three predictors. The primary coherent predictor (P) reduces the velocity error by $66.5\%$ and the acceleration error by $71.2\%$ compared to the polynomial predictor. When the secondary coherent neighbours are added to the pool (P\,+\,S), the reduction further increases to $74.2\%$ for the velocity error and $79.7\%$ for the acceleration error. This corresponds to an additional $23.0\%$ reduction in velocity error and $29.4\%$ reduction in acceleration error on top of what the primary neighbours already achieve. The ordering Poly $<$ P $<$ P\,+\,S holds for both velocity and acceleration and becomes more visible in high-acceleration regions, where the Lagrangian dynamics are complex and the polynomial extrapolation from history alone is the most penalised. The constraint of excluding the primary neighbours from the secondary set is necessary, as an overlapping set would dilute the additional signal brought by the secondary neighbours. This comparison confirms that the secondary coherent neighbours carry independent prior information about the target particle's future and that their inclusion improves the Lagrangian trajectory prediction beyond what the primary neighbours alone can provide.

Further robustness analyses on the 3D cylinder wake for different seeding densities, temporal resolutions, and noise levels are presented in Appendix~\ref{appB}. A Python implementation of the polynomial, primary, and primary plus secondary predictors compared in this subsection is available in the \texttt{01\_core\_predictor.ipynb} notebook of the companion repository (see Code availability statement).

\subsection{Confidence analyses}
\label{sec:p43}
In this section, we aim to assess the reliability and accuracy of Lagrangian prediction methods by comparing their performance against ground truth 3D wake flow trajectories at $Re = 3900$. We evaluate four schemes, including the DNS predictor, Wiener filter, polynomial predictor, and coherent predictor (see table~\ref{tab:p3}). In the following section, we analyse the position prediction error by decomposing it into two general sources, the bias error and the measurement uncertainties. We calculate the bias error from the deviation of the ground truth trajectories and the predicted positions estimated from perfect observations. So the bias error is caused by a misprediction of the flow dynamics. Then, to obtain the measurement uncertainty, we conduct Monte Carlo simulations.

Before proceeding to the bias and Monte Carlo analyses, we first quantify how reliable the predictor inputs are at the endpoint of a Lagrangian track, where the velocity and acceleration estimates are the most affected by the measurement noise. At the endpoint, only a backward-looking finite-difference stencil is available for $\dot{x}$ and $\ddot{x}$, while the mid-track estimates benefit from centred stencils. The resulting noise amplification can be computed analytically for unit positional noise $\sigma$. The $4$-point third-order backward stencil is $5.4$ times noisier than the $3$-point centred stencil on velocity and $2.8$ times noisier on acceleration. We evaluated this on the DNS $Re=3900$ wake test case over $11\,500$ prediction samples under a $10\%$ positional noise model. The empirical signal-to-noise ratios of the finite-difference estimates drop from mid-track to endpoint as expected, matching the theoretical stencil amplification to within $0.4\%$. This confirms that the endpoint acceleration is dominated by noise at our working noise level, which the LPT-challenge accuracy cited in \S~\ref{sec:p41} does not reflect.

The numbers here look larger than in \S~\ref{sec:p42b} because we are now testing the predictors against a polynomial baseline that suffers from realistic endpoint noise. The dataset is the same, and the order Poly, P, P\,+\,S still holds on acceleration. On velocity, the secondary constraint of P\,+\,S forces the predictor towards the next step, which is why P\,+\,S becomes worse than P near the trajectory endpoint. The primary coherent predictor (P) reduces the error relative to the polynomial baseline by $+77.6\%$ on velocity and $+79.0\%$ on acceleration over the full sample. Within the lowest-SNR quartile, the reduction remains $+75.5\%$ and $+77.3\%$, which is a spread of only $2$ percentage points between the full sample and the most degraded sub-population. The coherent-motion constraint was introduced for this purpose. Replacing the locally noisy finite difference with a weighted coherent-neighbour average makes the prediction insensitive to the endpoint noise level, since the coherent estimate's noise scales as $1/\sqrt{N_c}$ rather than with the stencil amplification factor. Adding the secondary term (P\,+\,S) provides a further $+3.5$ percentage points on acceleration, which is the quantity with the worst endpoint SNR. On velocity, P\,+\,S is substantially below P at the endpoint, reducing the error by $+51\%$ against $+78\%$ for P, because the secondary position constraint at $\tau_{n+1}$ pulls the predictor away from the measured endpoint position, which is precisely the reference against which the endpoint velocity is differenced. In the low-SNR sub-population this bias is absorbed by the dominant measurement noise and P and P\,+\,S perform the same. These results confirm that the coherent-motion constraint provides its largest benefit where the endpoint finite-difference estimates are worst, and they directly address the concern that the LPT-challenge noise model in \S~\ref{sec:p41} underestimates the endpoint degradation of velocity and acceleration in a live single-step predictor.


\subsubsection{Bias error}
\label{sec:p431}

The bias error increases due to the lack of accuracy in the prediction model. We defined the DNS predictor as a reference using the Euler integration method to transport particle positions by the ground truth DNS positions and velocities for every $20$ DNS time steps. In such a scenario, we can estimate the error achieved for this sparse temporal resolution and perfect observations, which can be considered the minimum bias. Figure~\ref{fig:p12} shows the normal probability density function (pdf) of the predicted position errors in the $x$~direction of four schemes. The bell curve distribution is obtained by spatiotemporal averaging of trajectory errors for all particle positions over $200$ time steps (i.e., time-averaged). Position error in $x$~direction shows that the deviations of the coherent predictor remain virtually below  $\epsilon/D=0.05$, where $D$ is the cylinder diameter. On the contrary, a significant number of particles are mispredicted in both polynomial and Wiener filter techniques. Similar significant improvements by using the coherent predictor are observed in $y$ and $z$~directions. The polynomial and Wiener filter predictors require nearly the same compute, while the coherent predictor takes about four times longer on a single CPU core. Figure~\ref{fig:p13} shows the instantaneous projected distribution of the bias error on the $xy$ plane for each predictor function. The prediction error is highly correlated with the flow topology in all schemes. 

\begin{figure}
\centering
  \includegraphics[width=1\textwidth]{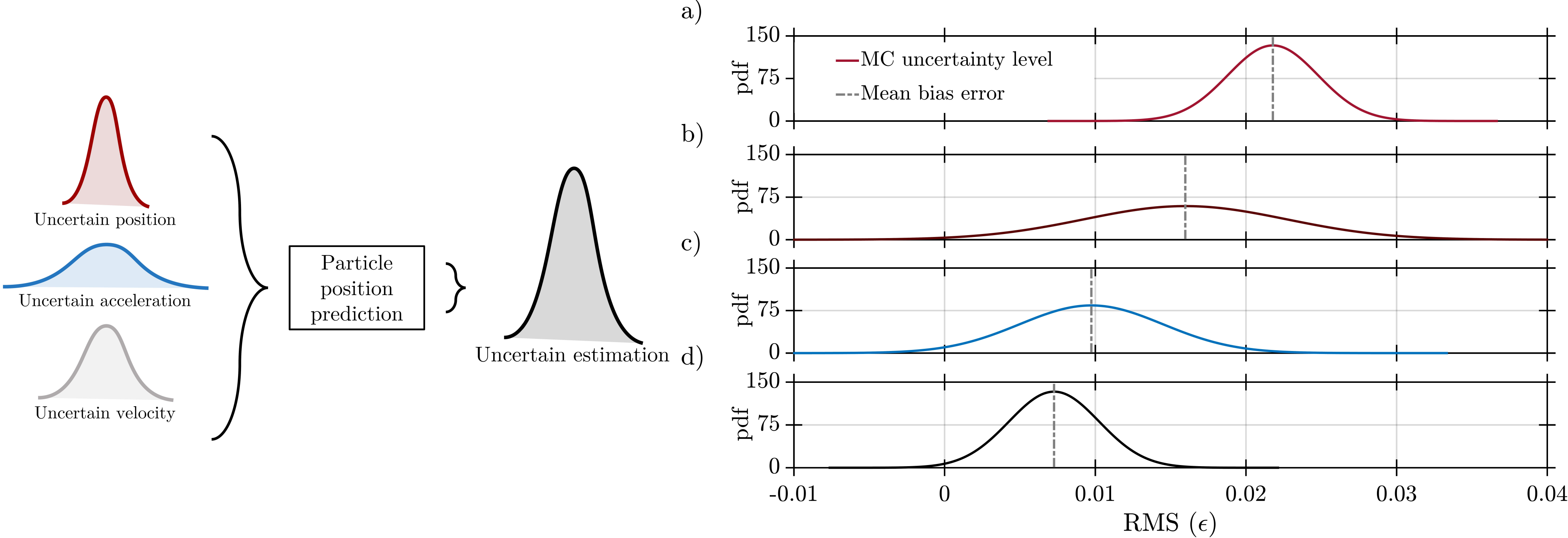}
\caption{Monte Carlo uncertainty quantification (MC-UQ). Left: schematic of the MC-UQ algorithm with distributed input parameters. Right: probability distribution around the bias error obtained from the MC-UQ of four predictor functions. (a) Second-order polynomial. (b) Third-order polynomial predictor. (c) Wiener filter. (d) Coherent predictor. Blue dashed lines represent the averaged bias error over $10000$ Lagrangian trajectories.}
\label{fig:p14}       
\end{figure}

In this classic wake flow case, the flow topology is captured by preserving the areas where motion gradients are produced. There are three such regions, the boundary layer, the two sideward shear layers at the edge of the formation region, and the wake. The boundary layer refers to the thin layer of fluid that forms along the surface of the cylinder. In this region, Lagrangian trajectories face strong deceleration from the free-stream velocity to zero at the cylinder's surface (due to the no-slip condition) as we move close to the cylinder's leading edge. A strong acceleration gradient also exists in the shear. Inside the wake, on the other hand, the complexity is different and arises from numerous small-scale motions and chaotic behaviour of Lagrangian trajectories. We observe direct signatures of these structures on the instantaneous error. Although the DNS predictor (see figure~\ref{fig:p13}.a) uses known ground truth velocity information, the travelling distance between two observations is large enough to introduce minor errors, particularly in the near wake region where it is chaotic. As shown in figure~\ref{fig:p13}.b, the third-order polynomial has the worst prediction error, which can be up to $\epsilon/D=0.1$ around the cylinder leading edge and inside the wake region. The polynomial prediction error distribution is thoroughly shaped by the flow motion (i.e., topology), meaning that any variations inside the flow create a huge estimation error. Overall and local performance of the Wiener filter is better than the polynomial predictor. The Wiener filter succeeded in reducing the prediction error in most of the peak regions (see figure~\ref{fig:p13}.b.c). The coherent predictor showed the best performance locally and globally compared to Wiener and polynomial predictors.

\smallskip
\noindent \textit{Region-conditioned error via the $Q$-criterion.} The spatial error maps in figure~\ref{fig:p13} show that the prediction error correlates with the local flow topology. To quantify this dependence, we partition the $12\,000$ prediction samples by the $Q$-criterion of the underlying Eulerian field. The $Q$-criterion is defined as $Q = \tfrac{1}{2}(\|\Omega_{ij}\|^2 - \|S_{ij}\|^2)$, where $\Omega_{ij}$ and $S_{ij}$ are the antisymmetric and symmetric parts of the velocity gradient tensor. Negative $Q$ identifies strain-dominated regions (braid and shear layers), while positive $Q$ identifies rotation-dominated regions (vortex cores). For each prediction sample, $Q$ is interpolated trilinearly from a representative DNS Eulerian snapshot onto the particle position at the prediction instant. We use the signed $Q$ rather than $|Q|$ because it distinguishes between the topological types (strain versus rotation) and the gradient intensity simultaneously, which is needed to test whether the coherent predictor advantage depends on the local flow regime.

\begin{figure}
\centering
  \includegraphics[width=1\textwidth]{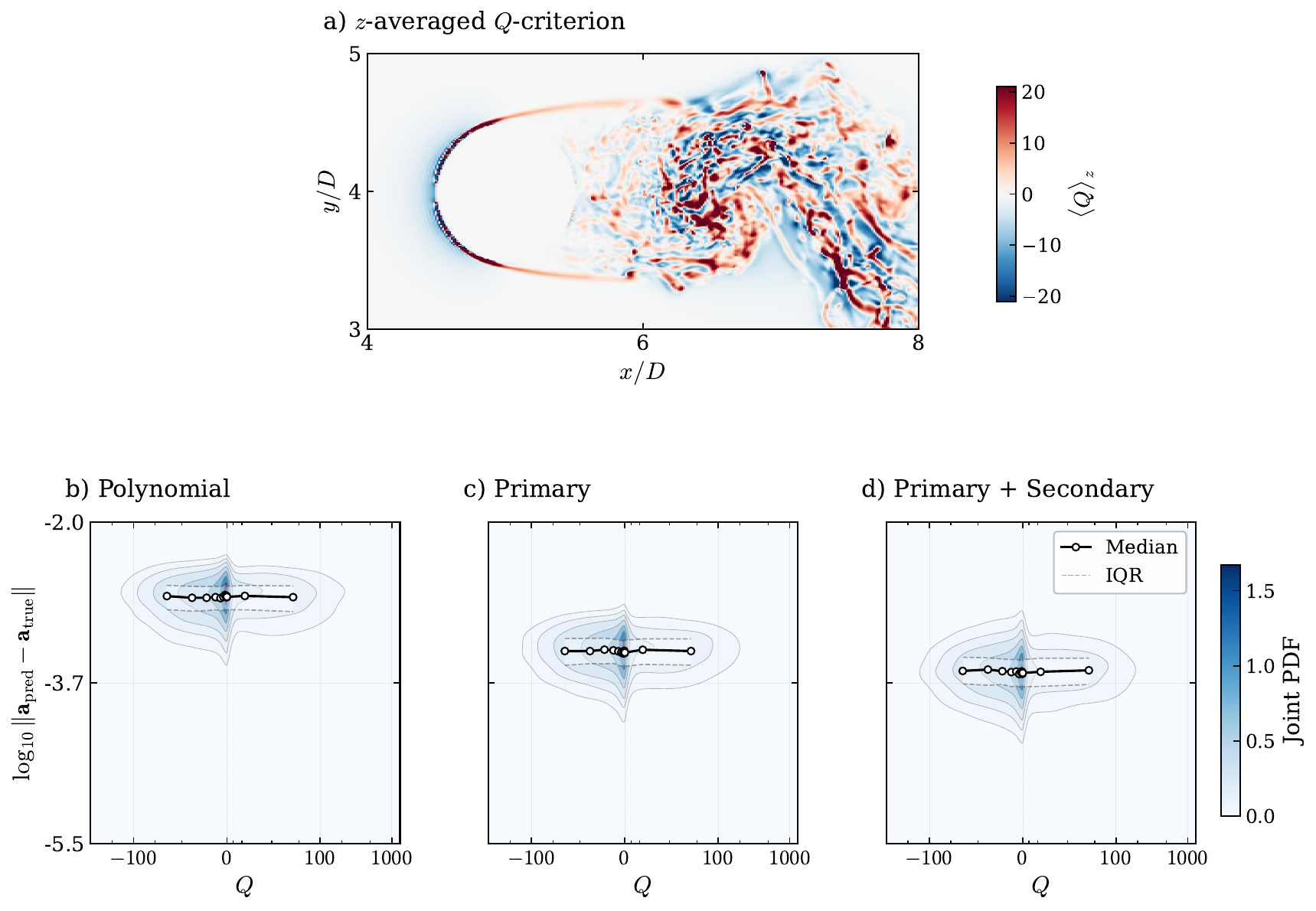}
\caption{Region-conditioned prediction error partitioned by the Eulerian $Q$-criterion. (a) $z$-averaged $Q$ field of the DNS $Re = 3900$ cylinder wake (representative snapshot). The wider $Q$ range in (b)-(d) reflects the spread of $Q$ values sampled at particle positions, including rare events in vortex cores and shear layers, beyond the bulk-field range shown in (a). Joint probability density of $Q$ and $\log_{10}\|\mathbf{a}_{\rm err}\|$ for the (b) Polynomial, (c) primary coherent (P), and (d) primary\,+\,secondary coherent (P\,+\,S) predictors. White dots show the binned median, dashed lines the interquartile range. The density cloud contracts from (b) to (d) while the median remains flat across the full $Q$ range, confirming topology-independent error reduction.}
\label{fig:p_qcrit}
\end{figure}

Figure~\ref{fig:p_qcrit} shows the joint density of the prediction error and $Q$ for the three predictors. The top panel displays the $z$-averaged $Q$ field for reference. The bottom panels show the joint probability density of $Q$ and the logarithmic acceleration error for the Polynomial, P, and P\,+\,S predictors. The density cloud contracts from Polynomial to P to P\,+\,S, and the binned-median acceleration error decreases monotonically from Polynomial to P to P\,+\,S in every $Q$ bin. The median's flatness across $Q$ in each panel shows the reduction is independent of the local flow topology. When the samples are divided into four $Q$-quartiles (from strong strain to strong rotation), the P\,+\,S RMS acceleration reduction relative to the polynomial baseline is $82.0\%$, $83.0\%$, $82.5\%$, and $82.1\%$, a total span of $1.0$ percentage point. The velocity reduction spans $9$ percentage points ($67.5$--$76.1\%$), with the extremes (strong strain and strong rotation) slightly harder than the intermediate regimes, as expected since the highest velocity gradients occur there. This analysis confirms that the coherent-motion advantage is practically independent of the local flow topology, from strain-dominated braid regions to rotation-dominated vortex cores, and that the error reduction reported in the preceding sections is not an artefact of averaging over different flow regions.

We note that \citet{Schanz2022PositionDensities} analyse the working range of particle position optimisation for Wiener-filter predictions at temporal spacings $\Delta t = 80$--$320\,t_{\rm DNS}$ on the same $Re = 3900$ cylinder-wake DNS used here. Their metric is the fraction of particles whose position can be corrected back to within $1\,\overline{px}$ of the true location, which decreases with $\Delta t$ as large prediction errors exceed the correction range. Our approach is complementary. Rather than correcting large mispredictions after the fact, the coherent cost function improves the prediction step itself, and the $\Delta t$ variation in Appendix~\ref{appB} shows that the coherent predictor is nearly constant ($+4\%$ over a $5\times$ increase) while the Wiener filter degrades by $57\%$.

\subsubsection{Monte Carlo uncertainty quantification (MC-UQ)}
\label{sec:p432}
For each individual trajectory, we conducted a Monte Carlo simulation to quantify the uncertainty level of the prediction function. In Monte Carlo uncertainty quantification (MC-UQ), the parameter distributions of models are sampled randomly, followed by statistics calculated on the output model distribution \citep[see][]{JointCommitteeforGuidesinMetrology2008EvaluationMethod}. Figure~\ref{fig:p14} schematically shows the probability distribution of the predicted positions as a function of distributed input uncertainties. To start the MC-UQ simulation, we need to quantify the uncertainty level of the input parameters that are fed into the predictor function. In the classic 4D-PTV process (see \S~\ref{sec:p32}), the prediction function receives positions without trajectories from \textit{iterative particle reconstruction} (IPR). The uncertainty level of IPR can be utilised as an input parameter for uncertainty quantification of the prediction function. This can be obtained from numerical or analytical IPR performance analyses reported by \citet{Wieneke2012} and \citet{Jahn2021AdvancedTracking}. However, these values are optimistic and might differ in practical conditions due to the IPR tuning parameters. Instead, similar to \S~\ref{sec:p41}, we average the estimation errors of all participants in the 1st LPT challenge \citep{Sciacchitano2021MainChallenge,Leclaire2021FirstBenchmark,Schroder20233DMechanics} to mimic practical and representative uncertainty levels that are introduced to a predictor function. Then, we can quantify the output uncertainty level of predictor functions using MC-UQ with estimated input normal distributions.  

MC-UQ process of four prediction functions, second-order polynomial, third-order polynomial, Wiener filter, and coherent predictor, for nearly $10000$ trajectories, are shown in figure~\ref{fig:p14}. We need to subtract the bias error from the predicted position error obtained from uncertain input parameters to decompose the impact of uncertainty with the impact of flow motion behaviour. MC-UQ requires nearly $10000$ iterations per trajectory to achieve a smooth Gaussian distribution in the output. As a result, the total number of $100$ million ($10000 \times 10000$) predictions was computed for each predictor function. Figure~\ref{fig:p14}.a.b shows that as the order of magnitudes in the polynomial predictor increases, the uncertainty level in position estimation rises while the bias error decreases. This suggests an inverse correlation between bias error and uncertainty level when using polynomial predictors. It can be concluded that the third-order polynomial maintains better overall accuracy and uncertainty performance compared to the second-order polynomial. The Wiener filter outperformed the third-order polynomial function in terms of both bias error reduction and narrower uncertainty distribution. The coherent predictor's uncertainty level was found to be minimum and equivalent to that of the second-order polynomial. When considering both bias error and uncertainty level, the coherent predictor provided an optimal balance in comparison to the other predictors.

\begin{figure}
\centering
  \includegraphics[width=1\textwidth]{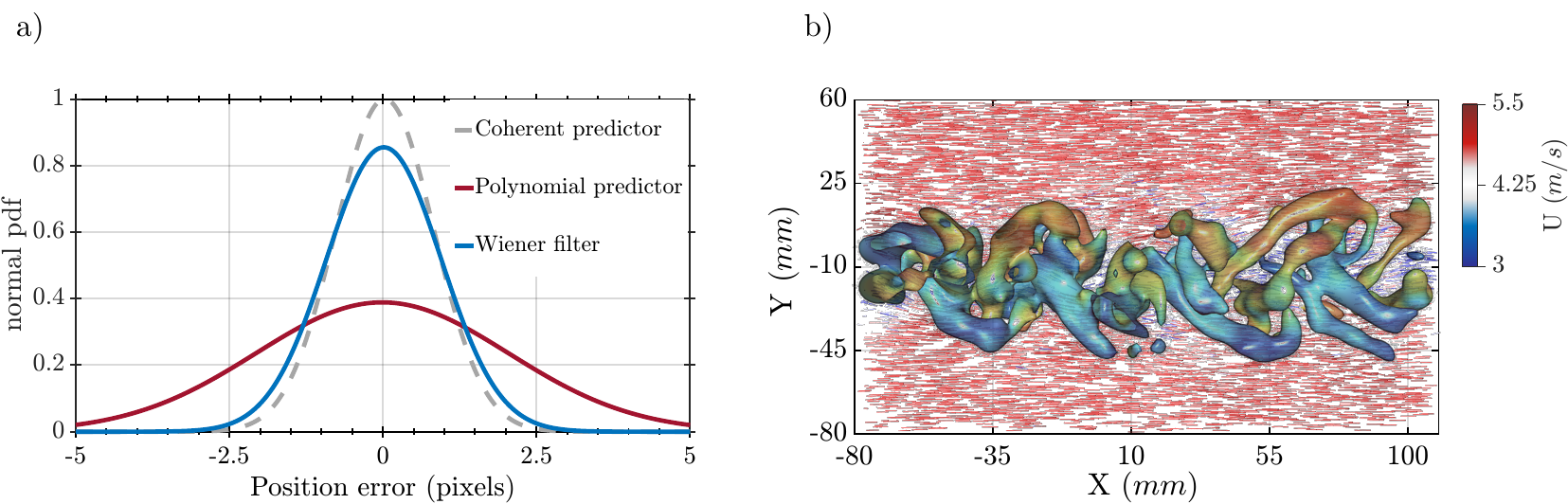}
\caption{The cylinder wake flow at Reynolds $3900$. (a) Experiment normal pdf results of particle position error in the $x$~direction of three predictors. Each predictor is compared with the final optimised positions of STB Davis. (b) Side view of particle trajectories superimposed by vorticity iso-surfaces.}
\label{fig:p16}       
\end{figure}

\subsection{Experimental demonstration}
\label{sec:p44}

To quantify the results of different schemes, we compared predictions with optimised positions obtained from STB Davis 10 (LaVision GmbH). Optimised positions refer to the ones acquired after the minimisation process, where the camera image residuals reach a minimum, commonly known as shaking. As a result of the experiment, STB successfully constructed nearly $12000$ particles. Lagrangian trajectories of the current experiment with superimposed vorticity iso-surfaces are shown in figure~\ref{fig:p16}.b. We successfully reconstructed long trajectories with a duration of up to $300$ time steps, with an average length of approximately $30$ time steps. The experimental results captured the wake structures, such as braid vortices and von Kármán vortex streets, that develop after the formation region behind the circular cylinder. These flow structures show the complexity of the flow field and the need for accurate prediction methods. Even in the instantaneous snapshot of the experiment (refer to figure~\ref{fig:p4}.b), coherent motions formed by passive transport barriers \citep{Haller2023TransportData} are readily visible to the naked eye. This observation from raw images shows that neighbour coherent particles follow similar dynamics and together form large-scale flow patterns while respecting the passive transport barriers. Therefore, incorporating the dynamics of these coherent neighbours can indeed contribute to improved predictions of Lagrangian trajectories.

We compared three techniques, polynomial, Wiener filter, and coherent predictors, with final optimised positions. For the coherent predictor, neighbour velocities and accelerations are obtained from a 5-point quadratic fit applied to each neighbour trajectory, the same smoothing applied to the target track. The deviation of position estimated of each technique is shown in figure~\ref{fig:p16}.a. The distribution shows that the coherent predictor has more accurate estimations within $1$ pixel deviation from the optimised positions. Position estimations of the Wiener filter and coherent predictors stay below $2.5$ pixels deviation for nearly all particles. On the contrary, the polynomial predictor has maximum deviation. The Wiener filter result is biased in this comparison, since the optimised positions (after shaking) in Davis are themselves obtained from a Wiener predictor.
We had to respect the original STB approach to obtain the optimised positions, which is why the Wiener filter performs better in this case than the synthetic case.  The coherent predictor gives its clearest advantages near strong shear regions where local separation is well-defined. When the flow is weakly deformed or tracer density is high, the gains over Wiener filtering are modest.

\begin{figure}
\centering
  \includegraphics[width=1\textwidth]{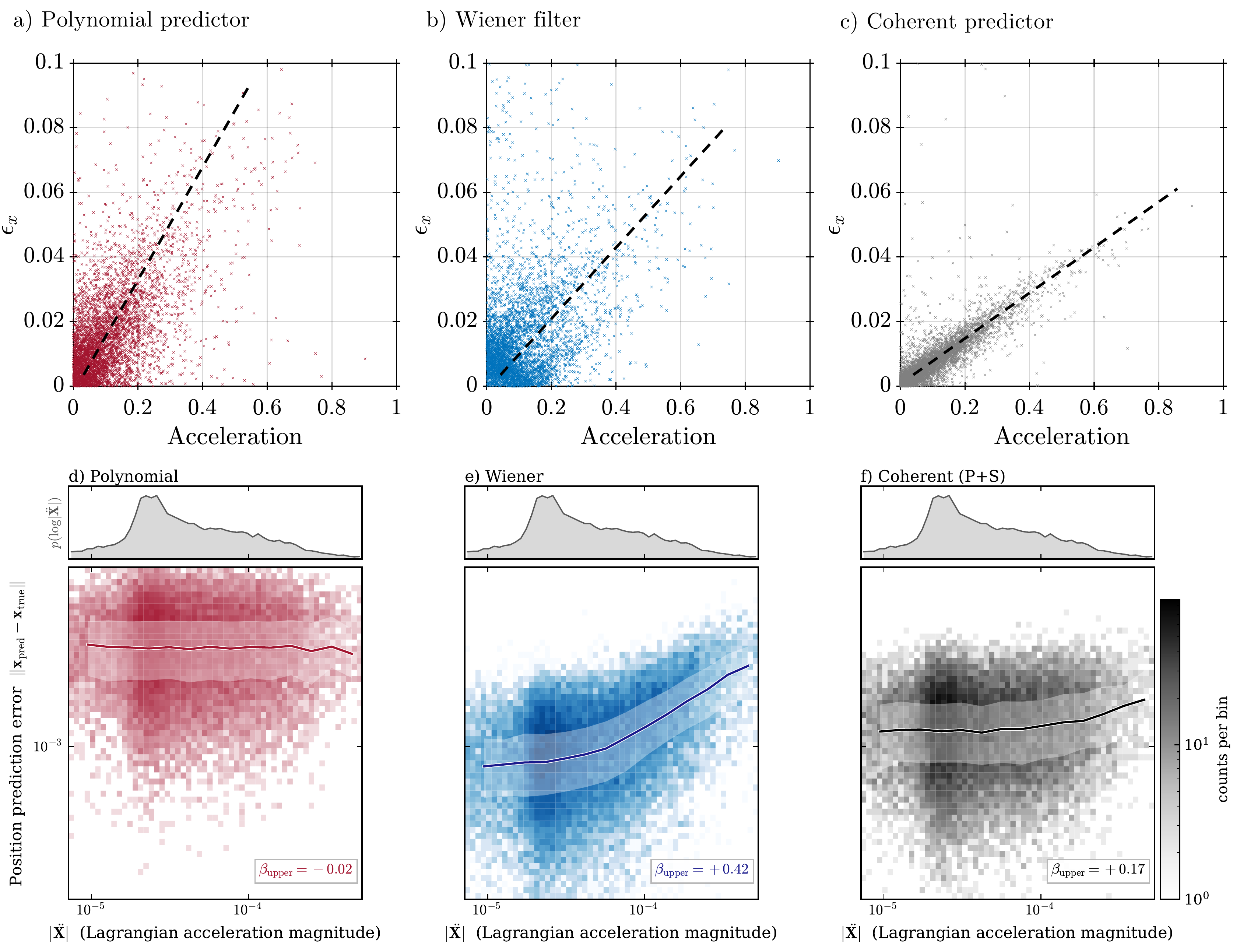}
\caption{Correlation between the magnitude of acceleration and the estimation error. Top row: experimental scatter of position-prediction error $\epsilon_x$ versus acceleration for (a) polynomial predictor, (b) Wiener filter, and (c) coherent predictor. Bottom row: joint density of position-prediction error and local acceleration magnitude for (d) polynomial predictor, (e) Wiener filter, and (f) coherent predictor on the DNS $Re=3900$ wake ($25\,650$ samples, $10\%$ positional noise). The Wiener filter density stretches diagonally into the high-$(|a|, \epsilon)$ corner, while the coherent predictor density remains nearly horizontal, visually confirming the $\beta$-slope separation.}
\label{fig:p17}       
\end{figure}

We can further assess the distribution of the estimation deviation as a function of Lagrangian acceleration. Figure~\ref{fig:p17} shows the correlation of predictor functions with acceleration. In agreement with the bell curve distribution in figure~\ref{fig:p16}, the coherent predictor has less deviation than the other two techniques. Both polynomial and Wiener filter predictors show scattered correlations against acceleration. The linear regression model for the polynomial predictor (the black dashed line in figure~\ref{fig:p17}.a) tends to explode as Lagrangian acceleration increases. Wiener filter succeeded in slightly controlling the regression slope. Coherent predictor, on the other hand, tends to fit inside narrow upper and lower bounds along with the regression line. This implies that the coherent predictor has a strong correlation with Lagrangian acceleration and effectively controls error amplification as the acceleration increases.

The marginal position-error distribution in figure~\ref{fig:p16} and the conditional error-versus-acceleration in figure~\ref{fig:p17} appear to give different impressions of the predictor performance. The marginal PDFs partially collapse, while the conditional plot reveals a clear separation. These two views are, in fact, projections of the same joint distribution $p(\epsilon, |a|)$, shown in the bottom row of figure~\ref{fig:p17}. Because the acceleration magnitude follows a heavy-tailed distribution where low-$|a|$ samples outnumber high-$|a|$ by approximately $10{:}1$, the marginal PDF is dominated by the regime where all predictors converge. This is why the marginal hierarchy appears compressed. On the other hand, projecting onto the conditional median $\epsilon_{50}(|a|)$ exposes how strongly each predictor couples with the local acceleration. To quantify this coupling, we fitted log--log regression slopes $\beta$ of the conditional median error over the upper half of the sampled $\log|a|$ range on the DNS $Re=3900$ wake ($25\,650$ valid samples, $10\%$ positional noise). The polynomial predictor gives $\beta_{\rm Poly} = -0.02$, confirming that its error is essentially acceleration-insensitive and noise-dominated. The Wiener filter gives $\beta_{\rm Wiener} = +0.42$, meaning that its error scales as $|a|^{0.42}$. This is a consequence of the linear-extrapolation failure in highly curved trajectories. The coherent predictor gives $\beta_{\rm Coherent} = +0.17$, which is $2.5$ times flatter than the Wiener filter. This shows that the coherent cost function, by imposing velocity and acceleration constraints, effectively controls the error growth with acceleration. This slope ratio describes how strongly each predictor couples with the local curvature, which is different from the absolute RMS gap. At the top $|a|$ decile, the absolute RMS ratio is $\text{RMS}_{\rm Wiener}/\text{RMS}_{\rm Coherent} = 1.20$, and both quantities should be considered together. As a further check, we computed the fraction of the total mean-squared error contributed by the top-$10\%$ acceleration tail. If the error were uniformly distributed, this fraction would be $10\%$. We found that the Wiener filter concentrates $27.0\%$ of its MSE in this tail (nearly $3$ times over-represented), while the coherent predictor concentrates only $14.1\%$ (close to uniform). These results explain the apparent discrepancy between figures~\ref{fig:p16} and~\ref{fig:p17}. The marginal PDF integrates the conditional over all acceleration values. Since the low-$|a|$ regime where all predictors converge dominates the integral, the marginal hierarchy is compressed.

\section{Conclusions and outlook}
\label{sec:p5}
In this study, we addressed the challenge of precise motion estimation of Lagrangian trajectories in 
experimental
fluid dynamics, particularly in the presence of temporal resolution limitations and high motion gradients. The main idea is to shift from traditional single-particle treatment to groups of coherent particles as they share the same dynamics. Inspired by the use of advective Lagrangian coherent structures, we developed a new physics-informed cost function called the \emph{coherent predictor} that takes into account the history of trajectories and local spatial and temporal coherent motions for less biased and lower uncertainty predictions compared to existing methods. We perform a local segmentation approach using FTLE to quantify primary and secondary coherent neighbours. Primary neighbours follow the target particle's path, while secondary neighbours exhibit a phase delay and are ahead in their history. Both carry useful information. Primary neighbours give an immediate signal, while secondary neighbours give prior knowledge about the target particle's future.

To assess the proposed approach, we performed the synthetic analysis of the 2D-HIT flow at a Reynolds number of $3000$ and the wake behind a smooth cylinder at Reynolds numbers equal to $3900$ and $300$, with the same measurement input uncertainty levels. The optimal solution of the cost function for all synthetic cases showed similar weighting configurations. The similar behaviour in all test cases led us to the conclusion that modelling the weighting parameters based on the measurement uncertainty is a feasible investigation. We performed further parametric studies and quantified a model that receives the measurement uncertainties to design the optimal cost function. Therefore, the optimal solution of the cost function is generic and applicable to other experimental test cases when the input uncertainty levels are known. The proposed coherent predictor outperformed the recent predictor functions by having a lower bias error, particularly in complex regions. The polynomial predictor showed maximum deviation from the ground truth data (i.e., bias error). To quantify the output uncertainty level of the proposed approach, we performed Monte Carlo simulations. The coherent predictor showed narrow output uncertainty distribution compared to the Wiener filter and third-order polynomial predictors. We also performed the 4D-PTV experiment of the wake far behind a smooth cylinder at a Reynolds number of $3900$. The predicted positions from the coherent predictor showed minimum deviation from the optimised positions compared with other predictor functions. It was found that the estimation error has a direct relation with Lagrangian acceleration. We found that the flow topology highly impacts the estimation error. In the synthetic case, the prediction error carries visible signatures of three main topologies, the leading-edge boundary layer, the sideward shears, and the vortex formation zone. In the experiment, we capture further downstream structures where braid vortices and von Kármán vortex streets exhibit. These structures are characterised by high acceleration and 3D directional motions. This independence in the cost function's behaviour might be attributed to the additional information provided by the coherent neighbours. In other words, coherent motions simplify the complexities associated with the prediction process by offering accurate direction and acceleration for Lagrangian trajectories.

The single-step accuracy improvements reported above have a direct consequence for trajectory reconstruction that goes beyond the per-step metrics. In sequential particle tracking, the expected tracklet length scales nonlinearly with the per-step correct-association probability, so even a moderate improvement in single-step prediction accumulates into longer unbroken trajectories. The 25--30\% additional one-step error reduction when secondary constraints are included on top of the primary predictor (\S\,\ref{sec:p42b}) therefore translates to considerably longer and less fragmented tracks than what the polynomial baseline can provide. As shown in \S\,\ref{sec:p42}, the advantage is sustained across all acceleration magnitudes, so the longer tracks are not biased towards low-acceleration regions. The coherent predictor therefore improves accuracy at each step and extends the Lagrangian analysis into the high-acceleration regimes where current methods fail.

Coherent prediction supports several downstream Lagrangian analyses in turbulence. It can contribute to studies requiring accurate, complex and long trajectory reconstructions, such as studies in Lagrangian physics and statistics \citep[see, e.g.,][]{Viggiano2021LagrangianJet}. Recent studies have focused on converting Lagrangian trajectories into Eulerian vector fields using data assimilation or interpolation techniques \citep{Jeon2022FineVIC+,Godbersen2020FunctionalTracking}. These studies typically search for a method to interpolate groups of trajectories within a cell to estimate Eulerian velocity. However, not all neighbour trajectories are coherent and interpolating them could lead to an over-smooth reconstruction through the transport barrier. The proposed approach can improve such research by interpolating only coherent velocity fields. In addition, coherent prediction can also be used to advect synthetic particles in Eulerian numerical simulations with sparse temporal resolution \citep[see, e.g.,][]{vanSebille2018LagrangianPractices}. Knowing that tracer particles respect advective transport barriers, interested readers can also implement other LCS stretching diagnostics \citep{Haller2023TransportData}. Finally, local segmentation of Lagrangian trajectories is a topic worth exploring in future research, to switch from a single particle treatment to groups of coherent particles. The structure of the proposed cost function parallels regularised loss functions in supervised learning. The history term serves as the data fidelity loss, and the coherent velocity and acceleration terms serve as physics-based regularisation penalties, similar to the physics-informed constraints used in neural network training \citep{Raissi2019PINN}. The weighting scheme that assigns importance to coherent neighbours based on their rate of separation and distance is functionally an attention mechanism, where the relevance of each neighbour is determined by physical criteria rather than learned parameters. We showed that the optimal regularisation weights can be modelled from the measurement uncertainties alone, which removes the need for expensive hyperparameter search when the noise characteristics of the data are known. Recent advances in deep-learning-based Lagrangian prediction \citep{Liang2024DeepLag} and physics-informed Lagrangian subgrid modelling \citep{Tian2023LagrangianLES} provide data-driven alternatives. Appendix~\ref{appC} shows that the data-plus-physics structure of the temporal-collocation cost function transfers to a sinusoidal representation network, giving lower per-particle velocity error than the polynomial P\,+\,S predictor and acceleration error comparable to it.



\par \textbf{Acknowledgements} {The authors gratefully acknowledge Sylvain Laizet from Imperial College London for the 3D wake DNS simulation. We also thank the LPT challenge committee, particularly Andrea Sciacchitano from TU Delft University, for the time-resolved LPT challenge results. Special thanks go to Philippe Georgeault and Johan Carlier, both from INRAE. Philippe Georgeault contributed to the experimental setup, and Johan Carlier provided the 2D DNS simulation data from the FLUID project.}

\par \textbf{Funding} {For the 3D wake DNS simulation, this work was supported by the EPSRC, providing computational time on the UK supercomputing facility ARCHER2 via the UK Turbulence Consortium (grant number EP/R029326/1). Additionally, the 2D HIT DNS simulation was funded by the FLUID European project under grant agreement ID 513663.}

\par \textbf{Declaration of interests}. The authors report no conflict of interest.

\par \textbf{Data availability statement}. The numerical 3D data supporting the findings of this study are openly available in the Data INRAE repository at \href{https://doi.org/10.15454/GLNRHK}{https://doi.org/10.15454/GLNRHK}. The experimental data supporting the findings of this study are available from the corresponding author upon reasonable request.

\par \textbf{Code availability statement}. A reference Python implementation of the coherent predictor and the SIREN-PINN variant \citep{coherent_predictor_2026} is openly available under the MIT licence and archived on Software Heritage through HAL at \url{https://hal.inrae.fr/hal-05609886}. The development version is hosted on GitHub at \url{https://github.com/AliRKhojasteh/coherent-predictor}. The repository has three runnable notebooks that reproduce the main results. The first one is a core predictor notebook for the polynomial, primary, and primary plus secondary variants of \S~\ref{sec:p4}. The second one is a SIREN-PINN notebook for the temporal-collocation predictor of Appendix~\ref{appC}. The third one is an FTLE evaluation notebook for the integration time and weighting analyses of \S~\ref{sec:p221}. The notebooks open directly in Google Colab and run on a small subset of the 2D HIT trajectories included with the repository.

\par \textbf{Author ORCIDs}. {AR. Khojasteh, \href{https://orcid.org/0000-0002-3545-8391}{https://orcid.org/0000-0002-3545-8391}; D. Heitz, \href{https://orcid.org/0000-0001-6295-2822}{https://orcid.org/0000-0001-6295-2822}}

\par \textbf{Author contributions}. {AR. Khojasteh, conceptualisation, methodology, computation, and authoring the original draft; D. Heitz, conceptualisation, methodology, supervision, reviewing, and editing.}

\appendix
\section{}\label{appA}
\label{Appendix_A}

This appendix describes how to minimise the non-dimensional cost function of the coherent predictor. We assume that the solution is a third-order polynomial with unknown $a_i$ coefficients as, 

\begin{equation}
\begin{aligned}
\sum_{i=1}^{n} \left(\sum_{j=0}^{\ell} a_{j} \cdot t_i^{j}\right) = \sum_{i=1}^{n} y_{i}.
\end{aligned}
\label{EqC1} 
\end{equation}

We can expand the solution coefficients into the cost function as follows,
\begin{equation}
\begin{aligned}
\mathcal{J}'=\frac{1}{n} \sum_{i=1}^{n}\left[\left(a_{0}+a_{1} t_{i}+a_{2} t_{i}^{2}+a_{3} t_{i}^{3}\right)-\mathbf{y}_{i}'\right]^{2}\\
+\alpha_{1}'\left[\left(a_{1}+2 a_{2} t_{n}+3 a_{3} t_{n}^{2}\right)-\dot{\mathbf{y}}_{{\rm c},n}'\right]^{2}\\
+\alpha_{2}'\left[\left(2 a_{2}+6 a_{3} t_{n}\right)-\ddot{\mathbf{y}}_{{\rm c},n}'\right]^{2}.
\end{aligned}
\label{EqC2}       
\end{equation}

If all three terms of~\eqref{Eq:p28} have exact weights, the linear solution of the prediction function is in three sets of equations as

\begin{equation}
\begin{aligned}
\left\{
    \begin{array}{l}
        \sum_{j=0}^{\ell}{a_{j}.t_i^{j}}=\mathbf{y}_{i}',\qquad i=1,\cdots,n, \\
        \\
        \sum_{j=1}^{\ell}{a_{j}.(j).t_n^{j-1}}=\dot{\mathbf{y}}_{{\rm c},n}', \\
        \\
        \sum_{j=2}^{\ell}{a_{j}.(j).(j-1).t_n^{j-2}}=\ddot{\mathbf{y}}_{{\rm c},n}'.
    \end{array}
\right.
\end{aligned}
\label{Eq67}       
\end{equation}

The first sets are rows of particle position history for $n$ time step observations. The second and third sets are additional coherency based constraints. Therefore, the solution for the cost function in~\eqref{Eq67} is not only smooth on the history of the target particle but also satisfies local coherent dynamics of the flow. The minimised solution should satisfy the partial derivative of the cost function over $a_i$ coefficients as,
\begin{equation}
\begin{aligned}
\frac{\partial \mathcal{J}'}{\partial a_i}=0\end{aligned}.
\label{EqC3}       
\end{equation}
Partials derivate of $a_{0}$ is given by,

\begin{equation}
\begin{aligned}
\frac{\partial \mathcal{J}'}{\partial a_{0}}=\frac{2}{n} \sum_{i=1}^{n}\left[\left(a_{0}+a_{1} t_{i}+a_{2} t_{i}^{2}+a_{3} t_{i}^{3}\right)-\mathbf{y}_{i}'\right]+0+0 \\
\sum_{i=1}^{n}\left[a_{0}+a_{1} t_{i}+a_{2} t_{i}^{2}+a_{3} t_{i}^{3}\right]=\sum_{i=1}^{n}\mathbf{y}_{i}'.
\end{aligned}
\label{EqC4}       
\end{equation}

Partial derivative of $a_1$ is,
\begin{equation}
\begin{aligned}
\frac{\partial \mathcal{J}'}{\partial a_{1}}=\frac{2}{n} \sum_{i=1}^{n}\left[\left(a_{0}+a_{1} t_{i}+a_{2} t_{i}^{2}+a_{3} t_{i}^{3}\right)-\mathbf{y}_{i}'\right] \cdot t_{i}\\
+2 \alpha_{1}'\cdot\left[\left(a_{1}+2 a_{2} t_{n}+3 a_{3} t_{n}^{2}\right)-\dot{\mathbf{y}}_{{\rm c},n}'\right]+0.
\end{aligned}
\label{EqC5}       
\end{equation}

Partial derivative of $a_2$ is, 

\begin{equation}
\begin{aligned}
\frac{\partial \mathcal{J}'}{\partial a_{2}}=\frac{2}{n} \sum_{i=1}^{n}\left[\left(a_{0}+a_{1} t_{i}+a_{2} t_{i}^{2}+a_{3} t_{i}^{3}\right)-\mathbf{y}_{i}'\right] \cdot t_{i}^{2}\\
+2 \alpha_{1}'\cdot\left[\left(a_{1}+2 a_{2} t_{n}+3 a_{3} t_{n}^{2}\right)-\dot{\mathbf{y}}_{{\rm c},n}'\right] \cdot\left(2 t_{n}\right)\\
+2 \alpha_{2}'\cdot\left[\left(2 a_{2}+6 a_{3} t_{n}\right)-\ddot{\mathbf{y}}_{{\rm c},n}'\right] \cdot(2).
\end{aligned}
\label{EqC6}       
\end{equation}

Partial derivative of $a_3$ is, 
\begin{equation}
\begin{aligned}
\frac{\partial \mathcal{J}'}{\partial a_{3}}=\frac{2}{n} \sum_{i=1}^{n}\left[\left(a_{0}+a_{1} t_{i}+a_{2} t_{i}^{2}+a_{3} t_{i}^{3}\right)-\mathbf{y}_{i}'\right] \cdot t_{i}^{3}\\
+2 \alpha_{1}'\cdot\left[\left(a_{1}+2 a_{2} t_{n}+3 a_{3} t_{n}^{2}\right)-\dot{\mathbf{y}}_{{\rm c},n}'\right] \cdot\left(3 t_{n}^{2}\right)\\
+2 \alpha_{2}'\cdot\left[\left(2 a_{2}+6 a_{3} t_{n}\right)-\ddot{\mathbf{y}}_{{\rm c},n}'\right] \cdot\left(6 t_{n}\right).
\end{aligned}
\label{EqC7}       
\end{equation}

To linearise the solution, we set each partial derivative to zero leading. A series of partial derivatives can be written in a matrix form as,  

\begin{equation}
\resizebox{\textwidth}{!}{$
\begin{aligned}
\left(\frac{1}{n} \sum_{i=1}^{n}\left[\begin{array}{cccc}
t_{i}^{6} & t_{i}^{5} & t_{i}^{4} & t_{i}^{3} \\
t_{i}^{5} & t_{i}^{4} & t_{i}^{3} & t_{i}^{2} \\
t_{i}^{4} & t_{i}^{3} & t_{i}^{2} & t_{i} \\
t_{i}^{3} & t_{i}^{2} & t_{i} & 1
\end{array}\right] + \alpha_{1}'\left[\begin{array}{cccc}
9 t_{n}^{4} & 6 t_{n}^{3} & 3 t_{n}^{2} & 0 \\
6 t_{n}^{3} & 4 t_{n}^{2} & 2 t_{n} & 0 \\
3 t_{n}^{2} & 2 t_{n} & 1 & 0 \\
0 & 0 & 0 & 0
\end{array}\right] + \alpha_{2}'\left[\begin{array}{cccc}
36 t_{n}^{2} & 12 t_{n} & 0 & 0 \\
12 t_{n} & 4 & 0 & 0 \\
0 & 0 & 0 & 0 \\
0 & 0 & 0 & 0
\end{array}\right]\right)\left[\begin{array}{l}
a_{3} \\
a_{2} \\
a_{1} \\
a_{0}
\end{array}\right] \\
=\left(\frac{1}{n} \sum_{i=1}^{n}\left[\begin{array}{l}
\mathbf{y}_{i}' t_{i}^{3} \\
\mathbf{y}_{i}' t_{i}^{2} \\
\mathbf{y}_{i}' t_{i} \\
\mathbf{y}_{i}'
\end{array}\right]+\alpha_{1}'\left[\begin{array}{l}
3 \dot{\mathbf{y}}_{{\rm c},n}' t_{n}^{2} \\
2 \dot{\mathbf{y}}_{{\rm c},n}' t_{n} \\
\dot{\mathbf{y}}_{{\rm c},n}' \\
0
\end{array}\right]+\alpha_{2}'\left[\begin{array}{l}
6 \ddot{\mathbf{y}}_{{\rm c},n}' t_{n} \\
2 \ddot{\mathbf{y}}_{{\rm c},n}' \\
0 \\
0
\end{array}\right]\right).
\end{aligned}
$}
\label{EqC8}       
\end{equation}

On the other hand,~\eqref{EqC8} is in the form of $AX=B$. So $a_i$ solutions can be solved by $X=A^{-1}B$.

\section{Robustness of the coherent predictor in the 3D cylinder wake}\label{appB}

This appendix presents the detailed robustness assessment summarised in \S~\ref{sec:p4}. The test case is the DNS $Re = 3900$ cylinder wake with $10\%$ positional noise unless otherwise stated.

\subsection{Seeding density in Kolmogorov units}

We investigated whether the coherent predictor's advantage depends on how densely the flow is seeded relative to its smallest scales. To answer this, we first computed the Kolmogorov length scale $\eta = (\nu^3/\varepsilon)^{1/4}$ for each test case (see table~\ref{tab:p1}). For the 3D cylinder wake at $Re = 3900$, the dissipation rate was obtained from the volume-averaged strain-rate tensor of the DNS Eulerian field, yielding $\eta = 0.0058\,D$. The mean nearest-neighbour distance $\langle d_{\rm nn}\rangle$ ranges from $1.8\,\eta$ (2D HIT, densest seeding) to $10.4\,\eta$ (experimental 4D-PTV, sparsest), confirming that the test cases span a physically meaningful range of particle densities.

To probe the predictor behaviour over an even wider range, we subsampled the $Re = 3900$ DNS trajectories at nine density fractions from $100\%$ down to $0.4\%$, spanning $\langle d_{\rm nn}\rangle/\eta$ from $8$ to $50$. At each fraction, we ran all four predictors, Polynomial ($\ell = 3$), Wiener ($M = 2$), Coherent P (primary only), and Coherent P\,+\,S (pooled primary and secondary), using the same cost function and optimised parameters as in \S~\ref{sec:p41}, with $10\%$ positional noise.

Figure~\ref{fig:p_robustness} shows the normalised RMS position-prediction error as a function of $\langle d_{\rm nn}\rangle/\eta$. The hierarchy Polynomial $>$ Wiener $\approx$ Coherent~P $>$ Coherent~P\,+\,S holds across the entire range, with no crossover or sudden failure. The P\,+\,S error reduction relative to the polynomial baseline decreases gradually from $52\%$ at $8\,\eta$ to $44\%$ at $50\,\eta$. At high seeding density ($\langle d_{\rm nn}\rangle/\eta \lesssim 10$), the Wiener filter converges to the P\,+\,S performance, which is expected since the AR filter performs well when neighbours are abundant and the trajectory is well-sampled. As density decreases, the coherent method's advantage grows, confirming that physics-based constraints are most valuable when the data become sparse.

\subsection{Temporal resolution and noise level}

The 2D HIT sensitivity analysis in \S~\ref{sec:p42} varied the temporal resolution and noise level in a two-dimensional flow. We now repeat this analysis on the 3D cylinder wake to verify that the conclusions carry over to the more complex, application-relevant configuration.

The DNS trajectories ($30\,000$ particles, $100$ time steps at $\Delta t_{\rm base} = 10\,\Delta t_{\rm DNS} = 0.0075\,D/U$) were subsampled at temporal strides $\{1,2,3,4,5\} \times \Delta t_{\rm base}$. At each stride, $10\%$ Gaussian noise relative to the characteristic displacement at that stride was added, and all three predictors, Polynomial ($\ell = 3$, $k = 5$), Wiener ($M = 2$, $k = 5$), and Coherent P\,+\,S, were evaluated over $800$ particles. The left inset of figure~\ref{fig:p_robustness} shows the normalised RMS position error as a function of the temporal stride. The polynomial and Wiener errors grow with coarser resolution, with the Wiener filter degrading by $57\%$ from stride~$1$ to stride~$5$. The coherent predictor, on the other hand, remains nearly flat, with the normalised error increasing from $0.296$ to $0.308$, a change of only $4\%$. This stability follows from the coherent cost function imposing velocity and acceleration constraints from spatial neighbours, which are independent of the temporal sampling rate.

In a separate test at the base stride, the noise level was varied from $5\%$ to $25\%$ of the characteristic displacement (right inset of figure~\ref{fig:p_robustness}). At the lowest noise level ($5\%$), the Wiener filter slightly outperforms the coherent predictor (normalised error $0.185$ versus $0.251$), consistent with the expectation that a linear autoregressive filter performs well when the signal-to-noise ratio is high. By $10\%$ noise, however, the coherent predictor overtakes the Wiener filter, and the gap widens steadily. At $25\%$ noise the coherent error is $0.52$ versus $1.22$ for the Wiener filter and $2.15$ for the polynomial. This crossover shows that the coherent predictor's advantage is largest in the noise-dominated regime that is characteristic of real particle tracking velocimetry experiments, where the spatial coherence constraints regularise the prediction against measurement noise.

\subsection{Summary}

\begin{figure}
\centering
  \includegraphics[width=0.7\textwidth]{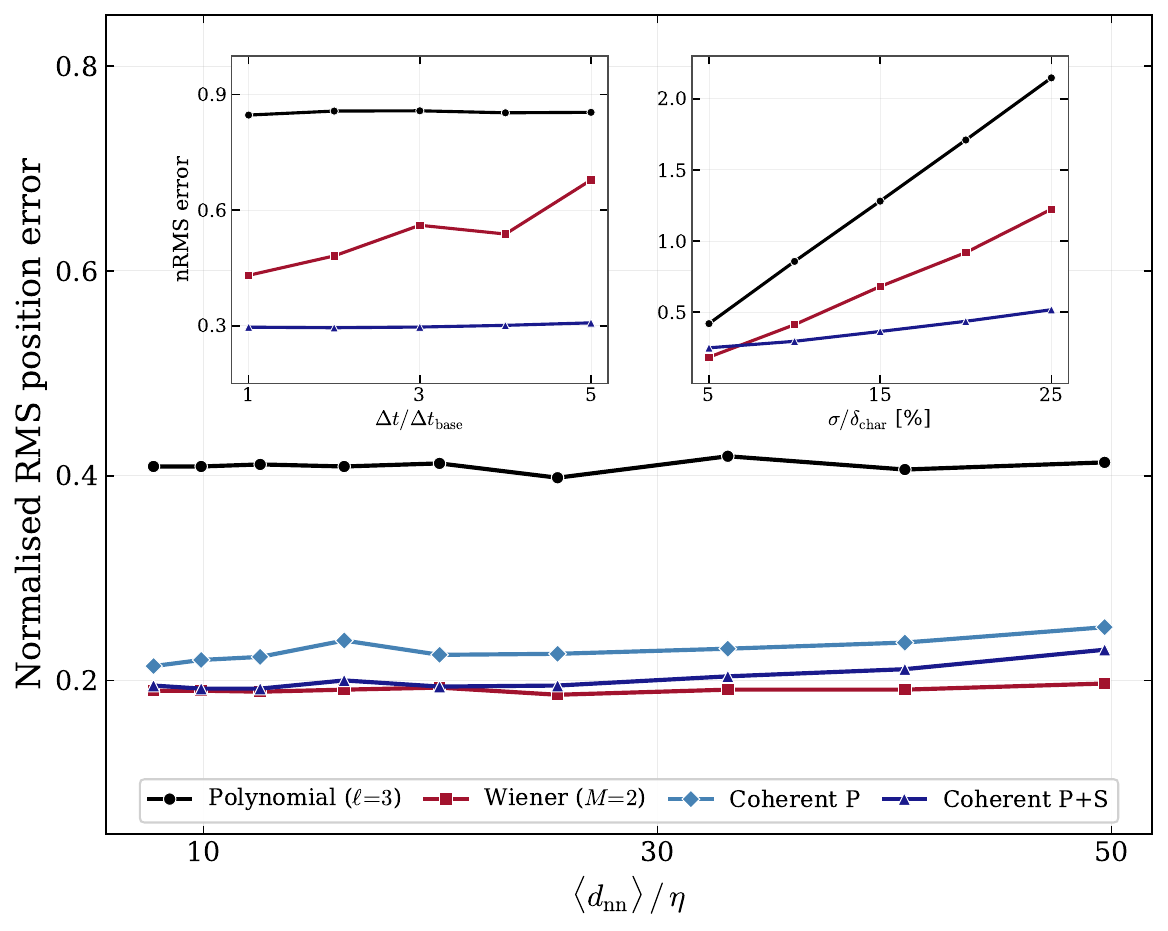}
\caption{Robustness of predictor performance across operating conditions, DNS $Re = 3900$ cylinder wake with $10\%$ positional noise unless otherwise stated. Normalised RMS position-prediction error versus mean inter-particle spacing $\langle d_{\rm nn}\rangle/\eta$, obtained by subsampling the particle set at nine density fractions. Left inset: normalised RMS error versus temporal stride $\Delta t/\Delta t_{\rm base}$ (noise fixed at $10\%$ of $\delta_{\rm char}$ at each stride, $800$ particles). Right inset: normalised RMS error versus noise level $\sigma/\delta_{\rm char}$ (stride fixed at $1\times$, $800$ particles).}
\label{fig:p_robustness}
\end{figure}

Taken together, the three analyses (figure~\ref{fig:p_robustness}) establish that the coherent predictor maintains its advantage over both polynomial and Wiener baselines across a wide range of operating conditions in the 3D cylinder wake, including seeding densities from $8\,\eta$ to $50\,\eta$, temporal strides up to $5\times$ the base resolution, and noise levels up to $25\%$ of the characteristic displacement. The predictor hierarchy is preserved throughout, with no abrupt failure threshold. The coherent method's advantage over the Wiener filter grows as conditions degrade (sparser seeding, coarser resolution, higher noise), which is the regime of practical 4D-PTV experiments where physics-based constraints are most valuable.

\section{Physics-informed neural network with temporal collocation}\label{appC}

This appendix examines whether the prediction gains reported in \S~\ref{sec:p4} originate from the polynomial basis or from the collocation cost function independently of the trajectory representation. We replace the polynomial by a sinusoidal representation network (SIREN) \citep{Sitzmann2020SIREN} equipped with a fixed Fourier feature embedding \citep{Tancik2020Fourier}, while retaining the same data-plus-physics structure used in the main body of the paper and extending it through temporal collocation of the coherent constraints over the full history window together with explicit secondary terms at $\tau_{n+1}$. The test cases and the primary coherent-neighbour identification are identical to those used in \S~\ref{sec:p4}, so that any difference in prediction error is attributable to the basis function and to the collocation structure of the cost function rather than to the problem set-up. The secondary neighbour set and the primary acceleration weight that accompany the collocation structure are specified in \S~\ref{appC1} and \S~\ref{appC3}.

\subsection{Formulation}\label{appC1}

Starting from the primary coherent cost function of \eqref{Eq:p28}, the physics-informed neural network (PINN) variant retains the same data-plus-physics structure and extends it in two directions, consistent with the P\,+\,S formulation of \S~\ref{sec:p42b}. First, the primary coherent constraints on velocity and acceleration are collocated at every history step $\tau_i$, rather than only at $\tau_n$. Second, secondary coherent constraints on position, velocity and acceleration are added at $\tau_{n+1}$. The trajectory is parametrised as a neural field $\mathbf{X}(\tau;\boldsymbol{\theta})$ with learnable parameters $\boldsymbol{\theta}$, and its derivatives are obtained by automatic differentiation rather than by analytic polynomial expressions. The cost function reads
\begin{equation}
\begin{aligned}
\mathcal{J}(\boldsymbol{\theta}) = \ &\frac{1}{k}\sum_{i=n-k+1}^{n}\left\|\mathbf{X}(\tau_i;\boldsymbol{\theta})-\mathbf{y}_i'\right\|^{2} \\
&+ \alpha_1'\sum_{i=n-k+1}^{n}\left\|\dot{\mathbf{X}}(\tau_i;\boldsymbol{\theta})-\dot{\mathbf{y}}_{{\rm c},i}'\right\|^{2} \\
&+ \alpha_2'\sum_{i=n-k+1}^{n}\left\|\ddot{\mathbf{X}}(\tau_i;\boldsymbol{\theta})-\ddot{\mathbf{y}}_{{\rm c},i}'\right\|^{2} \\
&+\beta_0'\left\|\mathbf{X}(\tau_{n+1};\boldsymbol{\theta})-\mathbf{y}_{\rm sec}'\right\|^{2} \\
&+\beta_1'\left\|\dot{\mathbf{X}}(\tau_{n+1};\boldsymbol{\theta})-\dot{\mathbf{y}}_{\rm sec}'\right\|^{2} \\
&+\beta_2'\left\|\ddot{\mathbf{X}}(\tau_{n+1};\boldsymbol{\theta})-\ddot{\mathbf{y}}_{\rm sec}'\right\|^{2},
\end{aligned}
\label{Eq:pinn_loss}
\end{equation}
where the first three terms enforce data fidelity and temporal collocation of the coherent velocity and acceleration at every history step $\tau_i$, and the last three terms impose the secondary coherent constraints on position, velocity and acceleration at $\tau_{n+1}$. The primary weights $\alpha_1'$ and $\alpha_2'$ follow the non-dimensional convention introduced in \eqref{Eq:p28}. The secondary weights $\beta_0'$, $\beta_1'$ and $\beta_2'$ are the dimensionless coefficients of the position, velocity and acceleration constraints at $\tau_{n+1}$, non-dimensionalised with the same velocity scale $U$ and length scale $D$ that set $\alpha_1'$ and $\alpha_2'$ in \eqref{Eq:p29}. The primary coherent acceleration targets entering \eqref{Eq:pinn_loss} are smoothed by a quadratic fit across the $k$ history points, which regularises the acceleration channel whose raw finite-difference estimate has a signal-to-noise ratio of approximately $0.17$. The primary coherent velocity targets retain their finite-difference form because their signal-to-noise ratio is roughly two orders of magnitude higher. The secondary velocity and acceleration targets at $\tau_{n+1}$ are obtained from a local five-point polynomial stencil applied to the noisy neighbour positions, so that the same noise regularisation is carried over to the secondary step at $\tau_{n+1}$. Unlike the phase-delayed FTLE identification of \S~\ref{sec:p221} used by the coherent predictor of the main body, the secondary neighbour set in the present PINN variant is taken as the complement of the primary coherent neighbours at $\tau_n$, so that the primary and secondary constraints draw from disjoint subsets of the particle population. Throughout this appendix, $\tau$ denotes the continuous time argument of the neural field $\mathbf{X}(\tau;\boldsymbol{\theta})$ and should not be confused with the phase-delay index of \S~\ref{sec:p221}. The latter does not appear in \eqref{Eq:pinn_loss}.

\subsection{Network architecture and training}\label{appC2}

\begin{figure}
\centering
  \includegraphics[width=0.85\textwidth]{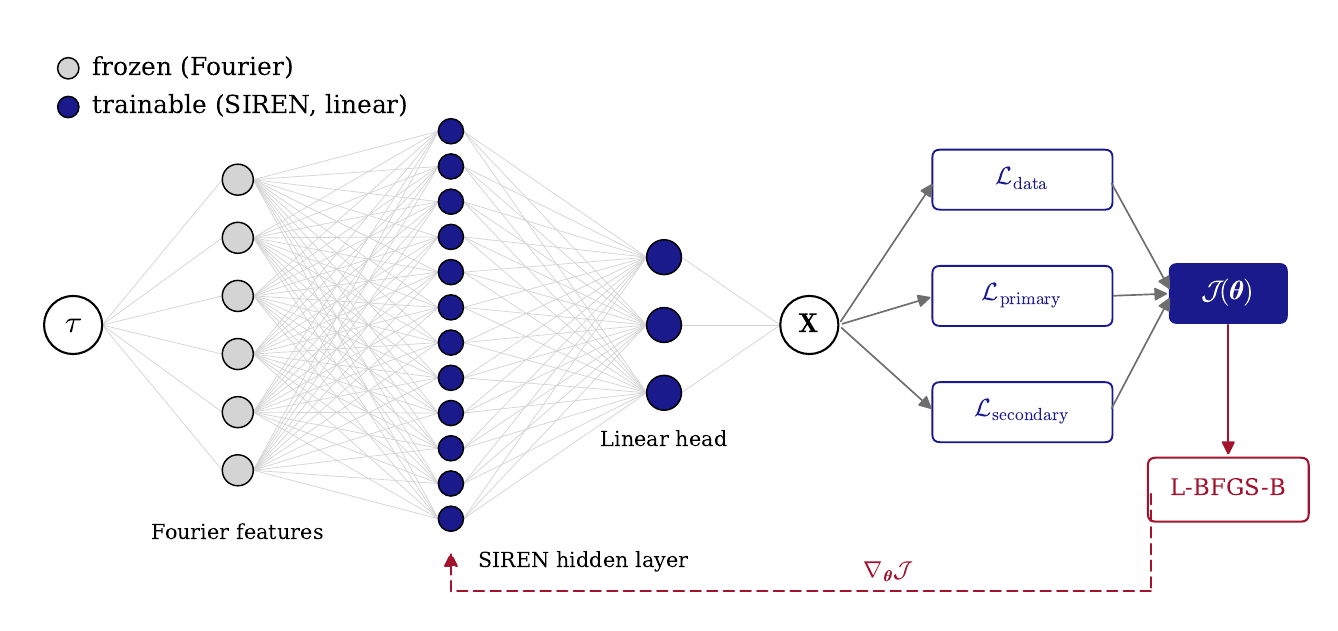}
\caption{SIREN network architecture. The scalar time $\tau$ is lifted by a fixed Fourier feature block ($K = 3$, shown in grey to mark the frozen parameters), passes through a SIREN hidden layer of $12$ sinusoidal neurons ($\omega_0 = 0.5$) and a linear output head (shown in navy to mark the trainable parameters), and yields the trajectory $\mathbf{X}(\tau)$. The cost $\mathcal{J}(\boldsymbol{\theta})$ is passed to the L-BFGS-B optimiser (solid red arrow), which uses the automatic-differentiation gradient $\nabla_{\boldsymbol{\theta}}\mathcal{J}$ to update the trainable parameters (dashed red arrow), closing the training loop.}
\label{fig:pinn_arch}
\end{figure}

The scalar time $\tau$ is lifted by a fixed Fourier feature block ($K = 3$, $6$ units, no trainable weights) with a base period of four times the history window to prevent aliasing. This vector feeds a single SIREN hidden layer of $12$ sinusoidal neurons ($\omega_0 = 0.5$) followed by a linear output head, giving $110$ trainable parameters in 2D and $123$ in 3D (figure~\ref{fig:pinn_arch}). At $\omega_0 = 0.5$ the sinusoidal activation operates close to its linear regime over the input range produced by the Fourier feature block, so the network behaves as a mildly nonlinear small MLP with the SIREN initialisation rather than as the high-frequency representation network of \citet{Sitzmann2020SIREN}. The same architecture and hyperparameters are used for both the 2D and 3D test cases without any case-specific tuning. Each particle is trained independently. Per particle, the hidden-layer weights are kept at their small random initialisation and the linear output head is fitted in closed form by least squares to a polynomial fit of the coherent constraints at the $k$ history points. The optimiser then reduces both the residual and the basis-mismatch error, because a single sinusoidal layer with a fixed Fourier lift cannot reproduce the polynomial target exactly outside the history points. The optimisation uses the limited-memory Broyden--Fletcher--Goldfarb--Shanno algorithm with box constraints (L-BFGS-B) of \citet{ByrdLuNocedalZhu1995LBFGSB} with automatic-differentiation gradients, capped at $100$ iterations with tolerances $f_{\rm tol}=10^{-15}$ and $g_{\rm tol}=10^{-10}$. The primary acceleration targets are smoothed by a local quadratic fit across the $k$ history points, while the primary velocity targets are kept as finite differences of the noisy positions. The secondary velocity and acceleration targets at $\tau_{n+1}$ are obtained from a local five-point polynomial stencil, as described in \S~\ref{appC1}. An ablation on a neighbouring variant ($\omega_0 \approx 0.75$, $150$ L-BFGS-B iterations, all other settings fixed) shows that activating the same quadratic smoothing on the primary acceleration targets raises the 3D acceleration error reduction from $63.8\%$ to $74.3\%$. Target smoothing is therefore the dominant factor for the acceleration channel. A reference implementation of this network and its training loop is provided in the \texttt{02\_siren\_pinn.ipynb} notebook of the companion repository (see Code availability statement).

\subsection{Results}\label{appC3}

We evaluated the SIREN-PINN coherent predictor on a sample of $500$ particles in the 2D HIT case and $3000$ particles in the 3D cylinder wake at $Re = 3900$, drawn from the full DNS field. Only the particles whose local frame contains at least one coherent neighbour after the FTLE percentile filter of \S~\ref{sec:p221} are retained for evaluation. The same retained set is used for the polynomial, P\,+\,S and SIREN-PINN predictors, so the per-particle comparison is paired and the set is independent of network convergence. The noise level, the history length $k$ and the primary FTLE-based coherent neighbour identification match those of the polynomial P\,+\,S predictor in \S~\ref{sec:p42b}. The secondary neighbour set at $\tau_{n+1}$ is taken as the complement of the primary coherent neighbours at $\tau_n$, as described in \S~\ref{appC1}. The dimensionless weights are $\alpha_1' = 0.5$ and $\alpha_2' = 15$ for the primary terms and $\beta_0' = 0.1$, $\beta_1' = 0.5$, $\beta_2' = 5$ for the secondary terms. The primary sums in \eqref{Eq:pinn_loss} run over all $k$ history points and are not divided by $k$, so the single-step equivalents against the polynomial P\,+\,S predictor of \S~\ref{sec:p42b} are $k\alpha_1' = 3.5$ on velocity and $k\alpha_2' = 105$ on acceleration, to be compared with $\alpha_1' = 0.1$ and $\alpha_2' = 20$ in that section. The PINN therefore applies a stronger effective penalty on both channels. The figure of merit reported here is the per-particle root mean square error averaged over the valid subset of trajectories, rather than the pooled single-step error used in \S~\ref{sec:p42b}, so the headline percentages are not directly comparable between the two sections.

\begin{figure}
\centering
  \includegraphics[width=1\textwidth]{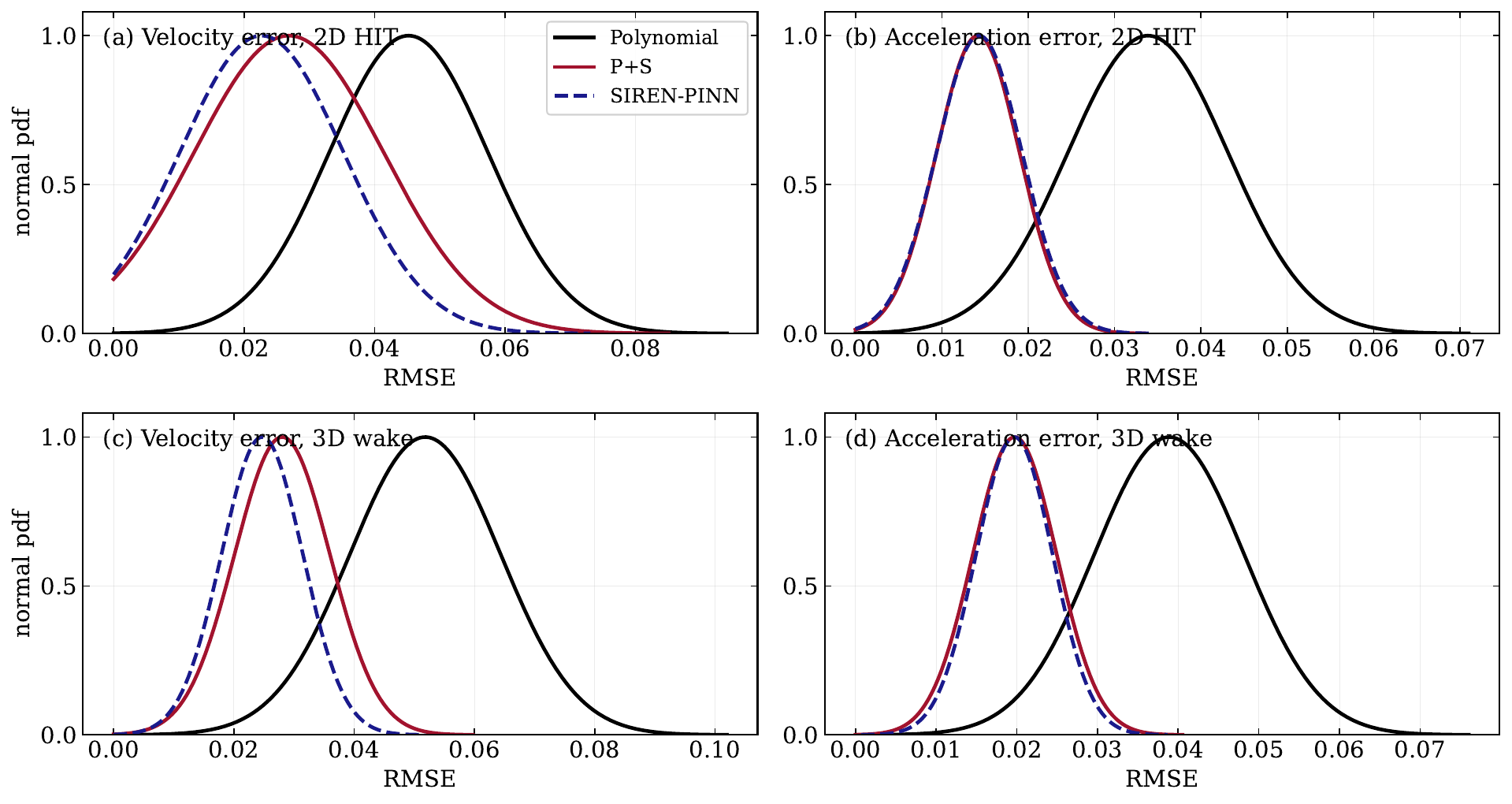}
\caption{Normal probability density function of the per-particle RMSE for the Polynomial, P\,+\,S, and SIREN-PINN predictors. Each curve is a Gaussian fit to the RMSE distribution, normalised so that the peak equals unity. (a,c) Velocity error in 2D HIT and 3D wake. (b,d) Acceleration error in 2D HIT and 3D wake. A narrower, left-shifted distribution indicates lower and more concentrated prediction errors.}
\label{fig:pinn_results}
\end{figure}

Figure~\ref{fig:pinn_results} shows the normal probability density of the per-particle root-mean-square error for the three predictors. On velocity (figure~\ref{fig:pinn_results}a,c), the SIREN-PINN distribution is visibly narrower and shifted towards lower errors compared with P\,+\,S, confirming that the improvement is systematic across the particle population. The mean velocity error reduction relative to the polynomial baseline is $69.3\%$ in 2D and $76.9\%$ in 3D for the SIREN-PINN, compared with $57.4\%$ and $70.1\%$ for P\,+\,S. On acceleration (figure~\ref{fig:pinn_results}b,d), the P\,+\,S and SIREN-PINN distributions overlap, indicating comparable performance. The mean acceleration error reductions are $81.4\%$ and $81.8\%$ in 2D and $74.3\%$ and $73.9\%$ in 3D for SIREN-PINN and P\,+\,S respectively. A per-particle comparison confirms that the SIREN-PINN yields lower velocity error for $81\%$ of particles in 2D and $79\%$ in 3D, while on 3D acceleration the SIREN-PINN produces lower error for $50\%$ of particles. On 2D acceleration the per-particle comparison is close to even at $48\%$, consistent with the statistical tie reported above.

\subsection{Discussion}\label{appC4}

The main finding is that the data-plus-physics structure of the temporal-collocation cost function transfers to a neural-field representation, giving lower per-particle velocity error than the P\,+\,S predictor of \S~\ref{sec:p42b} and acceleration error comparable to it. This indicates that the data-plus-physics structure of the cost function carries across trajectory representations, and the additional velocity gain observed with the SIREN-PINN confirms that the cost structure can be paired with representations beyond the polynomial family without compromising the acceleration channel. The structure of the cost function, where coherent velocity and acceleration are enforced at every history step rather than only at the endpoints, is conceptually related to the collocation of governing-equation residuals in physics-informed neural networks \citep{Raissi2019PINN}. In our formulation, the coherent fields from neighbouring particles play the role of soft physical constraints rather than known governing equations.

The smoothing applied to the acceleration targets, a quadratic fit across the full history window for the primary channel and a local five-point polynomial stencil for the secondary channel, suppresses the amplification of measurement noise on the finite-difference estimate and is the dominant noise-mitigation ingredient in the neural variant. The optimisation is budgeted at $100$ L-BFGS-B iterations with tight tolerances. Under this budget, the warm-started network converges on the coherent motion without observable overfitting on the test cases reported here. The transferable ingredient is the data-plus-physics structure of the cost function itself, which already carries the bulk of the improvement in the polynomial variant of \S~\ref{sec:p42b}.

\bibliographystyle{jfm}
\bibliography{references.bib}

\end{document}